\documentclass[10pt,twocolumn,superscriptaddress,amssymb,amsmath,aps,pra]{revtex4-2}
\usepackage{graphicx}
\usepackage{array}

\usepackage[normalem]{ulem}
\usepackage{color}

\begin{document}


\title{Covariant theory of light in a dispersive medium}
\date{August 13, 2021}
\author{Mikko Partanen}
\affiliation{Photonics Group, Department of Electronics and Nanoengineering, Aalto University, P.O. Box 13500, 00076 Aalto, Finland}
\author{Jukka Tulkki}
\affiliation{Engineered Nanosystems Group, School of Science, Aalto University, P.O. Box 12200, 00076 Aalto, Finland}

\begin{abstract}
The relativistic theory of the time- and position-dependent energy and momentum densities of light in glasses and other low-loss dispersive media, where different wavelengths of light propagate at different phase velocities, has remained a largely unsolved challenge until now. This is astonishing in view of the excellent overall theoretical understanding of Maxwell's equations and the abundant experimental measurements of optical phenomena in dispersive media. The challenge is related to the complexity of the interference patterns of partial waves and to the coupling of the field and medium dynamics by the optical force on the medium atoms. In this work, we use the mass-polariton theory of light [Phys.~Rev.~A \textbf{96}, 063834 (2017)] to derive the stress-energy-momentum (SEM) tensors of the field and the dispersive medium. Our starting point, the fundamental local conservation laws of energy and momentum densities in classical field theory, is close to that of a recent theoretical work on light in dispersive media by Philbin [Phys.~Rev.~A \textbf{83}, 013823 (2011)], which, however, excludes the power-conversion and force density source terms describing the coupling between the field and the medium. In the general inertial frame, we present the SEM tensors in terms of Lorentz scalars, four-vectors, and field tensors that reflect in a transparent way the Lorentz covariance of the theory. The SEM tensors of the field and the medium are symmetric, form-invariant for all inertial observers, and in full accordance with the covariance principle of the special theory of relativity. When the power-conversion and force density source terms are accounted for, there is no need to introduce asymmetric SEM tensors or heuristic symmetrization procedures even for the field and medium subsystems. Therefore, asymmetric SEM tensors based on strictly classical energy and momentum densities, which have been studied in previous literature, do not account for all aspects in the field-medium coupling of electromagnetic waves. However, being based on classical field theory, the present work prompts for further groundwork on the classical limit of the quantum mechanical spin of light in a medium. The SEM tensor of the coupled field-medium state of light also has zero four-divergence. Therefore, light in a dispersive medium has a well-defined four-momentum and rest frame. The volume integrals of the total energy and momentum densities of light agree with the model of mass-polariton quasiparticles having a nonzero rest mass. The coupled field-medium state of light drives forward an atomic mass density wave, which makes the constant center-of-energy velocity law of an isolated system---a fundamental conservation law of nature---satisfied. This provides strong evidence for the consistency of the theory. The predictions of our work, such as the atomic mass density wave associated with light, are accessible to experiments.
\end{abstract}

\maketitle


\section{Introduction}

Recent theoretical and experimental research of the reflection and transmission of light in dielectrics gives well-founded evidence that optical forces associated with light lead to atomic displacements that subsequently give rise to theoretically predictable and experimentally observable acoustic waves \cite{Pozar2018,Partanen2017c}. The generation of acoustic waves can be understood within the mass-polariton (MP) theory of light \cite{Partanen2017c,Partanen2019a,Partanen2019b,Partanen2017e,Partanen2018a,Partanen2018b}, which describes light in a dielectric as a coupled state of the field and the medium. The MP theory shows that atomic displacements resulting from the optical force form, in a bulk dielectric, an atomic mass density wave (MDW), which moves forward with light. The acoustic waves are generated when the atomic displacements related to the MDW are relaxed by elastic forces. When material interfaces are present, there is a separate optical-force-based displacement of atoms, which also generates acoustic waves that superimpose with those generated by the MDW.

Many previous theoretical works acknowledge that part of the total momentum of light in a dielectric is carried by the medium atoms \cite{Barnett2010b,Barnett2010a,Milonni2010,Hinds2009,Pfeifer2007,Brevik1979,Brevik2018b,Saldanha2017}. However, in distinction to the MP theory, they all neglect the position- and time-dependent \emph{displacements} of atoms by the optical force and, in particular, the related shift of the atomic \emph{rest energy} by the MDW. For common dielectric materials, this transferred rest energy is of the same order of magnitude as the energy of the field. Any theory of light that does not account for this transfer of rest energy cannot be in accordance with the principles of the special theory of relativity. Not accounting for the correct relativistic energy of the medium also breaks the constant center of energy velocity law of an isolated system (consisting of a light pulse and a medium block), and thus, violates the fundamental conservation laws of nature \cite{Partanen2017c,Partanen2019a}.

In addition to its unambiguous mathematical beauty, the MP theory of the coupled field-medium state of light has a simple phenomenological description and its predictions are directly accessible by experiments. The position- and time-dependent electromagnetic field associated with light first polarizes the medium atoms. Subsequently, these induced atomic dipoles are coupled to the electromagnetic field of light itself. As a result, there is an optical force on the atomic dipoles that alternatively accelerates and decelerates them in such a way that there will be a net displacement of atoms in the direction of the propagation of light.  A highly surprising feature of the MP theory of light is that, even though the atomic velocities resulting from the optical force are extremely small, the MDW  associated with the MP state propagates at the group velocity of light.

For appropriate combinations of the intensity and length of light pulses, the atomic displacements can be fractions of nanometers. The ultimate test of the MP theory would be the measurement of this predicted shift of atoms. The atomic displacements are real, and their measurement should be feasible using present photonic instruments. These measurements also provide a complementary approach to experimental studies of optical forces and the Abraham-Minkowski controversy \cite{Astrath2014,Ashkin1973,Casner2001,Pozar2018,Choi2017,Jones1954,Jones1978,Walker1975,She2008,Zhang2015,Campbell2005,Sapiro2009,Partanen2021a,Partanen2020b}. As shown in our recent preliminary work on the MP theory of light for dispersive media \cite{Partanen2017e}, the predictions of the MP theory for the momentum of light are in excellent agreement with the high-precision measurements of Jones \emph{et al.} \cite{Jones1978} for dispersive media.

In our previous works \cite{Partanen2017c,Partanen2019a,Partanen2019b}, we have presented the covariant stress-energy-momentum (SEM) tensor formulation of the MP theory of light for \emph{nondispersive} media. For \emph{dispersive} media, the MP theory has been previously applied by using a quasiparticle model and optomechanical continuum dynamics (OCD) simulations of the MDW \cite{Partanen2017e}. The goal of the present work is to develop a transparently covariant SEM tensor formulation of the MP theory of light for dispersive media. We also generalize the MP theory discussed in Refs.~\cite{Partanen2017c,Partanen2017e,Partanen2019a,Partanen2019b} to include the energy exchange between the field and the medium through the power-conversion source term. In the previous formulation of the MP theory, we accounted only for the optical-force-related momentum exchange between the field and the medium.

We base our derivations of the covariant SEM tensors of the field and the medium on the conservation laws of energy and momentum. Thus, our starting point is close to that of a recent theoretical work on light in dispersive media by Philbin \cite{Philbin2011}, which, however, excludes the coupling between the field and the medium. The form invariance of the conservation laws in different inertial frames is shown to lead to symmetric SEM tensors that are in full accordance with the covariance principle of the special theory of relativity. The splitting of the total MP SEM tensor of light into the field and the atomic MDW parts is also consistent with an unambiguous expression for the optical force in a dispersive medium. We also show how the total momentum of the coupled MP theory of light is split between the field and the medium. Therefore, the MP theory solves the Abraham-Minkowski controversy
\cite{Partanen2017c,Leonhardt2006a,Cho2010,Kemp2017,Kemp2015,Bliokh2017a,Bliokh2017b,Penfield1967,Kemp2011} for dispersive media. The SEM tensor formalism of the MP theory in dispersive media, presented in this work, is also fully consistent with the previously introduced MP quasiparticle model \cite{Partanen2017e}.

This paper is organized as follows: Section \ref{sec:foundations} describes theoretical foundations. Solution of Newton's equation of the medium in the presence of optical force density is described in Sec.~\ref{sec:Newtonsolution}. Section~\ref{sec:tensors} presents the formulation of the SEM tensors of the MP theory of light for dispersive media. Section \ref{sec:properties} investigates selected properties of the SEM tensors. Section \ref{sec:simulations} presents OCD simulations of the propagation of the MDW driven by a Gaussian light pulse. Section \ref{sec:observer} studies the energy and momentum of light measured by a general inertial observer moving with respect to the medium. The results are discussed and compared to previous theoretical works and experiments in Sec.~\ref{sec:discussion}. Finally, conclusions are drawn in Sec.~\ref{sec:conclusions}.

We have facilitated the reading of the paper with detailed description of the physical quantities and dynamical equations for both the electromagnetic field and the medium so that basic theoretical understanding of the electromagnetic fields, continuum dynamics, and the special theory of relativity would be enough for the understanding the key results of the paper. In the derivations, we have extensively used analytical mathematical models. Detailed proofs of the most important intermediate steps are provided in the Appendixes.

\section{\label{sec:foundations}Theoretical foundations}

\subsection{\label{sec:concepts}Concepts and approximations}

We assume that the medium is lossless, linear, and isotropic. In accordance with our previous works \cite{Partanen2019a,Partanen2019b}, we also assume that the medium is homogeneous having no material interfaces and limit our studies to light pulses having a narrow spectral range. Even with these restrictions, the theory covers a broad range of optical phenomena in solids, liquids, and gases relevant for photonics technologies. It is obvious that our theoretical formulations can also be extended to lossy, nonlinear, and anisotropic media. These extensions are, however, not discussed in the present work.

\subsubsection{Constitutive relations and inertial frames}

In linear and isotropic media, the constitutive relations of the electric field $\mathbf{E}$, magnetic field $\mathbf{H}$, electric flux density $\mathbf{D}$, and magnetic flux density $\mathbf{B}$ are given in the \emph{laboratory frame} (L frame) of the medium by the well-known frequency-domain expressions as $\mathbf{D}^\mathrm{(L)}=\varepsilon^\mathrm{(L)}(\omega)\mathbf{E}^\mathrm{(L)}$ and $\mathbf{B}^\mathrm{(L)}=\mu^\mathrm{(L)}(\omega)\mathbf{H}^\mathrm{(L)}$, where $\varepsilon^\mathrm{(L)}(\omega)$ and $\mu^\mathrm{(L)}(\omega)$ are the frequency-dependent permittivity and permeability of the medium in the L frame, and $\varepsilon_0$ and $\mu_0$ are the permittivity and permeability of vacuum. The L frame is defined as a frame where the medium atoms are at rest before the optical force starts to accelerate them. For lossless media studied in the present work, $\varepsilon^\mathrm{(L)}(\omega)$ and $\mu^\mathrm{(L)}(\omega)$ are real-valued. The frequency-dependent phase refractive index $n_\mathrm{p}^\mathrm{(L)}(\omega)$ is related to the permittivity and permeability of the medium as $\varepsilon^\mathrm{(L)}(\omega)\mu^\mathrm{(L)}(\omega)=\varepsilon_0\mu_0[n_\mathrm{p}^\mathrm{(L)}(\omega)]^2$. The group refractive index $n_\mathrm{g}^\mathrm{(L)}(\omega)$ is given by $n_\mathrm{g}^\mathrm{(L)}(\omega)=\frac{\partial}{\partial\omega}[\omega n_\mathrm{p}^\mathrm{(L)}(\omega)]$. Below, $n_\mathrm{p}^\mathrm{(L)}$ and $n_\mathrm{g}^\mathrm{(L)}$ without the angular frequency argument denote the phase and group refractive indices at the central angular frequency of the light pulse. All fields below are assumed to be given in the space-time domain. In a \emph{general inertial frame} (G frame), the constitutive relations between the field quantities are more complicated, but they are unambiguously tied to the relations in the L frame by the well-known Lorentz transformations of the fields described in Appendix \ref{apx:transformations}. The approximations of this section are effective throughout this work.

\subsubsection{\label{sec:elastic}Strain energy and elastic waves caused by the optical force}

By assuming that the medium is lossless, we neglect any optical absorption, but also any strain energies that are left in the medium by the light wave.  The strain energy left in the medium was found to be important in the description of the relaxation dynamics of the atomic displacements in the medium caused by the optical force  \cite{Partanen2017c,Partanen2017e,Partanen2018b}. However, the strain energy is vanishingly small in comparison with the field energy. The OCD simulations \cite{Partanen2017c} showed furthermore that, in realistic materials, the strain energy is also small in comparison to the kinetic energy given to the atoms by the field \cite{Partanen2017c}. It is evident that the description of strain energy could be added in the present analysis.

\subsubsection{\label{sec:kinetic1}Kinetic energy given to the medium by the optical force}

Since the strain energy of atoms resulting from the optical force is exceedingly small as discussed above, virtually all work done by the optical force goes to kinetic energy of the atoms. When the field strength is increasing, the work done by the optical force on the atoms is positive and the atoms are gaining kinetic energy. Correspondingly, when the field strength is decreasing, the work done by the optical force is negative and the kinetic energy of atoms is returned to the field.

Even for relatively large field strengths of the order of $10^{14}$ W/cm$^2$, the relative magnitude of the kinetic energy of atoms in comparison to the total field energy is in the L frame of the order of $10^{-8}$ for common dielectric materials \cite{Partanen2019b}. However, extending the MP theory to account for the exchange of energy between the field and the medium has fundamental theoretical interest in the study of covariance properties of light. Therefore, in the present work, we have explicitly included the kinetic energy in the SEM tensor of the medium and subtracted the work done by the optical force on the atoms from the SEM tensor of the field. This makes the SEM tensors of both the field and the medium symmetric in all inertial frames, which is a clear advantage regarding the completeness of the theory.

\subsection{\label{sec:summaryprevious}Generalized optical force density and the momentum density of the field}

As the starting point of the derivation, we need an expression of the optical force and the momentum density of the field. In previous literature, several forms of the optical force density have been suggested \cite{Brevik1979,Penfield1967,Shevchenko2010,Shevchenko2011,Brevik2018a,He2011,Mahdy2015}. Regarding the MP theory of light, in Ref.~\cite{Partanen2017e}, we found that the OCD simulations of the MP state of light for a Gaussian light pulse give a transferred mass of the MDW that is in agreement with the conservation of the center of energy velocity of an isolated system \emph{only} if we use for the optical force in the L frame an expression
\begin{equation}
 \mathbf{f}_\mathrm{opt}^\mathrm{(L)}=\dfrac{n_\mathrm{p}^\mathrm{(L)}n_\mathrm{g}^\mathrm{(L)}-1}{c^2}\dfrac{\partial}{\partial t}(\mathbf{E}^\mathrm{(L)}\times\mathbf{H}^\mathrm{(L)}).
 \label{eq:AbrahamforceL}
\end{equation}
We call this expression the generalized optical force. It differs from the conventional Abraham force by replacing the square of the refractive index of the nondispersive medium \cite{Brevik1979} by the product of the phase and group refractive indices of a dispersive medium. The narrow spectral range approximation discussed in Sec.~\ref{sec:concepts} applies to Eq.~\eqref{eq:AbrahamforceL}. The relation of Eq.~\eqref{eq:AbrahamforceL} to the MP quasiparticle model of a dispersive medium is briefly discussed in Appendix \ref{apx:quasiparticlemodel}. The derivation of Eq.~\eqref{eq:AbrahamforceL} from the principle of least action in the case of a nondispersive medium has been presented in Ref.~\cite{Partanen2019b}. Corresponding derivation is expected to be possible also in the present case of a dispersive medium. This derivation as well as the microscopic derivation of Eq.~\eqref{eq:AbrahamforceL} are left as interesting topics of future works.

For the momentum density of the field, we use an expression
\begin{equation}
 \mathbf{G}_\mathrm{field}^\mathrm{(L)}=\frac{\mathbf{E}^\mathrm{(L)}\times\mathbf{H}^\mathrm{(L)}}{c^2}.
 \label{eq:fieldmomentumdensityL}
\end{equation}
The Poynting vector $\mathbf{S}^\mathrm{(L)}=\mathbf{E}^\mathrm{(L)}\times\mathbf{H}^\mathrm{(L)}$, appearing in Eq.~\eqref{eq:fieldmomentumdensityL}, is continuous at material interfaces and equal to the electromagnetic energy flux in the L frame of a dispersive medium \cite{Jackson1999,Landau1984}. Thus, the momentum density of the field in Eq.~\eqref{eq:fieldmomentumdensityL} is also continuous at material interfaces. The momentum density of the field in Eq.~\eqref{eq:fieldmomentumdensityL} is also extensively used in the previous optics literature \cite{Jackson1999,Landau1984} and it can be proven to be in accordance with the MP quasiparticle model of Ref.~\cite{Partanen2017e}, see also Appendix \ref{apx:quasiparticlemodel}. Together with the approximations discussed in Sec.~\ref{sec:concepts}, Eqs.~\eqref{eq:AbrahamforceL} and \eqref{eq:fieldmomentumdensityL} can be considered as the starting points of the derivation of the SEM tensors below.

The expressions of the generalized optical force in Eq.~\eqref{eq:AbrahamforceL} and the momentum density of the field in Eq.~\eqref{eq:fieldmomentumdensityL} are both, within the approximations used in this work, inevitably interconnected with the conservation of the constant center-of-energy velocity of the field-medium system. They are also related to the existence of the rest frame of the coupled MP state of light. These aspects will be discussed in more detail in Secs.~\ref{sec:cev} and \ref{sec:observer}.

\subsection{Conservation laws and dynamical equations}

We start the derivation of the theory from the fundamental conservation laws of energy and momentum in the general inertial frame, the G frame. In this and all subsequent sections, the quantities not having a superscript indicating a special inertial frame, e.g., (L) for the L frame, are the quantities in the G frame. Using the energy density $W$, the momentum density $\mathbf{G}=(G^x,G^y,G^z)$, and the stress tensor $\boldsymbol{\mathcal{T}}$, with components $\mathcal{T}^{jk}$, where $j,k\in\{x,y,z\}$, the conservation laws of energy and momentum are written for an arbitrary continuous system as \cite{Penfield1967,Jackson1999,Landau1984}
\begin{equation}
 \frac{1}{c^2}\frac{\partial W}{\partial t}+\nabla\cdot\mathbf{G}=-\frac{\phi}{c^2},
 \label{eq:conservationphi}
\end{equation}
\begin{equation}
 \frac{\partial\mathbf{G}}{\partial t}+\nabla\cdot\boldsymbol{\mathcal{T}}=-\mathbf{f}.
 \label{eq:conservationf}
\end{equation}
Here $\mathbf{f}$ is the force density and $\phi$ is the power-conversion density. Both $\mathbf{f}$ and $\phi$ are zero for isolated systems, in which case Eq.~\eqref{eq:conservationphi} is equal to the conservation law of the total energy of the system and Eq.~\eqref{eq:conservationf} is equal to the conservation law of the total momentum. For interacting subsystems, the force density and power-conversion density are generally nonzero, which means that the subsystems are exchanging energy and momentum with each other.

\subsubsection{\label{sec:dynamicalfield}Dynamical equations of the field}

The dynamical equations of the fields are the Maxwell's equations, for which we use the standard presentation in terms of the fields $\mathbf{E}$, $\mathbf{D}$, $\mathbf{H}$, and $\mathbf{B}$ \cite{Jackson1999,Landau1984}. The energy density, momentum density, and stress tensor associated to the fields must obey the conservation laws in Eqs.~\eqref{eq:conservationphi} and \eqref{eq:conservationf}, written as
\begin{equation}
 \frac{1}{c^2}\frac{\partial W_\mathrm{field}}{\partial t}+\nabla\cdot\mathbf{G}_\mathrm{field}=-\frac{\phi_\mathrm{opt}}{c^2},
 \label{eq:continuityfieldG}
\end{equation}
\begin{equation}
 \frac{\partial\mathbf{G}_\mathrm{field}}{\partial t}+\nabla\cdot\boldsymbol{\mathcal{T}}_\mathrm{field}=-\mathbf{f}_\mathrm{opt}.
 \label{eq:momentumconservationfieldG}
\end{equation}
Here $\mathbf{f}_\mathrm{opt}$ is the generalized optical force in the G frame, obtained by the Lorentz transformation from its form in the L frame, given in Eq.~\eqref{eq:AbrahamforceL}, and $\phi_\mathrm{opt}$ is the related power-conversion density. Since the power conversion density changes only the kinetic energy of the medium, it must be equal to the dot product of the force density $\mathbf{f}_\mathrm{opt}$ and the atomic velocity $\mathbf{v}_\mathrm{a}$ \cite{Penfield1967,Landau1987} as
\begin{equation}
 \phi_\mathrm{opt}=\mathbf{f}_\mathrm{opt}\cdot\mathbf{v}_\mathrm{a}.
 \label{eq:powerconversion}
\end{equation}
The quantities $\mathbf{f}_\mathrm{opt}$ and $\phi_\mathrm{opt}$ form together the generalized optical force density four-vector as
\begin{equation}
 \boldsymbol{\mathcal{F}}_\mathrm{opt}=(\phi_\mathrm{opt}/c,\mathbf{f}_\mathrm{opt}).
 \label{eq:fourforce}
\end{equation}
For the Lorentz transformations of the force and power-conversion densities, see Appendix \ref{apx:transformations}.

Since we have included the kinetic-energy-related power-conversion density in Eq.~\eqref{eq:continuityfieldG}, it follows that the energy density of the field $W_\mathrm{field}$ accounts for the work done on the atoms by the optical force. This differs from the definition of the energy density of the field in previous works \cite{Partanen2017c,Partanen2017e,Partanen2019a,Partanen2019b}, where the power-conversion density was not included.

\subsubsection{\label{sec:dynamicalmedium}Dynamical equations of the medium}

From the conservation laws of the medium, we first write the conservation law of the number of atoms by using the atomic number density $n_\mathrm{a}$ and the atomic velocity field  $\mathbf{v}_\mathrm{a}$ as \cite{Penfield1967}
\begin{equation}
 \frac{\partial n_\mathrm{a}}{\partial t}+\nabla\cdot(n_\mathrm{a}\mathbf{v}_\mathrm{a})=0.
 \label{eq:numberconservationG}
\end{equation}
Since the field and the medium form together an isolated system, the source densities on the right hand sides of the energy and momentum conservation laws of the medium must be equal in magnitude but opposite in sign compared to the source densities on the right hand sides of the corresponding equations of the field in Eqs.~\eqref{eq:continuityfieldG} and \eqref{eq:momentumconservationfieldG}. The relativistic total energy density of the atom distribution is equal to $W_\mathrm{mat}=\rho_\mathrm{a}c^2=\gamma_{\mathbf{v}_\mathrm{a}}m_0c^2n_\mathrm{a}$ and the total relativistic momentum density of the medium is equal to $\mathbf{G}_\mathrm{mat}=\rho_\mathrm{a}\mathbf{v}_\mathrm{a}=\gamma_{\mathbf{v}_\mathrm{a}}m_0n_\mathrm{a}\mathbf{v}_\mathrm{a}$, where $\rho_a=\gamma_{\mathbf{v}_\mathrm{a}}m_0n_\mathrm{a}$ is the total mass density of atoms, $m_0$ is the rest mass of a single atom in the medium and $\gamma_{\mathbf{v}_\mathrm{a}}=1/\sqrt{1-|\mathbf{v}_\mathrm{a}|^2/c^2}$ is the Lorentz factor corresponding to $\mathbf{v}_\mathrm{a}$. By using these quantities in Eq.~\eqref{eq:conservationphi}, the conservation law of the energy of the medium is given by
\begin{equation}
 \frac{1}{c^2}\frac{\partial W_\mathrm{mat}}{\partial t}+\nabla\cdot\mathbf{G}_\mathrm{mat}
=\frac{\phi_\mathrm{opt}}{c^2}.
 \label{eq:masscontinuityG}
\end{equation}

Under the influence of the optical force, the atoms behave as classical particles, and in analogy to molecular dynamics, we can use Newton's equation to study their trajectories. Assuming that there are no other forces except the optical force, the relativistic Newton's equation of motion for the medium is given by \cite{Penfield1967}
\begin{equation}
 n_\mathrm{a}\frac{d\mathbf{p}_\mathrm{a}}{dt}
 =\mathbf{f}_\mathrm{opt},
 \label{eq:NewtonG}
\end{equation}
where $\mathbf{p}_\mathrm{a}=\gamma_{\mathbf{v}_\mathrm{a}}m_0\mathbf{v}_\mathrm{a}$ is the momentum of a single atom. Adding Eqs.~\eqref{eq:masscontinuityG} and \eqref{eq:NewtonG} side by side, using the material derivative, defined as $\frac{d}{dt}=\frac{\partial}{\partial t}+\mathbf{v}_\mathrm{a}\cdot\nabla$ \cite{Landau1987}, and applying the sum rule of differentiation gives the conservation law of the momentum density of the medium as
\begin{equation}
 \frac{\partial\mathbf{G}_\mathrm{mat}}{\partial t}+\nabla\cdot\boldsymbol{\mathcal{T}}_\mathrm{mat}
 =\mathbf{f}_\mathrm{opt}.
 \label{eq:momentumcontinuityG}
\end{equation}
Here the stress tensor of the medium is given by $\boldsymbol{\mathcal{T}}_\mathrm{mat}=\rho_\mathrm{a}\mathbf{v}_\mathrm{a}\otimes\mathbf{v}_\mathrm{a}=\gamma_{\mathbf{v}_\mathrm{a}}m_0n_\mathrm{a}\mathbf{v}_\mathrm{a}\otimes\mathbf{v}_\mathrm{a}$, where $\otimes$ denotes the outer product defined for two vectors $\mathbf{v}$ and $\mathbf{u}$ as a matrix with elements $(\mathbf{v}\otimes\mathbf{u})_{ij}=v_iu_j$. This form of the stress tensor of the medium is also well-known from previous literature \cite{Misner1973,Dirac1996,Penfield1967}.

\subsubsection{Energy exchange between the field and the medium}

The power-conversion density in Eqs.~\eqref{eq:continuityfieldG} and \eqref{eq:masscontinuityG} leads to the energy exchange between the field and the medium. We define the \emph{energy exchange density} $W_\phi$, which describes the work done on the medium by the optical force density, as
\begin{equation}
 W_\phi=\int_{-\infty}^t\phi_\mathrm{opt}dt'
 =\int_{-\infty}^t\mathbf{f}_\mathrm{opt}\cdot\mathbf{v}_\mathrm{a}dt'.
 \label{eq:Wphi}
\end{equation}
Here the power-conversion-density is given in Eq.~\eqref{eq:powerconversion}. By its definition, the energy exchange density $W_\phi$ is equal to the change of the kinetic energy of the medium by the optical force density. Accordingly, we define the \emph{exploitable energy density of light} as a sum of the energy density of the field and the energy exchange density as
\begin{equation}
 W_\mathrm{ex}=W_\mathrm{field}+W_\phi.
 \label{eq:Wex}
\end{equation}
The flux of the exploitable energy density is, in the L frame, equal to the electromagnetic energy flux given by the Poynting vector. The exploitable energy density of light is the effective energy density that is, e.g, usable in atomic transitions. Thus, we can think $W_\phi$ to have an electromagnetic origin in the same way as the polarization and magnetization energy densities. Between inertial frames, the exploitable energy density also experiences the experimentally verified Doppler shift as discussed in Sec.~\ref{sec:Doppler}.

The \emph{unexploitable} energy density components of the MP state of light do not originate from the power-conversion density $\phi_\mathrm{opt}=\mathbf{f}_\mathrm{opt}\cdot\mathbf{v}_\mathrm{a}$, This includes the rest energy density of the MDW and the kinetic energy density component of MDW atoms related to the kinetic energy of atoms in the equilibrium state of the medium. For detailed consideration of the kinetic and rest energy densities of the medium, see Sec.~\ref{sec:kineticMDW}.

\section{\label{sec:Newtonsolution}Solution of Newton's equation of motion in the L frame}

Newton's equation of motion of the medium in Eq.~\eqref{eq:NewtonG} can be solved for the generalized optical force density in Eq.~\eqref{eq:AbrahamforceL} once the fields in this equation are given. However, in the monochromatic field limit, the final result of the integrations carried out to solve Newton's equation can be presented as a function of the original fields $\mathbf{E}^\mathrm{(L)}$ and $\mathbf{H}^\mathrm{(L)}$ and harmonic cycle time average of $\frac{1}{2}(\mathbf{E}^\mathrm{(L)}\cdot\mathbf{D}^\mathrm{(L)}+\mathbf{H}^\mathrm{(L)}\cdot\mathbf{B}^\mathrm{(L)})$. The most important steps of the derivations are described in the appendixes.

The atomic displacement field in the L frame is denoted by $\mathbf{r}_\mathrm{a}^\mathrm{(L)}$ and related to the atomic velocity field $\mathbf{v}_\mathrm{a}^\mathrm{(L)}$ by $\mathbf{v}_\mathrm{a}^\mathrm{(L)}=\frac{d}{dt}\mathbf{r}_\mathrm{a}^\mathrm{(L)}$. The number density of the medium, $n_\mathrm{a}^\mathrm{(L)}$, has the following expression relating it to the position derivative of $\mathbf{r}_\mathrm{a}^\mathrm{(L)}$:
\begin{equation}
 n_\mathrm{a}^\mathrm{(L)}=\frac{n_\mathrm{a0}^\mathrm{(L)}}{1+\nabla\cdot\mathbf{r}_\mathrm{a}^\mathrm{(L)}}.
 \label{eq:rhoa}
\end{equation}
The expression of $n_\mathrm{a}^\mathrm{(L)}$ in Eq.~\eqref{eq:rhoa} emerges naturally from the contraction or expansion of the differential medium element that is directly proportional to $\nabla\cdot\mathbf{r}_\mathrm{a}^\mathrm{(L)}$.

One can obtain very accurate approximative analytic solutions of the velocity and mass density fields of the medium. As described in detail in Appendix \ref{apx:Newtonsolution}, the atomic number density and velocity distributions obtained as a solution to Newton's equation in Eq.~\eqref{eq:NewtonG} are given by
\begin{equation}
 n_\mathrm{a}^\mathrm{(L)}=\frac{n_\mathrm{a,min}^\mathrm{(L)}}{1-\dfrac{n_\mathrm{p}^\mathrm{(L)}|\mathbf{v}_\mathrm{a}^\mathrm{(L)}|}{c}}.
 \label{eq:rhoasol}
\end{equation}
\begin{equation}
 \mathbf{v}_\mathrm{a}^\mathrm{(L)}
 =\frac{\dfrac{(n_\mathrm{p}^\mathrm{(L)}n_\mathrm{g}^\mathrm{(L)}-1)W_\mathrm{ex,nd}^\mathrm{(L)}}{n_\mathrm{a,min}^\mathrm{(L)}m_0c^2}\mathbf{v}_\mathrm{p}^\mathrm{(L)}}{\sqrt{1+\Bigg[\dfrac{(n_\mathrm{p}^\mathrm{(L)}n_\mathrm{g}^\mathrm{(L)}-1)W_\mathrm{ex,nd}^\mathrm{(L)}}{n_\mathrm{p}^\mathrm{(L)}n_\mathrm{a,min}^\mathrm{(L)}m_0c^2}\Bigg]^2}}.
 \label{eq:atomicvelocitysol}
\end{equation}
Here $W_\mathrm{ex,nd}^\mathrm{(L)}$ is the conventional expression of \emph{the total exploitable energy density of light in a nondispersive medium}, given by
\begin{equation}
 W_\mathrm{ex,nd}^\mathrm{(L)}=\frac{1}{2}(\mathbf{E}^\mathrm{(L)}\cdot\mathbf{D}^\mathrm{(L)}+\mathbf{H}^\mathrm{(L)}\cdot\mathbf{B}^\mathrm{(L)}).
 \label{eq:Wexnd}
\end{equation}
There is a notation difference between the present work, including the power-conversion density in Eq.~\eqref{eq:continuityfieldG}, and the previous works \cite{Partanen2017c,Partanen2019a,Partanen2019b}. In Refs.~\cite{Partanen2017c,Partanen2019a,Partanen2019b}, the field energy density $W_\mathrm{field}^\mathrm{(L)}$ was defined to be the total exploitable energy density, and therefore, it corresponds to $W_\mathrm{ex,nd}^\mathrm{(L)}$ of the present work.

The phase velocity of light in Eq.~\eqref{eq:atomicvelocitysol} can be given in terms of the field vectors by
\begin{equation}
 \mathbf{v}_\mathrm{p}^\mathrm{(L)}=\frac{\mathbf{E}^\mathrm{(L)}\times\mathbf{H}^\mathrm{(L)}}{W_\mathrm{ex,nd}^\mathrm{(L)}}.
 \label{eq:phasevelocityL}
\end{equation}
In Eqs.~\eqref{eq:rhoasol} and \eqref{eq:atomicvelocitysol}, the quantity $n_\mathrm{a,min}^\mathrm{(L)}$ is the local minimum number density, which is an envelope function of $n_\mathrm{a}^\mathrm{(L)}$ formed from its minimum values over the harmonic cycle, obtained at points where $W_\mathrm{ex,nd}^\mathrm{(L)}$ and $\mathbf{v}_\mathrm{a}^\mathrm{(L)}$ are zero. As detailed in Appendix \ref{apx:Newtonsolution}, $n_\mathrm{a,min}^\mathrm{(L)}$ is given by
\begin{equation}
 n_\mathrm{a,min}^\mathrm{(L)}=n_\mathrm{a0}^\mathrm{(L)}+\frac{\big\langle W_\mathrm{ex,nd}^\mathrm{(L)}\big\rangle}{m_0c^2}(n_\mathrm{p}^\mathrm{(L)}n_\mathrm{g}^\mathrm{(L)}-1)\Big(\dfrac{n_\mathrm{g}^\mathrm{(L)}}{n_\mathrm{p}^\mathrm{(L)}}-1\Big),
 \label{eq:namin}
\end{equation}
where $n_\mathrm{a0}^\mathrm{(L)}$ is the number density of the equilibrium state of the medium in the absence of light and the angle brackets denote the time average over the harmonic cycle. The mass density corresponding to $n_\mathrm{a,min}^\mathrm{(L)}$ is given by $\rho_\mathrm{a,min}^\mathrm{(L)}=m_0 n_\mathrm{a,min}^\mathrm{(L)}$. Using Eq.~\eqref{eq:Wphi}, the energy exchange density $W_\phi^\mathrm{(L)}$ in the L frame is given by
\begin{equation}
 W_\phi^\mathrm{(L)}=\int_{-\infty}^t\mathbf{f}_\mathrm{opt}^\mathrm{(L)}\cdot\mathbf{v}_\mathrm{a}^\mathrm{(L)}dt'
 =(\gamma_{\mathbf{v}_\mathrm{a}}^\mathrm{(L)}-1)\rho_\mathrm{a,min}^\mathrm{(L)}c^2.
 \label{eq:WphiL}
\end{equation}
The calculation of the integral in Eq.~\eqref{eq:WphiL} is presented in Appendix \ref{apx:powerconversion}.

\section{\label{sec:tensors}SEM tensors in a dispersive medium}

The SEM tensor of a physical system describes the fluxes of energy and momentum in space-time. The general contravariant form of a SEM tensor in the Minkowski space-time is most conventionally defined by
$\mathbf{T}=T^{\alpha\beta}\mathbf{e}_\alpha\otimes\mathbf{e}_\beta$. Here the Einstein summation convention is used and $\mathbf{e}_\alpha$ and $\mathbf{e}_\beta$ are unit vectors of the four-dimensional Minkowski space-time. The Greek indices range over the four components of the space-time, i.e., $(ct,x,y,z)$. The corresponding matrix representation of $\mathbf{T}$ is given by \cite{Landau1989,Jackson1999,Misner1973}
\begin{equation}
 \mathbf{T}=
 \left[\begin{array}{cc}
  W & c\mathbf{G}^T\\
  c\mathbf{G} & \boldsymbol{\mathcal{T}}\\
 \end{array}\right]
 =\left[\begin{array}{cccc}
  W & cG^x & cG^y & cG^z\\
  cG^x & \mathcal{T}^{xx} & \mathcal{T}^{xy} & \mathcal{T}^{xz}\\
  cG^y & \mathcal{T}^{yx} & \mathcal{T}^{yy} & \mathcal{T}^{yz}\\
  cG^z & \mathcal{T}^{zx} & \mathcal{T}^{zy} & \mathcal{T}^{zz}
 \end{array}\right],
 \label{eq:emt}
\end{equation}
where the superscript $T$ denotes the transpose. 
All SEM tensors of the present work, including the SEM tensors of the field and medium subsystems, are strictly based on the classical definition in Eq.~\eqref{eq:emt}. Thus, they are all \emph{symmetric}.

In the literature, there exist \emph{asymmetric} definitions of the SEM tensors, such as the Minkowski SEM tensor \cite{Penfield1967,Brevik1979,Kemp2017,Griffiths2012}, which have been introduced to overcome certain difficulties of the theory. In the present work, we have found no need to introduce asymmetric SEM tensors. To put the SEM tensor definition of the present work into a wider perspective of the SEM tensor formulations in previous literature \cite{Schroder1968,Griffiths2012,Robinson1975,Penfield1967,Brevik1979,Mikura1976,Ramos2015}, we have presented a brief review of selected other SEM tensor formulations in Appendix \ref{apx:sem}. We also emphasize that, being based on the classical definition of the energy and momentum densities and stresses, the present work does not describe the quantum mechanical spin of light or its classical limit \cite{Bliokh2013b,Bliokh2017b,Mita2000}.

In the MP theory of light, the total SEM tensor $\mathbf{T}_\mathrm{tot}$ consists of the electromagnetic field part, $\mathbf{T}_\mathrm{field}$, and the medium part, $\mathbf{T}_\mathrm{mat}$, as
\begin{equation}
 \mathbf{T}_\mathrm{tot}=\mathbf{T}_\mathrm{field}+\mathbf{T}_\mathrm{mat}.
 \label{eq:tensorsum1}
\end{equation}
In distinction to previous works \cite{Partanen2017c,Partanen2017e,Partanen2019a,Partanen2019b}, in $\mathbf{T}_\mathrm{field}$, we have accounted for the work done on atoms by the optical force as discussed in Sec.~\ref{sec:kinetic1}. To describe the coupled MP state of light in the medium, we present the SEM tensor $\mathbf{T}_\mathrm{MDW}$ of the atomic MDW as the difference of the SEM tensors of the medium with and without the influence of the optical field as
\begin{equation}
 \mathbf{T}_\mathrm{MDW}=\mathbf{T}_\mathrm{mat}-\mathbf{T}_\mathrm{mat,0}.
 \label{eq:tensorsummdw}
\end{equation}
The SEM tensor $\mathbf{T}_\mathrm{mat,0}$ is called the SEM tensor of the equilibrium state of the medium. The SEM tensor of the coupled MP state of light, $\mathbf{T}_\mathrm{MP}$, is then given by
\begin{equation}
 \mathbf{T}_\mathrm{MP}=\mathbf{T}_\mathrm{field}+\mathbf{T}_\mathrm{MDW}.
 \label{eq:tensorsumMP}
\end{equation}
Using Eqs.~\eqref{eq:tensorsum1}--\eqref{eq:tensorsumMP}, we can write $\mathbf{T}_\mathrm{tot}$ in an alternative form, given by
\begin{equation}
 \mathbf{T}_\mathrm{tot}=\mathbf{T}_\mathrm{MP}+\mathbf{T}_\mathrm{mat,0}.
 \label{eq:tensorsum}
\end{equation}
Equation \eqref{eq:tensorsum} essentially means that $\mathbf{T}_\mathrm{MP}$ includes all terms that separate the total SEM tensor of the field and the medium from the SEM tensor of the equilibrium state of the medium. Therefore, $\mathbf{T}_\mathrm{MP}$ describes all energy and momentum terms of the coupled state of light.

The theory of relativity fundamentally requires that any physical quantity, including the SEM tensor, must be Lorentz-covariant. For SEM tensors, this requirement has two meanings, which are intimately linked to each other: First, the SEM tensors in different inertial frames must be related to each other by the Lorentz transformation of second-rank tensors (see Sec.~\ref{sec:Lorentzsecondrank}). Second, the elements of the SEM tensor must be writable in an unambiguous way in terms of the Lorentz-covariant dynamical variables, which transform between inertial frames according to their own Lorentz transformations (see Appendix \ref{apx:transformations}). This second requirement essentially means that the laws of physics are the same for all inertial observers.

\subsection{\label{sec:SEML}SEM tensors in the L frame}

In terms of the dynamical variables, the elements of the SEM tensor must have the same functional form in all \emph{general} inertial frames, but in \emph{special} inertial frames, such as the L frame, some dynamical variables can become zero. However, since the SEM tensor elements in different inertial frames are related to each other by the Lorentz transformation of second-rank tensors, the derivation of the SEM tensors in the L frame in this section, fixes the SEM tensors in all inertial frames as described in Sec.~\ref{sec:SEMG}.

\subsubsection{\label{sec:EMSEML}SEM tensor of the field in the L frame}

Due to the power-conversion density in Eq.~\eqref{eq:continuityfieldG}, the position- and time-dependent energy density of the field cannot be obtained from the momentum density in Eq.~\eqref{eq:fieldmomentumdensityL} by dividing it with the phase or group velocity of light, which is typically done in the literature, at least in the case of nondispersive media \cite{Jackson1999,Landau1984}. Instead, it must be solved from the conservation law of energy in Eq.~\eqref{eq:continuityfieldG}. By using the expression of the momentum density of the field in the L frame in Eq.~\eqref{eq:fieldmomentumdensityL} and integrating Eq.~\eqref{eq:continuityfieldG} over time, the energy density of the field is given by
\begin{equation}
 W_\mathrm{field}^\mathrm{(L)}=-\int_{-\infty}^t\nabla\cdot(\mathbf{E}^\mathrm{(L)}\times\mathbf{H}^\mathrm{(L)})dt'-W_\phi^\mathrm{(L)}.
 \label{eq:Wfieldintegral}
\end{equation}
Here we have used Eq.~\eqref{eq:WphiL} for the second term, which represents the work done on the atoms by the optical force.

The integral representation of the position- and time-dependent energy density of the field in Eq.~\eqref{eq:Wfieldintegral} is somewhat unconventional, but it is an unambiguous solution of the conservation law of the field energy in Eq.~\eqref{eq:continuityfieldG} for the definition of the momentum density, given in Eq.~\eqref{eq:fieldmomentumdensityL}. The Poynting vector and the conservation law of the energy of the field in Eq.~\eqref{eq:continuityfieldG} have been used to determine the energy density of the field in a dispersive medium also in some previous works, e.g., in Ref.~\cite{Philbin2011}, but the power-conversion density between the field and the medium has typically been set to zero.

To define the stress tensor, we start correspondingly from the conservation law of momentum in Eq.~\eqref{eq:momentumconservationfieldG}.
By integrating this equation with respect to the position in the direction of the propagation of light, we obtain the stress tensor of the field as
\begin{align}
 \boldsymbol{\mathcal{T}}_\mathrm{field}^\mathrm{(L)}
 &=-\int_{-\infty}^s\Big(\mathbf{f}_\mathrm{opt}^\mathrm{(L)}+\frac{\partial\mathbf{G}_\mathrm{field}^\mathrm{(L)}}{\partial t}\Big)ds'\otimes\hat{\mathbf{v}}_\mathrm{p}^\mathrm{(L)}\nonumber\\
 &=-\frac{n_\mathrm{p}^\mathrm{(L)}n_\mathrm{g}^\mathrm{(L)}}{c^2}\int_{-\infty}^s\frac{\partial}{\partial t}(\mathbf{E}^\mathrm{(L)}\times\mathbf{H}^\mathrm{(L)})ds'\otimes\hat{\mathbf{v}}_\mathrm{p}^\mathrm{(L)}.
 \label{eq:stressfieldintegral}
\end{align}
The variable $s$ is the position coordinate parallel to the direction of the propagation of light and $\hat{\mathbf{v}}_\mathrm{p}^\mathrm{(L)}$ denotes the unit vector parallel to $\mathbf{v}_\mathrm{p}^\mathrm{(L)}$.

The SEM tensor of the field, $\mathbf{T}_\mathrm{field}^\mathrm{(L)}$, can be obtained by calculating the integrals in Eqs.~\eqref{eq:Wfieldintegral} and \eqref{eq:stressfieldintegral}.
It follows that we can present the total SEM tensor of the field as a sum
\begin{equation}
 \mathbf{T}_\mathrm{field}^\mathrm{(L)}
 =\mathbf{T}_\mathrm{A}^\mathrm{(L)}+\mathbf{T}_\mathrm{int}^\mathrm{(L)}.
 \label{eq:EMSEML}
\end{equation}
Here $\mathbf{T}_\mathrm{A}^\mathrm{(L)}$ is equal to the Abraham SEM tensor in the L frame and $\mathbf{T}_\mathrm{int}^\mathrm{(L)}$ is the SEM tensor of the interaction. These SEM tensors are given by
\begin{widetext}
\begin{align}
 \mathbf{T}_\mathrm{A}^\mathrm{(L)}
 &=\left[\begin{array}{cc}
  \frac{1}{2}\Big(\mathbf{E}^\mathrm{(L)}\cdot\mathbf{D}^\mathrm{(L)}\!+\!\mathbf{H}^\mathrm{(L)}\cdot\mathbf{B}^\mathrm{(L)}\Big) & \frac{1}{c}(\mathbf{E}^\mathrm{(L)}\!\times\!\mathbf{H}^\mathrm{(L)})^T\\
  \frac{1}{c}\mathbf{E}^\mathrm{(L)}\!\times\!\mathbf{H}^\mathrm{(L)} & \frac{1}{2}\Big(\mathbf{E}^\mathrm{(L)}\cdot\mathbf{D}^\mathrm{(L)}\!+\!\mathbf{H}^\mathrm{(L)}\cdot\mathbf{B}^\mathrm{(L)}\Big)\mathbf{I}-\mathbf{E}^\mathrm{(L)}\!\otimes\!\mathbf{D}^\mathrm{(L)}-\mathbf{H}^\mathrm{(L)}\!\otimes\!\mathbf{B}^\mathrm{(L)}
 \end{array}\right],
 \label{eq:TAL}
\end{align}
\begin{equation}
    \mathbf{T}_\mathrm{int}^\mathrm{(L)} = 
\left[\begin{array}{cc}
  \big\langle W_\mathrm{ex,nd}^\mathrm{(L)}\big\rangle\big(n_\mathrm{g}^\mathrm{(L)}/n_\mathrm{p}^\mathrm{(L)}\!-\!1\big)-W_\phi^\mathrm{(L)}   & \mathbf{0} \\
\mathbf{0}     &   -\big(\big\langle W_\mathrm{ex,nd}^\mathrm{(L)}\big\rangle\!-\!W_\mathrm{ex,nd}^\mathrm{(L)}\big)\big(n_\mathrm{g}^\mathrm{(L)}/n_\mathrm{p}^\mathrm{(L)}\!-\!1\big)\hat{\mathbf{v}}_\mathrm{p}^\mathrm{(L)}\!\otimes\!\hat{\mathbf{v}}_\mathrm{p}^\mathrm{(L)}
\end{array}\right].
\label{eq:TintL}
\end{equation}
\end{widetext}
In Eq.~\eqref{eq:TAL}, $\mathbf{I}$ is the $3\times3$ unit matrix. How the time average over harmonic cycle in terms $\big\langle W_\mathrm{ex,nd}^\mathrm{(L)}\big\rangle$ in Eq.~\eqref{eq:TintL} results from the integrals in Eqs.~\eqref{eq:Wfieldintegral} and \eqref{eq:stressfieldintegral} is briefly described in appendixes \ref{apx:energydensity} and \ref{apx:stresstensor}.
For nondispersive media, the results for the SEM tensor of the field in the L frame presented in Refs.~\cite{Partanen2017c,Partanen2019a} are obtained from Eq.~\eqref{eq:EMSEML} by setting the phase and group refractive indices equal as $n_\mathrm{p}^\mathrm{(L)}=n_\mathrm{g}^\mathrm{(L)}$ and approximating the energy exchange density to zero as $W_\phi^\mathrm{(L)}\approx 0$.

The time averages of the energy density, momentum density, and stress tensor components of the SEM tensor of the field in the L frame in Eq.~\eqref{eq:EMSEML} can be shown to be equal to
\begin{align}
 \big\langle W_\mathrm{field}^\mathrm{(L)}\big\rangle
 &=\frac{1}{2}\Big[\frac{d(\omega_0^\mathrm{(L)}\epsilon^\mathrm{(L)})}{d\omega_0^\mathrm{(L)}}\big\langle\mathbf{E}^\mathrm{(L)}\!\cdot\!\mathbf{E}^\mathrm{(L)}\big\rangle\nonumber\\
 &\hspace{0.3cm}+\frac{d(\omega_0^\mathrm{(L)}\mu^\mathrm{(L)})}{d\omega_0^\mathrm{(L)}}\big\langle\mathbf{H}^\mathrm{(L)}\!\cdot\!\mathbf{H}^\mathrm{(L)}\big\rangle\Big]
 -\big\langle W_\phi^\mathrm{(L)}\big\rangle,
 \label{eq:energyfieldaverageL}
\end{align}
\vspace{-0.1cm}
\begin{equation}
 \big\langle\mathbf{G}_\mathrm{field}^\mathrm{(L)}\big\rangle
 =\frac{\big\langle\mathbf{E}^\mathrm{(L)}\times\mathbf{H}^\mathrm{(L)}\big\rangle}{c^2},
 \label{eq:momentumfieldaverageL}
\end{equation}
\begin{align}
 \big\langle\boldsymbol{\mathcal{T}}_\mathrm{field}^\mathrm{(L)}\big\rangle
 &=\frac{1}{2}\big(\big\langle\mathbf{E}^\mathrm{(L)}\cdot\mathbf{D}^\mathrm{(L)}\big\rangle+\big\langle\mathbf{H}^\mathrm{(L)}\cdot\mathbf{B}^\mathrm{(L)}\big\rangle\big)\mathbf{I}\nonumber\\
 &\hspace{0.5cm}-\big\langle\mathbf{E}^\mathrm{(L)}\otimes\mathbf{D}^\mathrm{(L)}\big\rangle-\big\langle\mathbf{H}^\mathrm{(L)}\otimes\mathbf{B}^\mathrm{(L)}\big\rangle.
\label{eq:stressfieldaverageL}
\end{align}
Here $\omega_0^\mathrm{(L)}$ is the central angular frequency of the pertinent narrow spectral range, $\Delta\omega^\mathrm{(L)}/\omega_0^\mathrm{(L)}$, where $\Delta\omega^\mathrm{(L)}$ is the standard deviation of the angular frequency. In previous literature, the form of the time-averaged energy density in Eq.~\eqref{eq:energyfieldaverageL} without the last energy exchange density term has been called the Brillouin form \cite{Landau1984,Jackson1999}. The magnitude of the last term of Eq.~\eqref{eq:energyfieldaverageL} is negligible and thus the time average of the energy density of the present work in the L frame is very accurately equal to the Brillouin form. The time average of the Poynting vector in Eq.~\eqref{eq:momentumfieldaverageL} is also widely used. The formula inside the brackets in Eq.~\eqref{eq:stressfieldaverageL} is formally the Maxwell stress tensor \cite{Panofsky1962,Rakich2010,Rakich2011}, and the applicability of its time average to a dispersive medium is also well documented in the optics literature \cite{Landau1984}.

\subsubsection{\label{sec:MATSEML}SEM tensor of the medium in the L frame}

The SEM tensor of the disturbed medium in the L frame, $\mathbf{T}_\mathrm{mat}^\mathrm{(L)}$ (field on), is unambiguously determined by the conservation laws in Eqs.~\eqref{eq:masscontinuityG} and \eqref{eq:momentumcontinuityG} and Newton's equation of motion in Eq.~\eqref{eq:NewtonG}. According to the relations of Sec.~\ref{sec:dynamicalmedium}, the energy density in $\mathbf{T}_\mathrm{mat}^\mathrm{(L)}$ is given by the conventional relativistic expression $W_\mathrm{mat}^\mathrm{(L)}=\rho_\mathrm{a}^\mathrm{(L)}c^2$, the momentum density is correspondingly $\mathbf{G}_\mathrm{mat}^\mathrm{(L)}=\rho_\mathrm{a}^\mathrm{(L)}\mathbf{v}_\mathrm{a}^\mathrm{(L)}$, and the stress tensor is given by $\boldsymbol{\mathcal{T}}_\mathrm{mat}^\mathrm{(L)}=\rho_\mathrm{a}^\mathrm{(L)}\mathbf{v}_\mathrm{a}^\mathrm{(L)}\otimes\mathbf{v}_\mathrm{a}^\mathrm{(L)}$. Thus, $\mathbf{T}_\mathrm{mat}^\mathrm{(L)}$ is written as \cite{Dirac1996,Misner1973,Penfield1967}
\begin{equation}
 \mathbf{T}_\mathrm{mat}^\mathrm{(L)}
 =\left[\begin{array}{cc}
\rho_\mathrm{a}^\mathrm{(L)}c^2 & \rho_\mathrm{a}^\mathrm{(L)}(\mathbf{v}_\mathrm{a}^\mathrm{(L)})^Tc\\
\rho_\mathrm{a}^\mathrm{(L)}\mathbf{v}_\mathrm{a}^\mathrm{(L)}c & \rho_\mathrm{a}^\mathrm{(L)}\mathbf{v}_\mathrm{a}^\mathrm{(L)}\otimes\mathbf{v}_\mathrm{a}^\mathrm{(L)}
\end{array}\right].
\label{eq:MATSEML}
\end{equation}
The atomic mass density $\rho_\mathrm{a}^\mathrm{(L)}=\gamma_{\mathbf{v}_\mathrm{a}}^\mathrm{(L)}m_0n_\mathrm{a}^\mathrm{(L)}$ and the atomic velocity $\mathbf{v}_\mathrm{a}^\mathrm{(L)}$ can be solved from Newton's equation in Eq.~\eqref{eq:NewtonG} as described in Sec.~\ref{sec:Newtonsolution}. Accordingly, $\rho_\mathrm{a}^\mathrm{(L)}$ and $\mathbf{v}_\mathrm{a}^\mathrm{(L)}$ can be unambiguously expressed in terms of the field amplitudes and the equilibrium atomic number density $n_\mathrm{a0}^\mathrm{(L)}$ using Eqs.~\eqref{eq:rhoasol} and \eqref{eq:atomicvelocitysol}.

The SEM tensor of the equilibrium state of the medium, $\mathbf{T}_\mathrm{mat,0}^\mathrm{(L)}$ (no field present), is also unambiguously determined by the conservation laws in Eqs.~\eqref{eq:masscontinuityG} and \eqref{eq:momentumcontinuityG}. In the equilibrium state of the medium, the atomic velocity in the L frame is $\mathbf{v}_\mathrm{a0}^\mathrm{(L)}=\mathbf{0}$. Accordingly, $\mathbf{T}_\mathrm{mat,0}^\mathrm{(L)}$ is obtained from the definition of the SEM tensor in Eq.~\eqref{eq:emt} by setting $W_\mathrm{mat,0}^\mathrm{(L)}=\rho_\mathrm{a0}^\mathrm{(L)}c^2$, $\mathbf{G}_\mathrm{mat,0}^\mathrm{(L)}=\rho_\mathrm{a0}^\mathrm{(L)}\mathbf{v}_\mathrm{a0}^\mathrm{(L)}=\mathbf{0}$, and $\boldsymbol{\mathcal{T}}_\mathrm{mat,0}^\mathrm{(L)}=\rho_\mathrm{a0}^\mathrm{(L)}\mathbf{v}_\mathrm{a0}^\mathrm{(L)}\otimes\mathbf{v}_\mathrm{a0}^\mathrm{(L)}=\mathbf{0}$, where $\rho_\mathrm{a0}^\mathrm{(L)}=m_0n_\mathrm{a0}^\mathrm{(L)}$. Thus, $\mathbf{T}_\mathrm{mat,0}^\mathrm{(L)}$ is given by
\begin{equation}
 \mathbf{T}_\mathrm{mat,0}^\mathrm{(L)}
 =\left[\begin{array}{cc}
\rho_\mathrm{a0}^\mathrm{(L)}c^2 & \mathbf{0}\\
\mathbf{0} & \mathbf{0}
\end{array}\right].
\label{eq:MATSEM0L}
\end{equation}

\subsubsection{\label{sec:MDWSEML}SEM tensor of the atomic MDW in the L frame}

The SEM tensor of the MDW, $\mathbf{T}_\mathrm{MDW}^\mathrm{(L)}=\mathbf{T}_\mathrm{mat}^\mathrm{(L)}-\mathbf{T}_\mathrm{mat,0}^\mathrm{(L)}$, defined in Eq.~\eqref{eq:tensorsummdw}, can be obtained from $\mathbf{T}_\mathrm{mat}^\mathrm{(L)}$ and $\mathbf{T}_\mathrm{mat,0}^\mathrm{(L)}$, given in Eqs.~\eqref{eq:MATSEML} and \eqref{eq:MATSEM0L}. It is instructive to write this SEM tensor in terms of the excess mass density of the medium  driven forward by the generalized optical force. We define the excess mass density of the medium by $\rho_\mathrm{MDW}^\mathrm{(L)}=\rho_\mathrm{a}^\mathrm{(L)}-\rho_\mathrm{a0}^\mathrm{(L)}$. Using this quantity, $\mathbf{T}_\mathrm{MDW}^\mathrm{(L)}$ is given by
\begin{equation}
 \mathbf{T}_\mathrm{MDW}^\mathrm{(L)}
 =\left[\renewcommand{\arraystretch}{0.7}\begin{array}{cc}
\rho_\mathrm{MDW}^\mathrm{(L)}c^2 & \rho_\mathrm{MDW}^\mathrm{(L)}(\mathbf{v}_\mathrm{MDW}^\mathrm{(L)})^Tc\\
\rho_\mathrm{MDW}^\mathrm{(L)}\mathbf{v}_\mathrm{MDW}^\mathrm{(L)}c & \rho_\mathrm{MDW}^\mathrm{(L)}\mathbf{v}_\mathrm{a}^\mathrm{(L)}\otimes\mathbf{v}_\mathrm{MDW}^\mathrm{(L)}
\end{array}\right].
\label{eq:MDWSEML}
\end{equation}
In Eq.~\eqref{eq:MDWSEML}, we have defined the local position- and time-dependent energy velocity of the MDW in the L frame by $\mathbf{v}_\mathrm{MDW}^\mathrm{(L)}=\rho_\mathrm{a}^\mathrm{(L)}\mathbf{v}_\mathrm{a}^\mathrm{(L)}/\rho_\mathrm{MDW}^\mathrm{(L)}$. This quantity is generally position- and time-dependent in a way that its mass-density-weighted time average over the harmonic cycle is equal to the group velocity as $\int_t^{t+\Delta t}\rho_\mathrm{MDW}^\mathrm{(L)}\mathbf{v}_\mathrm{MDW}^\mathrm{(L)}dt'/\int_t^{t+\Delta t}\rho_\mathrm{MDW}^\mathrm{(L)}dt'=\mathbf{v}_\mathrm{g}^\mathrm{(L)}$, where $\Delta t$ is the duration of the harmonic cycle. The position and time dependencies of $\mathbf{v}_\mathrm{MDW}^\mathrm{(L)}$ are described in more detail in Sec.~\ref{sec:simulations} dealing with the OCD simulations of a Gaussian light pulse.

\subsubsection{SEM tensor of the MP state of light in the L frame}

The SEM tensor of the MP state of light, $\mathbf{T}_\mathrm{MP}^\mathrm{(L)}=\mathbf{T}_\mathrm{field}^\mathrm{(L)}+\mathbf{T}_\mathrm{MDW}^\mathrm{(L)}$, defined in Eq.~\eqref{eq:tensorsumMP}, can be obtained from $\mathbf{T}_\mathrm{field}^\mathrm{(L)}$ and $\mathbf{T}_\mathrm{MDW}^\mathrm{(L)}$, given in Eqs.~\eqref{eq:EMSEML} and \eqref{eq:MDWSEML}. We have been able to give a surprisingly simple and transparent presentation for $\mathbf{T}_\mathrm{field}^\mathrm{(L)}$. Also, the physical quantities in $\mathbf{T}_\mathrm{MDW}^\mathrm{(L)}$ can be easily understood within the framework of the relativistic mechanics and conservation laws, through which they are unambiguously determined by the generalized optical force. However, both SEM tensors in Eqs.~\eqref{eq:EMSEML} and \eqref{eq:MDWSEML} include nuance-rich interference effects, which make the actual dependence of the SEM tensor components on the field amplitudes more sophisticated than in the case of a nondispersive medium.

\subsection{\label{sec:SEMG}SEM tensors in the G frame}

We next present the SEM tensors in the G frame. There has obviously been very few attempts to formulate a rigorously covariant theory of light in a dispersive medium starting \emph{ab-initio} from the principles of the special theory of relativity. Thus, our goal is to develop the SEM tensor formulation of the coupled MP state of light and its field and medium parts, which transform in a covariant way between arbitrary inertial frames and reduce in the L frame to the SEM tensors discussed above.

In the representation of the SEM tensors in the G frame below, we use the conventional electromagnetic field and displacement tensors and four-velocities $\mathbf{U}_\mathrm{g}$, $\mathbf{U}_\mathrm{p}$, $\mathbf{U}_\mathrm{a}$, and $\mathbf{U}_\mathrm{a0}$. The four-velocities are defined in terms of three-dimensional velocities and Lorentz factors as $\mathbf{U}_i=\gamma_{\mathbf{v}_i}(c,\mathbf{v}_i)$. Using these quantities enables a compact presentation of the SEM tensors. These quantities are also form-invariant in transformations between general inertial frames, and thus, reflect in a transparent way the covariance properties of the SEM tensors.

\subsubsection{\label{sec:Lorentzsecondrank}Lorentz transformation of the SEM tensors}

To define the Lorentz transformations between two general inertial frames, we assume that an arbitrary general inertial frame (G$'$ frame) is moving in another arbitrary general inertial frame (G frame) with a constant relative velocity $\mathbf{v}$. In this general case, the Lorentz boost can be written in the matrix form as
\begin{equation}
 \boldsymbol{\Lambda}=\left[\begin{array}{cc}
  \gamma_\mathbf{v} & -\gamma_\mathbf{v}\dfrac{\mathbf{v}^T}{c}\\
  -\gamma_\mathbf{v}\dfrac{\mathbf{v}}{c} & \;\mathbf{I}+(\gamma_\mathbf{v}-1)\hat{\mathbf{v}}\otimes\hat{\mathbf{v}}
 \end{array}\right],
 \label{eq:Lorentzboost}
\end{equation}
where $\gamma_\mathbf{v}=1/\sqrt{1-|\mathbf{v}|^2/c^2}$ is the Lorentz factor, and $\hat{\mathbf{v}}=\mathbf{v}/|\mathbf{v}|$ is the unit vector parallel to $\mathbf{v}$.

Using the Lorentz boost matrix in Eq.~\eqref{eq:Lorentzboost}, a second-rank tensor $\mathbf{T}$ in space-time transforms between the two inertial frames as
\begin{equation}
 \mathbf{T}'=\boldsymbol{\Lambda}\mathbf{T}\boldsymbol{\Lambda}^T.
\label{eq:LorentzT}
\end{equation}
This condition unambiguously relates the components of second-rank tensors in the G$'$ frame to those in the G frame. However, this condition alone does not make the tensor Lorentz-covariant as the tensor components must also correspond to the unambiguous physical definition of the components of the pertinent tensor.

\subsubsection{\label{sec:EMSEMG}SEM tensor of the field in the G frame}

The SEM tensor of the field in the G frame is unambiguously determined by the SEM tensor of the field in the L frame in Eq.~\eqref{eq:EMSEML} through the Lorentz transformation of the SEM tensors in Eq.~\eqref{eq:LorentzT}. After some algebra, we find that $\mathbf{T}_\mathrm{field}$ in the G frame is given in terms of the quantities of the G frame and split into the transformed Abraham and interaction SEM tensors as
\begin{equation}
 \mathbf{T}_\mathrm{field}=\mathbf{T}_\mathrm{A}+\mathbf{T}_\mathrm{int}.
 \label{eq:EMSEMG}
\end{equation}
The transformed Abraham SEM tensor is compactly written as
\begin{align}
 \mathbf{T}_\mathrm{A} &=\mathbf{F}\boldsymbol{g}\boldsymbol{\mathcal{D}}
 -\frac{1}{4}\boldsymbol{g}\mathrm{Tr}\Big(\mathbf{F}\boldsymbol{g}\boldsymbol{\mathcal{D}}\boldsymbol{g}\Big)\nonumber\\
 &\hspace{0.5cm}-\Big[\mathbf{F}\boldsymbol{g}\boldsymbol{\mathcal{D}}-\boldsymbol{\mathcal{D}}\boldsymbol{g}\mathbf{F}\Big]\boldsymbol{g}\frac{\mathbf{U}_\mathrm{a0}\otimes\mathbf{U}_\mathrm{a0}}{c^2},
 \label{eq:TAG}
\end{align}
where $\mathrm{Tr}(x)$ is the trace of a matrix and $\boldsymbol{g}=g_{\alpha\beta}\mathbf{e}^\alpha\otimes\mathbf{e}^\beta$, with $g_{00}=1$, $g_{11}=g_{22}=g_{33}=-1$, is the diagonal matrix representation of the Minkowski metric tensor. The electromagnetic field tensor $\mathbf{F}$ in Eq.~\eqref{eq:TAG} is given in the contravariant form $\mathbf{F}=F^{\alpha\beta}\mathbf{e}_\alpha\otimes\mathbf{e}_\beta$ as \cite{Jackson1999,Landau1989}
\begin{equation}
\mathbf{F}=\left[\begin{array}{cccc}
0 & -E_x/c & -E_y/c & -E_z/c\\
E_x/c & 0 & -B_z & B_y\\
E_y/c & B_z & 0 & -B_x\\
E_z/c & -B_y & B_x & 0
\end{array}\right],
\label{eq:Ftensor}
\end{equation}
and the electromagnetic displacement tensor $\boldsymbol{\mathcal{D}}$ is given in the contravariant form $\boldsymbol{\mathcal{D}}=\mathcal{D}^{\alpha\beta}\mathbf{e}_\alpha\otimes\mathbf{e}_\beta$ as \cite{Vanderlinde2004}
\begin{equation}
 \boldsymbol{\mathcal{D}}=\left[\begin{array}{cccc}
0 & -D_xc & -D_yc & -D_zc\\
D_xc & 0 & -H_z & H_y\\
D_yc & H_z & 0 & -H_x\\
D_zc & -H_y & H_x & 0
\end{array}\right].
\label{eq:Dtensor}
\end{equation}
The Abraham SEM tensor in the G frame in Eq.~\eqref{eq:TAG} does not appear in the same form in previous literature, including previous works on the MP theory of light \cite{Partanen2019a}. Instead, the Abraham SEM tensor is typically written in the G frame using its expression in the L frame in Eq.~\eqref{eq:TAL} and replacing only the field quantities of the L frame by those of the G frame. This has led to conclusions that the Abraham SEM tensor would not transform covariantly between inertial frames as it would not satisfy the Lorentz transformation in Eq.~\eqref{eq:LorentzT} \cite{Kemp2017,Partanen2019a}. However, this conclusion is no more valid when Eq.~\eqref{eq:TAG} is used. The essential feature of the form of the Abraham SEM tensor in Eq.~\eqref{eq:TAG} is that it depends on the four-velocity $\mathbf{U}_\mathrm{a0}$ of the equilibrium state of the medium. The dependence on $\mathbf{U}_\mathrm{a0}$ produces terms that make Eq.~\eqref{eq:LorentzT} satisfied. In some previous works on light in nondispersive media \cite{Makarov2011,Obukhov2008}, which, however, do not consider the MDW driven forward by the optical force, the Abraham SEM tensor has been suggested to have a form that corresponds to Eqs.~\eqref{eq:TAL} and \eqref{eq:TAG} with an additional symmetrization related to the order of the field quantities. This additional symmetrization is not needed in the present work.

The SEM tensor of the interaction in Eq.~\eqref{eq:EMSEMG} is given in the G frame by
\begin{align}
 &\mathbf{T}_\mathrm{int}\nonumber\\
 &\!=\dfrac{W_\mathrm{ex,nd}(1+\xi)}{c^2}\Big(\dfrac{1+\eta}{\gamma_{\mathbf{v}_\mathrm{g}}^2}+\dfrac{\eta}{\gamma_{\mathbf{v}_\mathrm{a0}}^2}-\dfrac{2\eta C_\mathrm{a0,g}}{\gamma_{\mathbf{v}_\mathrm{g}}\gamma_{\mathbf{v}_\mathrm{a0}}}\Big)\mathbf{U}_\mathrm{a0}\otimes\mathbf{U}_\mathrm{a0}\nonumber\\
 &\hspace{0.4cm}+\dfrac{\xi\eta\big(\big\langle W_\mathrm{ex,nd}\big\rangle\!-\!W_\mathrm{ex,nd}\big)}{\gamma_{\mathbf{v}_\mathrm{a0}}^2c^2}\Big(\dfrac{C_\mathrm{p,a0}C_\mathrm{g,a0}}{C_\mathrm{p,g}}\!-\!1\Big)\!\Big[\mathbf{U}_\mathrm{a0}\otimes\mathbf{U}_\mathrm{a0}\nonumber\\
 &\hspace{0.4cm}-\dfrac{C_\mathrm{g,a0}^2}{C_\mathrm{g,a0}^2-1}\Big(\dfrac{\mathbf{U}_\mathrm{g}}{C_\mathrm{g,a0}}-\mathbf{U}_\mathrm{a0}\Big)\otimes\Big(\dfrac{\mathbf{U}_\mathrm{g}}{C_\mathrm{g,a0}}-\mathbf{U}_\mathrm{a0}\Big)\Big]\nonumber\\
 &\hspace{0.4cm}-W_\phi\frac{\mathbf{U}_\mathrm{a0}\otimes\mathbf{U}_\mathrm{a0}}{\gamma_{\mathbf{v}_\mathrm{a0}}^2c^2}.
 \label{eq:TintG}
\end{align}
Here $W_\mathrm{ex,nd}$ is the conventionally used expression of the total exploitable energy density of the field in a nondispersive medium in the G frame, given by
\begin{equation}
 W_\mathrm{ex,nd}=\frac{1}{2}(\mathbf{E}\cdot\mathbf{D}+\mathbf{H}\cdot\mathbf{B}).
\end{equation}
Corresponding to Eq.~\eqref{eq:WphiL} for the L frame, the energy exchange density $W_\phi$ in Eq.~\eqref{eq:TintG} is transformed to the G frame by 
\begin{equation}
 W_\phi=\int_{-\infty}^t\!\!\!\mathbf{f}_\mathrm{opt}\cdot\mathbf{v}_\mathrm{a}dt'
 =\frac{\gamma_{\mathbf{v}_\mathrm{a}}-\gamma_{\mathbf{v}_\mathrm{a0}}}{\gamma_{\mathbf{v}_\mathrm{a0}}}\rho_\mathrm{a,min}c^2.
 \label{eq:WphiG}
\end{equation}
In Eq.~\eqref{eq:TintG}, for the sake of the compactness of the presentation, we have defined the normalized Minkowski inner product of four-velocities by the symbol $C_{i,j}=\mathbf{U}_i\boldsymbol{g}\mathbf{U}_j/c^2$. We have also introduced the \emph{dispersion parameter} $\xi$ and the \emph{coupling parameter} $\eta$ as given in Table \ref{tbl:parameters}. In the L frame, these parameters obtain simple forms $\xi^\mathrm{(L)}=n_\mathrm{g}^\mathrm{(L)}/n_\mathrm{p}^\mathrm{(L)}-1$ and $\eta^\mathrm{(L)}=n_\mathrm{p}^\mathrm{(L)}n_\mathrm{g}^\mathrm{(L)}-1$. Thus, the dispersion parameter is zero for nondispersive media and the coupling parameter is zero if the product of the phase and group refractive indices is unity.

The phase and group velocities and the atomic velocity of the equilibrium state of the medium are related to the field amplitudes and the dispersion parameter in the G frame as
\begin{equation}
 \mathbf{v}_\mathrm{p}=\frac{\mathbf{E}\times\mathbf{H}}{W_\mathrm{ex,nd}},\hspace{0.5cm}
 \mathbf{v}_\mathrm{p}+\xi\mathbf{v}_\mathrm{a0}=\mathbf{v}_\mathrm{g}(1+\xi)
 \label{eq:phasevelocity}
\end{equation}
Using these relations, the SEM tensor of the field in Eq.~\eqref{eq:EMSEMG} can be written as a sum of outer products of four-velocities multiplied by scalar prefactors. This representation of $\mathbf{T}_\mathrm{field}$ is given in Table \ref{tbl:SEMsummary}. The prefactors appearing in Table \ref{tbl:SEMsummary} are given in Table \ref{tbl:prefactors}. These prefactors can be proven to be \emph{Lorentz scalars}. Therefore, since $\mathbf{T}_\mathrm{field}$ is expressed solely in terms of four-velocities and Lorentz scalars, it is transparently covariant.

\begin{table}
 \setlength{\tabcolsep}{4.5pt}
 \renewcommand{\arraystretch}{2.0}
 \caption{\label{tbl:parameters}
 Representations of the dispersion and coupling parameters and the atomic MDW quantities in the G and L frames. The symbol $\ominus$ denotes the relativistic velocity subtraction, defined for two arbitrary velocity vectors $\mathbf{v}$ and $\mathbf{u}$ by $\mathbf{v}\ominus\mathbf{u}=(\mathbf{v}-\mathbf{u}_\mathrm{\perp}/\gamma_\mathbf{v}-\mathbf{u}_\mathrm{\parallel})/(1-\mathbf{v}\cdot\mathbf{u}/c^2)$ \cite{Jackson1999}. Here the subscripts $\parallel$ and $\perp$ denote components parallel, $\mathbf{u}_\parallel=(\mathbf{u}\cdot\hat{\mathbf{v}})\hat{\mathbf{v}}$, and perpendicular, $\mathbf{u}_\perp=\mathbf{u}-(\mathbf{u}\cdot\hat{\mathbf{v}})\hat{\mathbf{v}}$, to the velocity $\mathbf{v}$, for which $\gamma_\mathbf{v}=1/\sqrt{1-|\mathbf{v}|^2/c^2}$ is the Lorentz factor.}
\begin{tabular}{ccc}
   \hline\hline
   \hspace{-2mm}Quantity\hspace{-8mm} & G frame & L frame \\[2pt]
   \hline\\[-18pt]
$\xi$ & $\dfrac{\dfrac{|\mathbf{v}_\mathrm{a0}\ominus\mathbf{v}_\mathrm{p}|}{|\mathbf{v}_\mathrm{a0}\ominus\mathbf{v}_\mathrm{g}|}-1}{1-\dfrac{(\mathbf{v}_\mathrm{a0}\ominus\mathbf{v}_\mathrm{p})\cdot\mathbf{v}_\mathrm{a0}}{c^2}}$ & $\dfrac{n_\mathrm{g}^\mathrm{(L)}}{n_\mathrm{p}^\mathrm{(L)}}-1$\\[17pt]
$\eta$ & $\dfrac{\dfrac{c^2}{|\mathbf{v}_\mathrm{a0}\ominus\mathbf{v}_\mathrm{p}||\mathbf{v}_\mathrm{a0}\ominus\mathbf{v}_\mathrm{g}|}-1}{1-\dfrac{(\mathbf{v}_\mathrm{a0}\ominus\mathbf{v}_\mathrm{p})\cdot\mathbf{v}_\mathrm{a0}}{|\mathbf{v}_\mathrm{a0}\ominus\mathbf{v}_\mathrm{p}|^2}}$ & $n_\mathrm{p}^\mathrm{(L)}n_\mathrm{g}^\mathrm{(L)}-1$\\[17pt]
$\kappa$ & $\dfrac{\dfrac{c^2}{|\mathbf{v}_\mathrm{a0}\ominus\mathbf{v}_\mathrm{p}||\mathbf{v}_\mathrm{a0}\ominus\mathbf{v}_\mathrm{g}|}-1}{\dfrac{c^2}{|\mathbf{v}_\mathrm{a0}\ominus\mathbf{v}_\mathrm{p}|^2}-1}$ & $\dfrac{n_\mathrm{p}^\mathrm{(L)}n_\mathrm{g}^\mathrm{(L)}-1}{(n_\mathrm{p}^\mathrm{(L)})^2-1}$\\[21pt]
$\rho_\mathrm{MDW}$ & $\rho_\mathrm{a}-\rho_\mathrm{a0}$ & $\rho_\mathrm{a}^\mathrm{(L)}-\rho_\mathrm{a0}^\mathrm{(L)}$\\[4pt]
$\rho_\mathrm{MDW,min}$ & $\rho_\mathrm{a,min}-\rho_\mathrm{a0}$ & $\rho_\mathrm{a,min}^\mathrm{(L)}-\rho_\mathrm{a0}^\mathrm{(L)}$\\[4pt]
$\mathbf{v}_\mathrm{MDW}$ & $\dfrac{\rho_\mathrm{a}\mathbf{v}_\mathrm{a}-\rho_\mathrm{a0}\mathbf{v}_\mathrm{a0}}{\rho_\mathrm{a}-\rho_\mathrm{a0}}$ & $\dfrac{\rho_\mathrm{a}^\mathrm{(L)}\mathbf{v}_\mathrm{a}^\mathrm{(L)}-\rho_\mathrm{a0}^\mathrm{(L)}\mathbf{v}_\mathrm{a0}^\mathrm{(L)}}{\rho_\mathrm{a}^\mathrm{(L)}-\rho_\mathrm{a0}^\mathrm{(L)}}$\\[6pt]
   \hline\hline
 \end{tabular}
\end{table}

\begin{table*}
 \setlength{\tabcolsep}{5.0pt}
 \renewcommand{\arraystretch}{2.0}
 \caption{\label{tbl:SEMsummary}
 Expressions of the SEM tensors of the MP theory of light in a general inertial frame, the G frame, in terms of four-velocities $\mathbf{U}_\mathrm{g}$, $\mathbf{U}_\mathrm{a}$, and $\mathbf{U}_\mathrm{a0}$ and Lorentz invariant scalar prefactors $\rho_\mathrm{a}/\gamma_{\mathbf{v}_\mathrm{a}}^2$, $\rho_\mathrm{a0}/\gamma_{\mathbf{v}_\mathrm{a0}}^2$, $\rho_{\mathrm{f}i}$, and $\rho_{\mathrm{m}j}$, where $i=1,2,...,5$ and $j=1,2,3$. The Lorentz invariant prefactors $\rho_{\mathrm{f}i}$ and $\rho_{\mathrm{m}j}$ are given in Table \ref{tbl:prefactors}.}
\begin{tabular}{cc}
   \hline\hline
   SEM tensor & Expression \\[2pt]
   \hline\\[-20pt]
$\mathbf{T}_\mathrm{field}$ & $(\rho_\mathrm{f1}+\rho_\mathrm{f3})\mathbf{U}_\mathrm{g}\otimes\mathbf{U}_\mathrm{g}+(\rho_\mathrm{f2}+\rho_\mathrm{f3})\mathbf{U}_\mathrm{a0}\otimes\mathbf{U}_\mathrm{a0}-(\rho_\mathrm{f4}+\rho_\mathrm{f5})(\mathbf{U}_\mathrm{g}\otimes\mathbf{U}_\mathrm{a0}+\mathbf{U}_\mathrm{a0}\otimes\mathbf{U}_\mathrm{g})$\\[0pt]
 $\mathbf{T}_\mathrm{mat}$ & $\dfrac{\rho_\mathrm{a}}{\gamma_{\mathbf{v}_\mathrm{a}}^2}\mathbf{U}_\mathrm{a}\otimes\mathbf{U}_\mathrm{a}$\\[4pt]
 $\mathbf{T}_\mathrm{mat,0}$ & $\dfrac{\rho_\mathrm{a0}}{\gamma_{\mathbf{v}_\mathrm{a0}}^2}\mathbf{U}_\mathrm{a0}\otimes\mathbf{U}_\mathrm{a0}$\\[4pt]
 $\mathbf{T}_\mathrm{MDW}=\mathbf{T}_\mathrm{mat}-\mathbf{T}_\mathrm{mat,0}$ & $\rho_\mathrm{m1}\mathbf{U}_\mathrm{g}\otimes\mathbf{U}_\mathrm{g}+\rho_\mathrm{m2}\mathbf{U}_\mathrm{a0}\otimes\mathbf{U}_\mathrm{a0}+\rho_\mathrm{m3}(\mathbf{U}_\mathrm{g}\otimes\mathbf{U}_\mathrm{a0}+\mathbf{U}_\mathrm{a0}\otimes\mathbf{U}_\mathrm{g})$ \\[0pt]
$\mathbf{T}_\mathrm{MP}=\mathbf{T}_\mathrm{field}+\mathbf{T}_\mathrm{MDW}$ & $(\rho_\mathrm{f1}+\rho_\mathrm{f3}+\rho_\mathrm{m1})\mathbf{U}_\mathrm{g}\otimes\mathbf{U}_\mathrm{g}+(\rho_\mathrm{f2}+\rho_\mathrm{f3}+\rho_\mathrm{m2})\mathbf{U}_\mathrm{a0}\otimes\mathbf{U}_\mathrm{a0}$ \\[-10pt]
 & $+(\rho_\mathrm{m3}-\rho_\mathrm{f4}-\rho_\mathrm{f5})(\mathbf{U}_\mathrm{g}\otimes\mathbf{U}_\mathrm{a0}+\mathbf{U}_\mathrm{a0}\otimes\mathbf{U}_\mathrm{g})$ \\[4pt]
  \hline\hline
 \end{tabular}
\end{table*}

\begin{table}
 \setlength{\tabcolsep}{3.5pt}
 \renewcommand{\arraystretch}{2.0}
 \caption{\label{tbl:prefactors}
 Lorentz invariant scalar prefactors of the outer products of four-velocities in the SEM tensors given in Table \ref{tbl:SEMsummary}. The prefactors are given in terms of $W_\mathrm{ex,nd}$, $W_\phi$, the phase velocity $\mathbf{v}_\mathrm{p}$, the group velocity $\mathbf{v}_\mathrm{g}$, the atomic velocity $\mathbf{v}_\mathrm{a}$, and the equilibrium atomic velocity $\mathbf{v}_\mathrm{a0}$. The other quantities are the local MDW energy velocity $\mathbf{v}_\mathrm{MDW}$, the dispersion parameter $\xi$, and the coupling parameter $\eta$, see Table \ref{tbl:parameters}. The symbols $C_{i,j}$ are defined by $C_{i,j}=\mathbf{U}_i\boldsymbol{g}\mathbf{U}_j/c^2$.}
\begin{tabular}{cc}
   \hline\hline
   Quantity & Expression \\[2pt]
   \hline\\[-18pt]
$\rho_\mathrm{f1}$ & $\dfrac{W_\mathrm{ex,nd}(1+\xi)(1+\eta)}{\gamma_{\mathbf{v}_\mathrm{g}}^2c^2}$\\[4pt]
$\rho_\mathrm{f2}$ & $\dfrac{W_\mathrm{ex,nd}(1+\xi)\eta-W_\phi}{\gamma_{\mathbf{v}_\mathrm{a0}}^2c^2}$\\[4pt]
$\rho_\mathrm{f3}$ & $\dfrac{\xi\eta\big(\big\langle W_\mathrm{ex,nd}\big\rangle-W_\mathrm{ex,nd}\big)}{\gamma_{\mathbf{v}_\mathrm{a0}}^2c^2}\dfrac{\dfrac{C_\mathrm{p,a0}C_\mathrm{g,a0}}{C_\mathrm{p,g}}-1}{1-C_\mathrm{g,a0}^2}$\\[4pt]
$\rho_\mathrm{f4}$ & $\dfrac{\xi\eta\big(\big\langle W_\mathrm{ex,nd}\big\rangle-W_\mathrm{ex,nd}\big)}{\gamma_{\mathbf{v}_\mathrm{a0}}^2c^2}\dfrac{\dfrac{C_\mathrm{p,a0}C_\mathrm{g,a0}}{C_\mathrm{p,g}}-1}{(1-C_\mathrm{g,a0}^2)/C_\mathrm{g,a0}}$\\[4pt]
$\rho_\mathrm{f5}$ & $\dfrac{W_\mathrm{ex,nd}(1+\xi)\eta}{\gamma_{\mathbf{v}_\mathrm{g}}\gamma_{\mathbf{v}_\mathrm{a0}}c^2}$\\[4pt]
$\rho_\mathrm{m1}$ & $\dfrac{\rho_\mathrm{MDW}}{\gamma_{\mathbf{v}_\mathrm{g}}^2}\dfrac{|\mathbf{v}_\mathrm{MDW}-\mathbf{v}_\mathrm{a0}|}{|\mathbf{v}_\mathrm{g}-\mathbf{v}_\mathrm{a0}|}\Big(1-\dfrac{|\mathbf{v}_\mathrm{g}-\mathbf{v}_\mathrm{a}|}{|\mathbf{v}_\mathrm{g}-\mathbf{v}_\mathrm{a0}|}\Big)$\\[4pt]
$\rho_\mathrm{m2}$ & $\dfrac{\rho_\mathrm{MDW}}{\gamma_{\mathbf{v}_\mathrm{a0}}^2}\Big[1-\dfrac{|\mathbf{v}_\mathrm{MDW}-\mathbf{v}_\mathrm{a0}|}{|\mathbf{v}_\mathrm{g}-\mathbf{v}_\mathrm{a0}|}\Big(1+\dfrac{|\mathbf{v}_\mathrm{g}-\mathbf{v}_\mathrm{a}|}{|\mathbf{v}_\mathrm{g}-\mathbf{v}_\mathrm{a0}|}\Big)\Big]$\\[5pt]
$\rho_\mathrm{m3}$ & $\dfrac{\rho_\mathrm{MDW}}{\gamma_{\mathbf{v}_\mathrm{g}}\gamma_{\mathbf{v}_\mathrm{a0}}}\dfrac{|\mathbf{v}_\mathrm{MDW}-\mathbf{v}_\mathrm{a0}|}{|\mathbf{v}_\mathrm{g}-\mathbf{v}_\mathrm{a0}|}\dfrac{|\mathbf{v}_\mathrm{g}-\mathbf{v}_\mathrm{a}|}{|\mathbf{v}_\mathrm{g}-\mathbf{v}_\mathrm{a0}|}$\\[6pt]
   \hline\hline
 \end{tabular}
\end{table}

\subsubsection{\label{sec:MATSEMG}SEM tensor of the medium in the G frame}

In the SEM tensor $\mathbf{T}_\mathrm{mat}$ of the disturbed medium, the total energy and momentum densities and the stress tensor in terms of the atomic mass density $\rho_\mathrm{a}$ and atomic velocity $\mathbf{v}_\mathrm{a}$ are obtained from the conservation laws in Eqs.~\eqref{eq:masscontinuityG} and \eqref{eq:momentumcontinuityG}. The conventional expressions $W_\mathrm{mat}=\rho_\mathrm{a}c^2$, $\mathbf{G}_\mathrm{mat}=\rho_\mathrm{a}\mathbf{v}_\mathrm{a}$ and $\boldsymbol{\mathcal{T}}_\mathrm{mat}=\rho_\mathrm{a}\mathbf{v}_\mathrm{a}\otimes\mathbf{v}_\mathrm{a}$ apply also in the G frame and thus $\mathbf{T}_\mathrm{mat}$ is given by
\begin{equation}
 \mathbf{T}_\mathrm{mat}
 =\left[\begin{array}{cc}
\rho_\mathrm{a}c^2 & \rho_\mathrm{a}\mathbf{v}_\mathrm{a}^Tc\\
\rho_\mathrm{a}\mathbf{v}_\mathrm{a}c & \rho_\mathrm{a}\mathbf{v}_\mathrm{a}\otimes\mathbf{v}_\mathrm{a}
\end{array}\right]
=\dfrac{\rho_\mathrm{a}}{\gamma_{\mathbf{v}_\mathrm{a}}^2}\mathbf{U}_\mathrm{a}\otimes\mathbf{U}_\mathrm{a}.
\label{eq:MATSEMG}
\end{equation}

The SEM tensor $\mathbf{T}_\mathrm{mat,0}$ of the equilibrium state of the medium in terms of the equilibrium atomic mass density $\rho_\mathrm{a0}$ and the equilibrium atomic velocity $\mathbf{v}_\mathrm{a0}$ is of the known conventional form, given by \cite{Misner1973,Dirac1996}.
\begin{equation}
 \mathbf{T}_\mathrm{mat,0}
 =\left[\begin{array}{cc}
\rho_\mathrm{a0}c^2 & \rho_\mathrm{a0}\mathbf{v}_\mathrm{a0}^Tc\\
\rho_\mathrm{a0}\mathbf{v}_\mathrm{a0}c & \rho_\mathrm{a0}\mathbf{v}_\mathrm{a0}\otimes\mathbf{v}_\mathrm{a0}
\end{array}\right]
=\dfrac{\rho_\mathrm{a0}}{\gamma_{\mathbf{v}_\mathrm{a0}}^2}\mathbf{U}_\mathrm{a0}\otimes\mathbf{U}_\mathrm{a0}.
\label{eq:MATSEM0G}
\end{equation}
The scalar prefactors $\rho_\mathrm{a}/\gamma_{\mathbf{v}_\mathrm{a}}^2$ and $\rho_\mathrm{a0}/\gamma_{\mathbf{v}_\mathrm{a0}}^2$ of the outer products of four-velocities in Eqs.~\eqref{eq:MATSEMG} and \eqref{eq:MATSEM0G} are both Lorentz invariants since they are equal to the mass densities in the local rest frames of the disturbed and equilibrium states of the medium, respectively.

\subsubsection{\label{sec:MDWSEMG}SEM tensor of the atomic MDW in the G frame}

The SEM tensor of the atomic MDW in the G frame, $\mathbf{T}_\mathrm{MDW}$, is obtained as the difference of $\mathbf{T}_\mathrm{mat}$ and $\mathbf{T}_\mathrm{mat,0}$, given in Eqs.~\eqref{eq:MATSEMG} and \eqref{eq:MATSEM0G}. Using the excess mass density of the medium driven forward by the generalized optical force in the G frame, defined by $\rho_\mathrm{MDW}=\rho_\mathrm{a}-\rho_\mathrm{a0}$, and the local position- and time-dependent energy velocity of the MDW, defined by $\mathbf{v}_\mathrm{MDW}=(\rho_\mathrm{a}\mathbf{v}_\mathrm{a}-\rho_\mathrm{a0}\mathbf{v}_\mathrm{a0})/\rho_\mathrm{MDW}$, $\mathbf{T}_\mathrm{MDW}$ is given by
\begin{align}
 &\mathbf{T}_\mathrm{MDW}\nonumber\\
 &\!\!=\!\!\left[\renewcommand{\arraystretch}{0.7}\begin{array}{cc}
\rho_\mathrm{MDW}c^2 & \rho_\mathrm{MDW}\mathbf{v}_\mathrm{MDW}^Tc\\
\!\!\rho_\mathrm{MDW}\mathbf{v}_\mathrm{MDW}c & \rho_\mathrm{MDW}[\mathbf{v}_\mathrm{a}\!\!\otimes\!\mathbf{v}_\mathrm{MDW}\!+\!(\!\mathbf{v}_\mathrm{MDW}\!\!-\!\!\mathbf{v}_\mathrm{a}\!)\!\otimes\!\mathbf{v}_\mathrm{a0}]\!\!\!
\end{array}\right]\!\!.
\label{eq:MDWSEMG}
\end{align}
Comparison of Eqs.~\eqref{eq:MDWSEML} and \eqref{eq:MDWSEMG} shows that the presentations of the MDW SEM tensor in the L and G frames are formally identical apart from the stress tensor term $\rho_\mathrm{MDW}(\mathbf{v}_\mathrm{MDW}-\mathbf{v}_\mathrm{a})\otimes\mathbf{v}_\mathrm{a0}$. This term is a consequence of the form invariance of $\mathbf{T}_\mathrm{mat}$ and $\mathbf{T}_\mathrm{mat,0}$ in Eqs.~\eqref{eq:MATSEMG} and \eqref{eq:MATSEM0G}, and it is zero in the L frame, for which $\mathbf{v}_\mathrm{a0}^\mathrm{(L)}=\mathbf{0}$.
The Lorentz transformations of the quantities $\rho_\mathrm{MDW}$ and $\mathbf{v}_\mathrm{MDW}$ are determined by the Lorentz transformations of $\rho_\mathrm{a}$, $\rho_\mathrm{a0}$, $\mathbf{v}_\mathrm{a}$, and $\mathbf{v}_\mathrm{a0}$, given in Appendix \ref{apx:transformations}.

The MDW velocity $\mathbf{v}_\mathrm{MDW}$ and the atomic velocity $\mathbf{v}_\mathrm{a}$ can be written as linear combinations of the group velocity $\mathbf{v}_\mathrm{g}$ and the equilibrium atomic velocity $\mathbf{v}_\mathrm{a0}$ as
\begin{equation}
 \mathbf{v}_\mathrm{MDW}=\frac{|\mathbf{v}_\mathrm{MDW}-\mathbf{v}_\mathrm{a0}|}{|\mathbf{v}_\mathrm{g}-\mathbf{v}_\mathrm{a0}|}\mathbf{v}_\mathrm{g}
 +\Big(1-\frac{|\mathbf{v}_\mathrm{MDW}-\mathbf{v}_\mathrm{a0}|}{|\mathbf{v}_\mathrm{g}-\mathbf{v}_\mathrm{a0}|}\Big)\mathbf{v}_\mathrm{a0},
\end{equation}
\begin{equation}
 \mathbf{v}_\mathrm{a}=\Big(1-\frac{|\mathbf{v}_\mathrm{g}-\mathbf{v}_\mathrm{a}|}{|\mathbf{v}_\mathrm{g}-\mathbf{v}_\mathrm{a0}|}\Big)\mathbf{v}_\mathrm{g}
 +\frac{|\mathbf{v}_\mathrm{g}-\mathbf{v}_\mathrm{a}|}{|\mathbf{v}_\mathrm{g}-\mathbf{v}_\mathrm{a0}|}\mathbf{v}_\mathrm{a0}.
\end{equation}
Using these relations, the expression of the SEM tensor of the MDW in Eq.~\eqref{eq:MDWSEMG} can be expanded to a sum of outer products of four-velocities multiplied by scalar prefactors. This form of $\mathbf{T}_\mathrm{MDW}$ is given in Table \ref{tbl:SEMsummary} with scalar prefactors given in Table \ref{tbl:prefactors}. All these scalar prefactors can be proven to be Lorentz scalars. Thus, the form of $\mathbf{T}_\mathrm{MDW}$ in Table \ref{tbl:SEMsummary} is transparently covariant. The covariance of $\mathbf{T}_\mathrm{MDW}$ is also evident from its definition as the difference of the two covariant tensors $\mathbf{T}_\mathrm{mat}$ and $\mathbf{T}_\mathrm{mat,0}$ in agreement with Eq.~\eqref{eq:tensorsummdw}.

\subsubsection{SEM tensor of the MP state of light in the G frame}

The SEM tensor of the MP state of light in the G frame, $\mathbf{T}_\mathrm{MP}$, is, according to Eq.~\eqref{eq:tensorsumMP}, equal to the sum of $\mathbf{T}_\mathrm{field}$ and $\mathbf{T}_\mathrm{MDW}$ given in Eqs.~\eqref{eq:EMSEMG} and \eqref{eq:MDWSEMG}. The MP SEM tensor also has an expression in terms of four-velocities and Lorentz scalars as presented in Table \ref{tbl:SEMsummary}. All prefactors of the outer products of four-velocities in the MP SEM tensor in Table \ref{tbl:SEMsummary} are Lorentz scalars. Therefore, since the MP SEM tensor is expressed solely in terms of four-vectors and Lorentz scalars, it is transparently covariant.

\section{\label{sec:properties}Selected properties of the SEM tensors}

\subsection{\label{sec:amt}Definition of the angular momentum tensor}

In addition to the energy and momentum, the SEM tensor is also directly related to the description of angular momentum. The definitions of the angular momentum density (AMD) tensor and the related integrated quantity, the angular momentum (AM) tensor, used below, are general and not specific to the MP theory of light. We reiterate that, since the present work is based on the classical definition of the energy, momentum, and angular momentum densities, the present work does not describe the quantum mechanical spin of light. Using the index notation, the angular momentum density with respect to the position four-vector $\bar{x}^\alpha$ is described by the AMD tensor, which is a third-rank tensor, given by \cite{Jackson1999,Landau1989,Misner1973}
\begin{equation}
 \mathcal{M}^{\alpha\beta\gamma}=(x^\alpha-\bar{x}^\alpha)T^{\beta\gamma}-(x^\beta-\bar{x}^\beta)T^{\alpha\gamma}.
 \label{eq:angularmomentumdensity}
\end{equation}
By its definition in Eq.~\eqref{eq:angularmomentumdensity}, the AMD tensor is antisymmetric with respect to the indices $\alpha$ and $\beta$. The boundary integral of the AMD tensor over the three-dimensional space-time hypersurface $\partial\Omega$, which is the boundary of a four-dimensional space-time volume $\Omega$, produces the AM tensor that is a second-rank tensor given by \cite{Misner1973}
\begin{equation}
 M^{\alpha\beta}=\frac{1}{c}\oint_{\partial\Omega}\mathcal{M}^{\alpha\beta\gamma}d\Sigma_\gamma.
\end{equation}
The differential volume element vector $d\Sigma_\gamma$ is normal to $\partial\Omega$ and the integral is taken over the space-time coordinates $x^\alpha$. When the AMD tensor is known to become zero at infinity and the space-time hypersurface $\partial\Omega$ is chosen to be a spacelike surface of constant time with $\gamma=0$ and $d\Sigma_0=dxdydz=d^3r$, we obtain the AM tensor as \cite{Misner1973}
\begin{equation}
 M^{\alpha\beta}=\frac{1}{c}\int\mathcal{M}^{\alpha\beta 0}d^3r.
 \label{eq:angularmomentumtensor1}
\end{equation}

By separately using the time component and the spatial three-dimensional position vector $\mathbf{r}$, one can define the conventional three-dimensional angular momentum density $\boldsymbol{\mathcal{J}}$ \cite{Andrews2013,Piccirillo2013,Allen1992b,Jackson1999,Landau1989} and the boost momentum density $\boldsymbol{\mathcal{N}}$ \cite{Bliokh2013b,Bliokh2018a,Smirnova2018,Partanen2019a} as
\begin{equation}
 \boldsymbol{\mathcal{J}}=(\mathbf{r}-\bar{\mathbf{r}})\times\mathbf{G},
\end{equation}
\begin{equation}
 \boldsymbol{\mathcal{N}} =\frac{W}{c^2}(\mathbf{r}-\bar{\mathbf{r}})+\mathbf{G}(t-\bar{t}).
\end{equation}
The boost momentum is a quantity whose conservation ensures the rectilinear motion of the energy centroid of light in the assumed homogeneous medium. In terms of $\boldsymbol{\mathcal{J}}$ and $\boldsymbol{\mathcal{N}}$, the AM tensor in Eq.~\eqref{eq:angularmomentumtensor1} can be written in the contravariant form as a matrix, given by \cite{Fayngold2008}
\begin{align}
 \mathbf{M} &=
 \int\left[\begin{array}{cc}
  0 & -c\boldsymbol{\mathcal{N}}^T\\
  c\boldsymbol{\mathcal{N}} & (\mathbf{r}-\bar{\mathbf{r}})\wedge\mathbf{G}\\
 \end{array}\right]d^3r\nonumber\\
 &=\int\left[\begin{array}{cccc}
  0 & -c\mathcal{N}^x & -c\mathcal{N}^y & -c\mathcal{N}^z\\
  c\mathcal{N}^x & 0 & \mathcal{J}^z & -\mathcal{J}^y\\
  c\mathcal{N}^y & -\mathcal{J}^z & 0 & \mathcal{J}^x\\
  c\mathcal{N}^z & \mathcal{J}^y & -\mathcal{J}^x & 0
 \end{array}\right]d^3r.
 \label{eq:angularmomentumtensor2}
\end{align}
Here $\wedge$ denotes the exterior product. The AM tensor of an isolated system is form-invariant and transforms according to the Lorentz transformation of second-rank tensors, discussed in Sec.~\ref{sec:Lorentzsecondrank}.

\subsection{\label{sec:conservation}Compact presentation of the conservation laws}

The conservation laws of energy, momentum, and angular momentum of light can be written in terms of the MP SEM tensor and the MP AMD tensor as
\begin{equation}
 \partial_\alpha T_\mathrm{MP}^{\alpha\beta}=0,\hspace{0.5cm} \partial_\gamma\mathcal{M}_\mathrm{MP}^{\alpha\beta\gamma}=0.
 \label{eq:conservation1}
\end{equation}
The first equation of Eq.~\eqref{eq:conservation1} corresponds to the conservation laws of energy and momentum in Eqs.~\eqref{eq:conservationphi} and \eqref{eq:conservationf}, where the right-hand sides can be set to zero since the MP is an isolated system. Together with the conservation laws of energy and momentum, the conservation law of angular momentum in the second equation of Eq.~\eqref{eq:conservation1} corresponds to the diagonal symmetry of the SEM tensor \cite{Jackson1999,Landau1989,Misner1973}. Therefore, all the conservation laws of energy, momentum, and angular momentum of light can be compactly expressed using the MP SEM tensor as \cite{Jackson1999}
\begin{equation}
 \partial_\alpha T_\mathrm{MP}^{\alpha\beta}=0,\hspace{0.5cm} T_\mathrm{MP}^{\alpha\beta}=T_\mathrm{MP}^{\beta\alpha}.
 \label{eq:conservation2}
\end{equation}
Here the equation on the right describes the diagonal symmetry that is needed to fulfill the conservation law of angular momentum in Eq.~\eqref{eq:conservation1}. Thus, the total MP SEM tensor in Table \ref{tbl:SEMsummary} fulfills all the conservation laws of energy, momentum, and angular momentum. Consequently, this tensor is the total conserved Poincar\'e current of light \cite{Partanen2019b}. This is a strong argument for the ultimate consistency of the MP state of light. The conservation laws in Eqs.~\eqref{eq:conservation1} and \eqref{eq:conservation2} also naturally apply to $\mathbf{T}_\mathrm{mat,0}$, whose elements are constants, and to $\mathbf{T}_\mathrm{tot}$, which is the total conserved Poincar\'e current of the full system of the field and the medium \cite{Partanen2019b}.

\subsection{\label{sec:balance}Law of action and reaction between the field and the MDW and the generalized optical force}

The SEM tensors of the field and the MDW satisfy the conservation laws of energy and momentum in Eqs.~\eqref{eq:conservationphi} and \eqref{eq:conservationf} with the nonzero source terms $\phi_\mathrm{opt}/c^2$ and $\mathbf{f}_\mathrm{opt}$. The four-divergences of these SEM tensors can then be written using the generalized optical four-force density $\boldsymbol{\mathcal{F}}_\mathrm{opt}$ in Eq.~\eqref{eq:fourforce} as
\begin{equation}
 \partial_\beta(T_\mathrm{field})^{\alpha\beta}=-(\mathcal{F}_\mathrm{opt})^\alpha,
 \label{eq:conservationfield}
\end{equation}
\begin{equation}
 \partial_\beta(T_\mathrm{MDW})^{\alpha\beta}=(\mathcal{F}_\mathrm{opt})^\alpha,
 \label{eq:conservationmdw}
\end{equation}
From Eqs.~\eqref{eq:conservationfield} and \eqref{eq:conservationmdw}, it follows that the four-divergence of the total MP SEM tensor of light is zero in accordance with Eq.~\eqref{eq:conservation2} since the opposite source terms in Eqs.~\eqref{eq:conservationfield} and \eqref{eq:conservationmdw} cancel each other.

The generalized optical force density, which is one of the basic foundations of the theory, is given for the L frame in Eq.~\eqref{eq:AbrahamforceL}. Using the Lorentz transformation of the force and power-conversion densities, given in Appendix \ref{apx:transformations}, $\mathbf{f}_\mathrm{opt}$ is given in the G frame by
\begin{align}
 \mathbf{f}_\mathrm{opt} &=\Big(\frac{\partial}{\partial t}+\mathbf{v}_\mathrm{a0}\cdot\nabla\Big)\Big[\kappa\Big(\mathbf{D}\times\mathbf{B}-\frac{\mathbf{E}\times\mathbf{H}}{c^2}\Big)
 +\frac{W_\phi\mathbf{v}_\mathrm{a0}}{c^2}\Big].
 \label{eq:AbrahamforceGeneral}
\end{align}
The coefficient $\kappa$ is given in Table \ref{tbl:parameters}. The expression of the power-conversion density in the G frame is obtained from $\mathbf{f}_\mathrm{opt}$ in Eq.~\eqref{eq:AbrahamforceGeneral} using Eq.~\eqref{eq:powerconversion}.

\subsection{\label{sec:covarianceeqtot}Lorentz transformations of the total energy, momentum, and angular momentum of light}

The energy density $W_\mathrm{MP}$ and the momentum density $\mathbf{G}_\mathrm{MP}$ of the MP state of light in the G frame are obtained from the corresponding components of $\mathbf{T}_\mathrm{MP}$, derived above. Using these quantities, we can furthermore define the angular momentum and boost momentum densities of the MP state of light in the G frame by
\begin{equation}
 \boldsymbol{\mathcal{J}}_\mathrm{MP} =(\mathbf{r}-\bar{\mathbf{r}})\times\mathbf{G}_\mathrm{MP},
 \label{eq:Jmp}
\end{equation}
\vspace{-0.5cm}
\begin{equation}
 \boldsymbol{\mathcal{N}}_\mathrm{MP} =\frac{W_\mathrm{MP}}{c^2}(\mathbf{r}-\bar{\mathbf{r}})+\mathbf{G}_\mathrm{MP}(t-\bar{t})
 \label{eq:Nmp}.
\end{equation}

The total energy, momentum, angular momentum, and boost momentum of light are obtained as volume integrals of the pertinent densities as
\begin{align}
 E_\mathrm{MP} &=\int W_\mathrm{MP}d^3r,\label{eq:EMP}\hspace{0.5cm} \mathbf{p}_\mathrm{MP} =\int \mathbf{G}_\mathrm{MP}d^3r,\\
 \mathbf{J}_\mathrm{MP} &=\int \boldsymbol{\mathcal{J}}_\mathrm{MP}d^3r,\hspace{0.4cm} \mathbf{N}_\mathrm{MP} =\int \boldsymbol{\mathcal{N}}_\mathrm{MP}d^3r\label{eq:NMP}.
\end{align}
Since the MP is an isolated system, it is described by the MP SEM tensor that has \emph{zero four-divergence} as presented in Eq.~\eqref{eq:conservation2}. Therefore, the volume-integrated quantities in Eqs.~\eqref{eq:EMP}--\eqref{eq:NMP} obey the conventional Lorentz transformations of the energy-momentum four-vector and the angular momentum and boost momentum vectors, given by
\begin{align}
 E_\mathrm{MP}' &=\gamma_\mathbf{v}(E_\mathrm{MP}-\mathbf{v}\cdot\mathbf{p}_\mathrm{MP}),\label{eq:intEmp}\\
 \mathbf{p}_\mathrm{MP}' &=\mathbf{p}_{\mathrm{MP},\perp}+\gamma_\mathbf{v}(\mathbf{p}_{\mathrm{MP},\parallel}-\frac{1}{c^2}E_\mathrm{MP}\mathbf{v}),\label{eq:intPmp}\\
 \mathbf{J}_\mathrm{MP}' &=\mathbf{J}_{\mathrm{MP},\parallel}+\gamma_\mathbf{v}(\mathbf{J}_{\mathrm{MP},\perp}+\mathbf{v}\times\mathbf{N}_\mathrm{MP}),\label{eq:intJmp}\\
 \mathbf{N}_\mathrm{MP}' &=\mathbf{N}_{\mathrm{MP},\parallel}+\gamma_\mathbf{v}(\mathbf{N}_{\mathrm{MP},\perp}-\frac{1}{c^2}\mathbf{v}\times\mathbf{J}_\mathrm{MP})\label{eq:intNmp}.
\end{align}
That these Lorentz transformations are satisfied for the MP state of light is a strong argument for the full consistency of the SEM tensors of the present work and the results of the previously introduced MP quasiparticle model for dispersive media \cite{Partanen2017e}. Note that the volume-integrated quantities corresponding to the field or the MDW SEM tensors do not \emph{separately} satisfy the Lorentz transformations in Eqs.~\eqref{eq:intEmp}--\eqref{eq:intNmp} due to their coupling through the generalized optical force and power-conversion densities.

\subsection{\label{sec:kineticMDW}Kinetic and rest energies of the MDW}

The kinetic energy density $W_\mathrm{MDW,kin}$ and the rest energy density $W_\mathrm{MDW,rest}$ of the MDW can be defined in terms of the differences of the corresponding terms of the SEM tensors $\mathbf{T}_\mathrm{mat}$ and $\mathbf{T}_\mathrm{mat,0}$ as
\begin{align}
 W_\mathrm{MDW,kin} &=W_\mathrm{mat,kin}-W_\mathrm{mat,0,kin}\nonumber\\
 &=(\gamma_{\mathbf{v}_\mathrm{a}}-1)m_0c^2n_\mathrm{a}-(\gamma_{\mathbf{v}_\mathrm{a0}}-1)m_0c^2n_\mathrm{a0}\nonumber\\
 &=\frac{\gamma_{\mathbf{v}_\mathrm{a}}-1}{\gamma_{\mathbf{v}_\mathrm{a}}}\rho_\mathrm{a}c^2-\frac{\gamma_{\mathbf{v}_\mathrm{a0}}-1}{\gamma_{\mathbf{v}_\mathrm{a0}}}\rho_\mathrm{a0}c^2,
 \label{eq:Wmdwkin}
\end{align}
\begin{align}
 W_\mathrm{MDW,rest} &=W_\mathrm{mat,rest}-W_\mathrm{mat,0,rest}\nonumber\\
 &=m_0c^2n_\mathrm{a}-m_0c^2n_\mathrm{a0}\nonumber\\
 &=\frac{\rho_\mathrm{a}c^2}{\gamma_{\mathbf{v}_\mathrm{a}}}-\frac{\rho_\mathrm{a0}c^2}{\gamma_{\mathbf{v}_\mathrm{a0}}}.
 \label{eq:Wmdwrest}
\end{align}
Equation \eqref{eq:Wmdwrest} can be used to define the MDW rest mass density as
\begin{equation}
 \rho_\mathrm{MDW,rest}=\frac{W_\mathrm{MDW,rest}}{c^2}
 =m_0(n_\mathrm{a}-n_\mathrm{a0})
 =\frac{\rho_\mathrm{a}}{\gamma_{\mathbf{v}_\mathrm{a}}}-\frac{\rho_\mathrm{a0}}{\gamma_{\mathbf{v}_\mathrm{a0}}}.
\end{equation}
The integral of this rest mass density over the light pulse gives the total transferred medium rest mass as
\begin{equation}
 \delta M=\int\rho_\mathrm{MDW,rest}d^3r.
 \label{eq:dM}
\end{equation}
This transferred rest mass is Lorentz invariant scalar meaning that the integral on the right-hand side gives the same value for all inertial frames. For the volume-integrated rest energy of the MDW, we obtain $E_\mathrm{MDW,rest}=\int W_\mathrm{MDW,rest}c^2d^3r=\delta Mc^2$.

The kinetic energy density of the MDW, as defined in Eq.~\eqref{eq:Wmdwkin}, does not fully belong to the exploitable energy of light since it includes the equilibrium state kinetic energy density of the medium corresponding to the MDW rest mass density, given by
$W_\mathrm{MDW,kin,0}=(\gamma_{\mathbf{v}_\mathrm{a0}}-1)\rho_\mathrm{MDW,rest}c^2$. This kinetic energy is obtained from the equilibrium state of the medium, and it is carried forward in a form of flux by the MDW. As we will see below, the total kinetic energy of the MDW in Eq.~\eqref{eq:Wmdwkin} also includes other terms that behave as flux.

Multiplying Eq.~\eqref{eq:numberconservationG} by $m_0$ and subtracting it from Eq.~\eqref{eq:masscontinuityG} side by side, multiplying both sides of the resulting equation by $c^2$, integrating over time from negative infinity to $t$, and rearranging the terms gives
\begin{align}
 &W_\mathrm{MDW,kin}\nonumber\\
&=\int_{-\infty}^t\mathbf{f}_\mathrm{opt}\cdot\mathbf{v}_\mathrm{a}dt'-\int_{-\infty}^t\nabla\cdot(W_\mathrm{mat,kin}\mathbf{v}_\mathrm{a})dt'\nonumber\\
&=W_\phi+W_\mathrm{MDW,kin,0}+(\gamma_{\mathbf{v}_\mathrm{a}}-\gamma_{\mathbf{v}_\mathrm{a0}})\Big(\frac{\rho_\mathrm{a}}{\gamma_{\mathbf{v}_\mathrm{a}}}-\frac{\rho_\mathrm{a,min}}{\gamma_{\mathbf{v}_\mathrm{a0}}}\Big)c^2.
 \label{eq:Wmdwkin2}
\end{align}
Here the first term on the right-hand side is the energy exchange density. It converts field energy into kinetic energy of atoms in regions where the field energy density is increasing and returns this energy to the field in regions where the field energy density is decreasing. Thus, $W_\phi$ is part of the \emph{exploitable energy density of light}.

The second term on the right-hand side of the first line of Eq.~\eqref{eq:Wmdwkin2} describes the \emph{flux of the kinetic energy density}. A detailed consideration of this term shows that $W_\mathrm{MDW,kin,0}$ generally dominates the flux part. The MDW transfers this kinetic energy through the medium. Since this kinetic energy moves in the medium in a form of flux, there should not be a power-conversion density source term in the dynamical equations related to this part of the kinetic energy density. Therefore, we call it an \emph{unexploitable part of the kinetic energy density of the MDW}.

\subsection{\label{sec:Doppler}Exploitable energy and Doppler shift}

The total exploitable energy density of light is the sum of the electromagnetic energy density $W_\mathrm{field}$ and the exploitable part of the kinetic energy density of the MDW in Eq.~\eqref{eq:Wmdwkin2}, which equals the energy exchange density $W_\phi$, given in Eq.~\eqref{eq:Wphi}. Therefore, we can write the total exploitable energy of light as given in Eq.~\eqref{eq:Wex}. The total exploitable energy of light is obtained as a volume integral of $W_\mathrm{ex}$ in Eq.~\eqref{eq:Wex} over the light pulse as $E_\mathrm{ex}=\int W_\mathrm{ex}d^3r$. From the Lorentz transformations of the quantities of Eq.~\eqref{eq:Wex}, given in Appendix \ref{apx:transformations}, it follows that the total exploitable energies of light in the G and G$'$ frames are related by
\begin{equation}
 E_\mathrm{ex}'=\gamma_\mathbf{v}\Big(1-\frac{\mathbf{v}\cdot\hat{\mathbf{v}}_\mathrm{p}}{|\mathbf{v}_\mathrm{p}|}\Big)E_\mathrm{ex},
 \label{eq:Eex}
\end{equation}
For a monochromatic field, $\hat{\mathbf{v}}_\mathrm{p}/|\mathbf{v}_\mathrm{p}|=\mathbf{k}/\omega$, where $\omega$ is the angular frequency and $\mathbf{k}$ is the wave number of light in the medium. Then, by comparing Eq.~\eqref{eq:Eex} with the Lorentz transformation of the frequency, given in Appendix \ref{apx:transformations}, we see that $E_\mathrm{ex}'/\omega'=E_\mathrm{ex}/\omega$. Thus, the transformation of the energy $E_\mathrm{ex}$ in Eq.~\eqref{eq:Eex} corresponds to the accurately experimentally verified Doppler shift of the frequency of light in a dispersive medium \cite{ChenJi2011}.

\subsection{Lorentz invariance of the SEM tensor contractions}

The contraction is a tensor operation that, in the case of a second-rank tensor, produces a scalar quantity. This scalar quantity is known to be Lorentz invariant for a tensor satisfying Eq.~\eqref{eq:LorentzT} \cite{Misner1973}. The contraction of a SEM tensor $T^{\alpha\beta}$ with a matrix presentation $\mathbf{T}$ in the Minkowski spacetime is defined as
$T^\alpha_{\;\alpha}=T^{\alpha\beta}g_{\alpha\beta}=\mathrm{Tr}(\mathbf{T}\boldsymbol{g})$. The Lorentz invariant contractions of the SEM tensors of the present work are given by
\begin{equation}
 \mathrm{Tr}(\mathbf{T}_\mathrm{A}\boldsymbol{g})=0,
 \label{eq:traceAbraham}
\end{equation}
\begin{align}
 &\mathrm{Tr}(\mathbf{T}_\mathrm{int}\boldsymbol{g})\nonumber\\
 &=W_\mathrm{ex,nd}(1+\xi)\Big(\dfrac{1+\eta}{\gamma_{\mathbf{v}_\mathrm{g}}^2}+\dfrac{\eta}{\gamma_{\mathbf{v}_\mathrm{a0}}^2}-\dfrac{2\eta C_\mathrm{a0,g}}{\gamma_{\mathbf{v}_\mathrm{g}}\gamma_{\mathbf{v}_\mathrm{a0}}}\Big)
 -\frac{W_\phi}{\gamma_{\mathbf{v}_\mathrm{a0}}^2}\nonumber\\
  &\hspace{0.5cm}
  +\frac{2\xi\eta\big(\langle W_\mathrm{ex,nd}\rangle
 -W_\mathrm{ex,nd}\big)}{\displaystyle\gamma_{\mathbf{v}_\mathrm{a0}}^2}\Big(\dfrac{C_\mathrm{p,a0}C_\mathrm{g,a0}}{C_\mathrm{p,g}}\!-\!1\Big),
 \label{eq:traceint}
\end{align}
\begin{equation}
 \mathrm{Tr}(\mathbf{T}_\mathrm{mat}\boldsymbol{g})=\frac{\rho_\mathrm{a}c^2}{\gamma_{\mathbf{v}_\mathrm{a}}^2},
 \label{eq:tracemat}
\end{equation}
\begin{equation}
 \mathrm{Tr}(\mathbf{T}_\mathrm{mat,0}\boldsymbol{g})=\frac{\rho_\mathrm{a0}c^2}{\gamma_{\mathbf{v}_\mathrm{a0}}^2},
 \label{eq:tracemat0}
\end{equation}
\begin{equation}
 \mathrm{Tr}(\mathbf{T}_\mathrm{field}\boldsymbol{g})=\mathrm{Tr}(\mathbf{T}_\mathrm{A}\boldsymbol{g})+\mathrm{Tr}(\mathbf{T}_\mathrm{int}\boldsymbol{g}),
 \label{eq:tracefield}
\end{equation}
\begin{equation}
 \mathrm{Tr}(\mathbf{T}_\mathrm{MDW}\boldsymbol{g})=\mathrm{Tr}(\mathbf{T}_\mathrm{mat}\boldsymbol{g})-\mathrm{Tr}(\mathbf{T}_\mathrm{mat,0}\boldsymbol{g}),
 \label{eq:traceMDW}
\end{equation}
\begin{equation}
 \mathrm{Tr}(\mathbf{T}_\mathrm{MP}\boldsymbol{g})=\mathrm{Tr}(\mathbf{T}_\mathrm{field}\boldsymbol{g})+\mathrm{Tr}(\mathbf{T}_\mathrm{MDW}\boldsymbol{g}).
 \label{eq:traceMP}
\end{equation}
The contractions of the SEM tensors of the disturbed and equilibrium states of the medium in Eq.~\eqref{eq:tracemat} and \eqref{eq:tracemat0} are equal to the pertinent rest energy densities of the medium. The physical interpretation of the contraction of the MP SEM tensor is discussed in the next subsection.

\subsection{Lagrangian and Hamiltonian of the MP state of light}

 The volume integral of the MP SEM tensor contraction, $\mathrm{Tr}(\mathbf{T}_\mathrm{MP}\boldsymbol{g})$, normalized by the photon number $N_0$ and a negative sign in the G frame is equal to $L=-\int\mathrm{Tr}(\mathbf{T}_\mathrm{MP}\boldsymbol{g})d^3r/N_0=-m_\mathrm{MP}c^2\sqrt{1-v_\mathrm{g}^2/c^2}$, where $m_\mathrm{MP}=n_\mathrm{p}^\mathrm{(L)}\sqrt{(n_\mathrm{g}^\mathrm{(L)})^2-1}\,\hbar\omega_0^\mathrm{(L)}/c^2$ is the MP quasiparticle rest mass \cite{Partanen2017e}. Thus, this integral is the Lagrangian of a relativistic quasiparticle with a rest mass $m_\mathrm{MP}$ moving at velocity $v_\mathrm{g}$ \cite{Landau1989}. The corresponding Hamiltonian is given by $H=v_\mathrm{g}\partial L/\partial v_\mathrm{g}-L=\gamma_\mathrm{\mathbf{v}_\mathrm{g}}m_\mathrm{MP}c^2$, and it is the relativistic total energy of the MP quasiparticle \cite{Partanen2017e}.

\subsection{\label{sec:cev}Constant center-of-energy velocity of an isolated system}

The center of energy of an isolated system moves with a constant velocity. This is a fundamental law of nature that any consistent physical theory must fulfill. As discussed below, this conservation law is a synthesis of the conservation laws of energy and momentum of an isolated system. We will show that this law is fulfilled for the MP theory of light in a dispersive medium by considering a Gaussian light pulse having in vacuum the field energy $E_\mathrm{field}^\mathrm{(L)}$, which enters a material body having a rest mass $M_\mathrm{body}^\mathrm{(L)}$. To simplify the model, we assume no reflection at the interface.

When the light pulse and the related field energy are still in the vacuum, the velocity of the center of energy is given by
\begin{equation}
 v_\mathrm{CEV}^\mathrm{(L)}=\frac{\sum_iE_i^\mathrm{(L)}v_i^\mathrm{(L)}}{\sum_iE_i^\mathrm{(L)}}=\frac{E_\mathrm{field}^\mathrm{(L)}c}{E_\mathrm{field}^\mathrm{(L)}+M_\mathrm{body}^\mathrm{(L)}c^2}.
 \label{eq:cev1}
\end{equation}
After the light pulse has entered the body, the total energy of the system consists of two components: (1) The energy of the coupled field-medium MP state of light is given by $E_\mathrm{MP}^\mathrm{(L)}=n_\mathrm{p}^\mathrm{(L)}n_\mathrm{g}^\mathrm{(L)}E_\mathrm{field}^\mathrm{(L)}$. This energy includes the rest energy of the transferred mass of the MP state equal to $\delta M^\mathrm{(L)}=\int\rho_\mathrm{MDW}^\mathrm{(L)}d^3r=(n_\mathrm{p}^\mathrm{(L)}n_\mathrm{g}^\mathrm{(L)}-1)E_\mathrm{field}^\mathrm{(L}/c^2$. (2) The energy of the body is equal to $M_\mathrm{r}^\mathrm{(L)}c^2$, where $M_\mathrm{r}^\mathrm{(L)}=M_\mathrm{body}^\mathrm{(L)}-\delta M^\mathrm{(L)}$ is the recoil mass. The recoil mass takes the recoil following from the momentum change of light from $p_\mathrm{vac}^\mathrm{(L)}=\mathrm{E}_\mathrm{field}^\mathrm{(L)}/c$ to $p_\mathrm{MP}^\mathrm{(L)}=n_\mathrm{p}^\mathrm{(L)}E_\mathrm{field}^\mathrm{(L)}/c$ when it enters the medium. The mass $M_\mathrm{r}$ moves with the recoil velocity $v_\mathrm{r}^\mathrm{(L)}$ after the field has entered the medium. The kinetic energy of the recoil is negligibly small in comparison to the field energy by the same argument as the kinetic energy of the MDW is extremely small in the L frame. Therefore, it is enough to account for the recoil momentum of the body, which is not small.

From the conservation law of momentum at the interface, we obtain $v_\mathrm{r}^\mathrm{(L)}=(1-n_\mathrm{p}^\mathrm{(L)})E_\mathrm{field}^\mathrm{(L)}/(M_\mathrm{r}c)$. Thus, the center of energy velocity of the system, when light is propagating inside the body, is given by
\begin{equation}
 v_\mathrm{CEV}^\mathrm{(L)}=\frac{\sum_iE_i^\mathrm{(L)}v_i^\mathrm{(L)}}{\sum_iE_i^\mathrm{(L)}}=\frac{n_\mathrm{p}^\mathrm{(L)}n_\mathrm{g}^\mathrm{(L)}E_\mathrm{field}^\mathrm{(L)}v_\mathrm{g}^\mathrm{(L)}+M_\mathrm{r}^\mathrm{(L)}c^2v_\mathrm{r}^\mathrm{(L)}}{n_\mathrm{p}^\mathrm{(L)}n_\mathrm{g}^\mathrm{(L)}E_\mathrm{field}^\mathrm{(L)}+M_\mathrm{r}^\mathrm{(L)}c^2}.
 \label{eq:cev2}
\end{equation}
By expressing $M_\mathrm{r}^\mathrm{(L)}$ and $v_\mathrm{r}^\mathrm{(L)}$ in terms of $M_\mathrm{body}^\mathrm{(L)}$ and $\delta M^\mathrm{(L)}=(n_\mathrm{p}^\mathrm{(L)}n_\mathrm{g}^\mathrm{(L)}-1)E_\mathrm{field}^\mathrm{(L}/c^2$, given above, one can directly see the equivalence of Eqs.~\eqref{eq:cev1} and \eqref{eq:cev2}. Thus, the constant center-of-energy-velocity law of an isolated system is satisfied in the MP theory of light. The equality of the numerators of Eqs.~\eqref{eq:cev1} and \eqref{eq:cev2} divided by $c^2$ corresponds to the conservation law of momentum and the equality of the denominators is the conservation law of energy. In the case of nondispersive media, the corresponding analysis of the velocity of the center of energy is presented in Ref.~\cite{Partanen2017c} and, for a different geometry, in Ref.~\cite{McClymer2020}.

\section{\label{sec:simulations}Simulation of the atomic MDW in a dispersive medium}

The analytic solution of Newton's equation in Sec.~\ref{sec:Newtonsolution} can be used to calculate the position and time dependence of the medium dynamics for a given optical field. In this section, we illustrate the behavior of selected key physical variables of the field and the medium for a realistic Gaussian light pulse in a dispersive medium. Here we solve the coupled dynamical equations of the field and the medium numerically as a function of the position and time using the finite-difference-based OCD method described in Ref.~\cite{Partanen2017c}. This provides an independent cross-check of the numerical and analytical results directly on the basis of Newton's equation.

The only assumptions in the simulations are the values of the material parameters and the generalized optical force density calculated using Eq.~\eqref{eq:AbrahamforceL} for the given Gaussian light pulse. In this section, we present the results of these simulations for the atomic MDW and the local energy velocity of light. The numerical accuracy of the simulations is approximately seven digits. The comparison between the analytical results and numerical simulations is partly limited by the finite spectral width used in the simulations. Within the numerical accuracy of the simulations and the limited accuracy due to the finite spectral width, all results obtained in this section are in agreement with the analytic results derived above and the MP quasiparticle model of a dispersive medium presented in Ref.~\cite{Partanen2017e}. In the figures of this section, we have presented the results of the numerical simulations. The analytic results agree with these results so accurately that they could not be seen as separate graphs.

\subsection{\label{sec:dispersion}Dispersion relation}

Independently of the actual form of the dispersion relation of a lossless dispersive medium, for a narrow spectral range, it can be approximated by a linear form. The linearized dispersion relation is a good approximation if the actual dispersion relation does not have sharp variations due to resonances, and if the distance traveled by light is not very long. The linearized dispersion relation around
the central frequency $\omega_0^\mathrm{(L)}$
contains the zeroth and first order terms of the Taylor expansion of $\omega^\mathrm{(L)}(k)$ as
\begin{equation}
\omega^\mathrm{(L)}(k)\approx\omega_0^\mathrm{(L)}+(c/n_\mathrm{g}^\mathrm{(L)})(k-k_\mathrm{0,med}^\mathrm{(L)}).
\label{eq:lineardispersion}
\end{equation}
Here $k_\mathrm{0,med}^\mathrm{(L)}=n_\mathrm{p}^\mathrm{(L)}k_0^\mathrm{(L)}$ is the wave number of light in the medium, in which $k_0^\mathrm{(L)}=\omega_0^\mathrm{(L)}/c$ is the wave number in vacuum at $\omega_0^\mathrm{(L)}$ and
$n_\mathrm{p}^\mathrm{(L)}$ is the phase refractive index of the medium at $\omega_0^\mathrm{(L)}$.
The group refractive index $n_\mathrm{g}^\mathrm{(L)}$ is constant.
For a more accurate description, one should also account for higher-order terms in the Taylor expansion of $\omega^\mathrm{(L)}(k)$.

The linear dispersion relation in Eq.~\eqref{eq:lineardispersion} can be used to define the frequency-dependent phase refractive index in the L frame as
\begin{equation}
 n_\mathrm{p}^\mathrm{(L)}(\omega)=n_\mathrm{g}^\mathrm{(L)}+(n_\mathrm{p}^\mathrm{(L)}-n_\mathrm{g}^\mathrm{(L)})\frac{\omega_0^\mathrm{(L)}}{\omega}.
\end{equation}
The linear form of the dispersion relation in Eq.~\eqref{eq:lineardispersion} 
does not lead to distortion of the \emph{light pulse envelope} while the pulse is propagating.

\subsection{Electric and magnetic fields}

The most general form of the electric field
of a one-dimensional
light pulse, linearly polarized in the $x$ direction and propagating in the $z$ direction in a dispersive
medium, is written as \cite{Griffiths1998}
\begin{equation}
 \mathbf{E}^\mathrm{(L)}(\mathbf{r},t)
 =\mathrm{Re}\Big[\int_{-\infty}^\infty \tilde E^\mathrm{(L)}(k)e^{i[kz-\omega^\mathrm{(L)}(k) t]}dk\Big]\hat{\mathbf{x}},
 \label{eq:electricfieldgeneral}
\end{equation}
Here $\hat{\mathbf{x}}$ is the unit vector parallel to the $x$ axis and $\tilde E^\mathrm{(L)}(k)$ is the Fourier component of the field. Using Maxwell's equations and the constitutive relations given in Sec.~\ref{sec:concepts}, expressions corresponding to Eq.~\eqref{eq:electricfieldgeneral} apply also for the other field vectors with the exception that the magnetic field and magnetic flux density are parallel to the $y$ axis.

We assume that the electric field of the light pulse has a Gaussian form with
$\tilde E^\mathrm{(L)}(k)=\frac{\omega_0^\mathrm{(L)}\mathcal{E}}{\sqrt{2\pi}\,n_\mathrm{g}^\mathrm{(L)}\Delta k_0^\mathrm{(L)}}e^{-[(k-n_\mathrm{p}^\mathrm{(L)}k_0^\mathrm{(L)})/(n_\mathrm{g}^\mathrm{(L)}\Delta k_0^\mathrm{(L)})]^2/2}$,
where $\mathcal{E}$ is a normalization factor
and $\Delta k_0^\mathrm{(L)}$ is the standard deviation
of the wave number in vacuum. The pulse width
in the $z$ direction is given by $\Delta z=1/(\sqrt{2}n_\mathrm{g}^\mathrm{(L)}\Delta k_0^\mathrm{(L)})$.
The corresponding width of the pulse in time is then
$\Delta t=n_\mathrm{g}^\mathrm{(L)}\Delta z/c=1/(\sqrt{2}\Delta k_0^\mathrm{(L)}c)$
and the full width at half maximum is given by $\Delta t_\mathrm{FWHM}=2\sqrt{2\ln 2}\,\Delta t$.
Using Eq.~\eqref{eq:electricfieldgeneral} with the linear dispersion
relation in Eq.~\eqref{eq:lineardispersion},
the electric field becomes
\vspace{-0.1cm}
\begin{align}
 \mathbf{E}^\mathrm{(L)}(\mathbf{r},t)
 & =\omega_0^\mathrm{(L)}\mathcal{E}\cos\Big[n_\mathrm{p}^\mathrm{(L)}k_0^\mathrm{(L)}\Big(z-\frac{ct}{n_\mathrm{p}^\mathrm{(L)}}\Big)\Big]\nonumber\\
 &\hspace{0.5cm}\times e^{-(n_\mathrm{g}^\mathrm{(L)}\Delta k_0^\mathrm{(L)})^2(z-ct/n_\mathrm{g}^\mathrm{(L)})^2/2}\hat{\mathbf{x}}.
 \label{eq:electricfield}
\end{align}
The normalization factor $\mathcal{E}$ in Eq.~\eqref{eq:electricfield}
can be determined
so that the integral of the corresponding instantaneous
energy density over $z$ gives $E_\mathrm{ex}/A$,
where $A$ is the cross-sectional area.

In the calculation of the energy density and Poynting vector in the monochromatic field limit, i.e., for small $\Delta k_0^\mathrm{(L)}/k_0^\mathrm{(L)}=\Delta\omega^\mathrm{(L)}/\omega_0^\mathrm{(L)}$, we can approximate the Fourier components of the field vectors $\mathbf{D}^\mathrm{(L)}$, $\mathbf{H}^\mathrm{(L)}$, and $\mathbf{B}^\mathrm{(L)}$ as $\tilde D^\mathrm{(L)}(k)\approx\varepsilon^\mathrm{(L)}\tilde E^\mathrm{(L)}(k)$, $\tilde H^\mathrm{(L)}(k)\approx\sqrt{\varepsilon^\mathrm{(L)}/\mu^\mathrm{(L)}}\tilde E^\mathrm{(L)}(k)$, and $\tilde B^\mathrm{(L)}(k)\approx\mu^\mathrm{(L)}\tilde H^\mathrm{(L)}(k)$, where $\varepsilon^\mathrm{(L)}=\varepsilon^\mathrm{(L)}(\omega_0^\mathrm{(L)})$ and $\mu^\mathrm{(L)}=\mu^\mathrm{(L)}(\omega_0^\mathrm{(L)})$. With these approximations, the electric field of the Gaussian light pulse in Eq.~\eqref{eq:electricfield}, and the corresponding expressions for the other field vectors approach an exact solution of Maxwell's equations in the monochromatic field limit.

\subsection{Simulation parameters}

We apply the OCD model to illustrate the node structure of the
MDW and the atomic displacements
resulting from the optical force of a Gaussian light pulse of Eq.~\eqref{eq:electricfield}
in a material where the dispersion is approximately linear near the central frequency.
We assume that the vacuum wavelength of light is $\lambda_0^\mathrm{(L)}=800$ nm corresponding to the central frequency of $\omega_0^\mathrm{(L)}=2\pi c/\lambda_0^\mathrm{(L)}$. The total exploitable energy of the Gaussian light pulse is $E_\mathrm{ex}^\mathrm{(L)}=500$ nJ (recall the definition of the exploitable energy used in this work).
This energy is incident to a circular cross-sectional area $A=\pi(d/2)^2$ of diameter $d=100$ $\mu$m. The resulting maximum value $4.3\times10^{11}$ W/cm$^2$
of the Poynting vector averaged over the harmonic cycle is below the
bulk value of the breakdown threshold irradiance of many common materials, e.g.,
$5.0\times10^{11}$ W/cm$^2$ reported for fused silica \cite{Smith2007}.
The relative spectral width
of the pulse, in our example, is $\Delta\omega^\mathrm{(L)}/\omega_0^\mathrm{(L)}=\Delta k_0^\mathrm{(L)}/k_0^\mathrm{(L)}=0.05$, which
corresponds to the FWHM of $\Delta t_\mathrm{FWHM}^\mathrm{(L)}=15$ fs. The FWHM is fixed to this close to feasibility limit value
to make the node structure of the MDW visible in the same scale with the Gaussian envelope of the field.
In our simulations, we use space discretization of
$h_z=\lambda/80$ and time discretization of $h_t=2\pi/(80\omega_0)$
that are dense compared to the scale of the harmonic cycle.
The computational details of the simulation are described in more detail
in Appendix C of Ref.~\cite{Partanen2017c}.

For visualization needs, we assume here an artificial dielectric material for which the phase refractive index is $n_\mathrm{p}^\mathrm{(L)}=1.2$ and the group refractive index is $n_\mathrm{g}^\mathrm{(L)}=2$.
The chosen phase refractive index is below typical values for glasses and the group refractive index is larger by a suitable amount to enable a clear visual separation of the phase velocity dynamics of the nodes inside the Gaussian envelope. The equilibrium density of the medium is assumed to be $\rho_\mathrm{a0}^\mathrm{(L)}=2500$ kg/m$^3$, which is close to typical values for glasses. The medium is assumed to be isotropic, and we use the value $B=50$ GPa for the bulk modulus and $G=25$ GPa for the shear modulus. These values are close to typical values of
the corresponding quantities for glasses. The elasticity parameters of the medium have, however, no visible effect on the dynamics of the field of the very short Gaussian light pulse. We neglect losses like phonon scattering altogether.

\begin{figure}
\centering
\includegraphics[width=0.95\columnwidth]{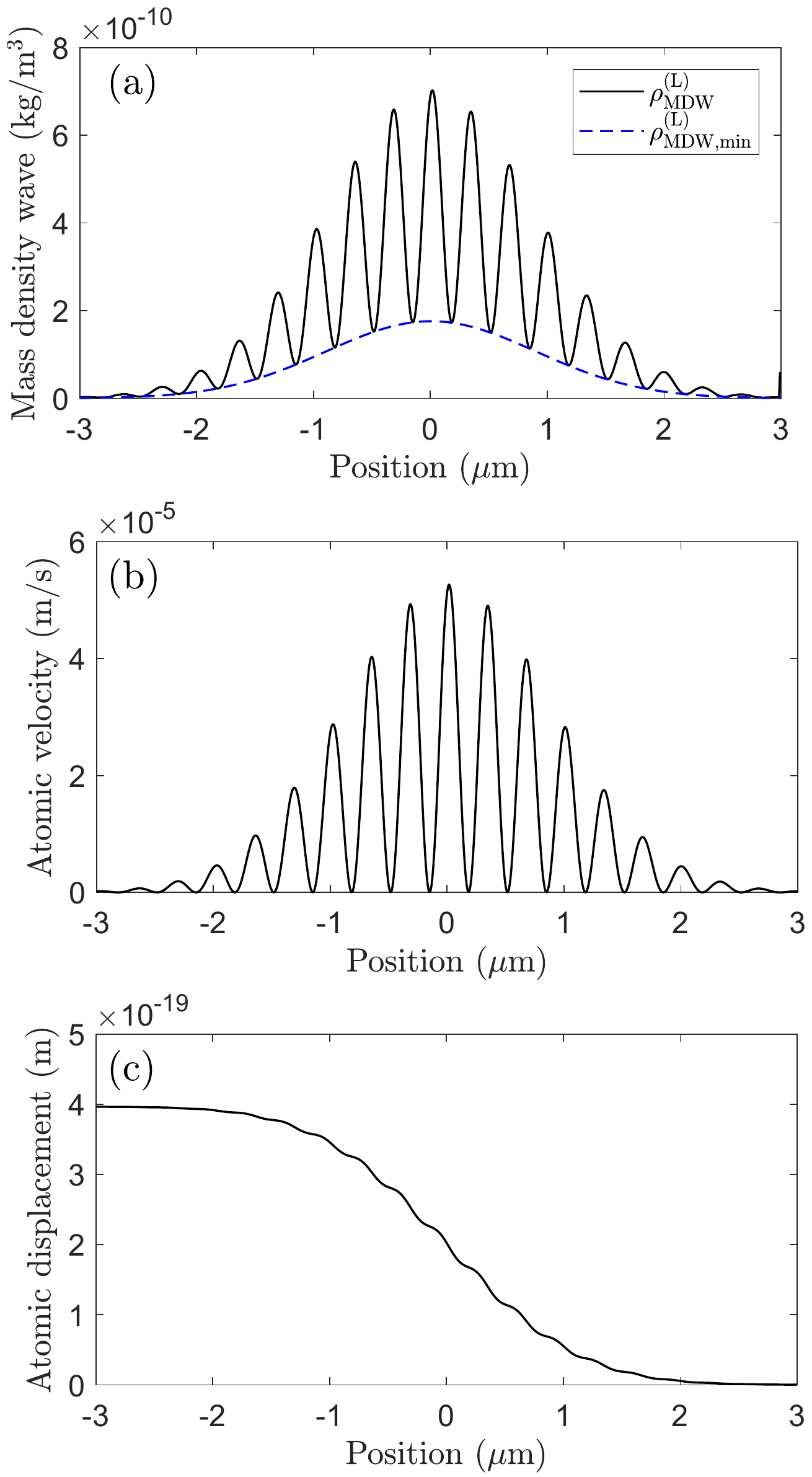}
\vspace{-0.2cm}
\caption{\label{fig:simulation}
Simulation of the atomic MDW in the L frame of a dispersive dielectric medium, where the dispersion is linear around the central frequency with the phase refractive index $n_\mathrm{p}^\mathrm{(L)}=1.2$ and the group refractive index $n_\mathrm{g}^\mathrm{(L)}=2$. (a) The mass density of the atomic MDW and its lower envelope, (b) the atomic velocity, and (c) the atomic displacement as a function of the position. All figures correspond to the same instance of time when the center of the Gaussian light pulse is at $z=0$ $\mu$m. The Gaussian light pulse has a central wavelength of $\lambda_0^\mathrm{(L)}=800$ nm, a field energy $E_\mathrm{field}^\mathrm{(L)}=500$ nJ per circular cross-sectional area of diameter $d=100$ $\mu$m, and a temporal FWHM of $\Delta t_\mathrm{FWHM}^\mathrm{(L)}=14$ fs.}
\vspace{-0.2cm}
\end{figure}

\subsection{Simulation results}

Figure \ref{fig:simulation}(a) shows the simulated atomic MDW as a function of the position
when the light pulse is propagating at the position $z=0$ $\mu$m.
The time-dependent simulation is presented
as a video file in the Supplemental Material \cite{supplementaryvideo}.
The MDW mass density $\rho_\mathrm{MDW}$ is the difference of the disturbed mass
density $\rho_\mathrm{a}^\mathrm{(L)}$
and the equilibrium mass density $\rho_\mathrm{a0}^\mathrm{(L)}$
of the medium and it is obtained by solving
Newton's equation of motion with the optical force density in Eq.~\eqref{eq:AbrahamforceL} and using Eq.~\eqref{eq:rhoa}.
Figure \ref{fig:simulation}(a) also presents the lower envelope function of the atomic MDW, given by $\rho_\mathrm{MDW,min}^\mathrm{(L)}=\rho_\mathrm{a,min}^\mathrm{(L)}-\rho_\mathrm{a0}^\mathrm{(L)}$.
The higher and lower envelopes of the MDW clearly follow the Gaussian
form of the pulse as expected. As the light pulse
is not very long compared to the harmonic cycle,
the node structure of the MDW can be seen in the
same scale with the Gaussian envelope.
The energy density of the field has the same node structure as the MDW in Fig.~\ref{fig:simulation}. The nonzero lower envelope function is a result of the interference of partial waves. This same interference pattern is seen in Fig.~1 of Philbin \cite{Philbin2011} since the exclusion of the power-conversion density does not lead to observable differences in the L frame, where it is negligibly small.

When we integrate the MDW mass density
in Fig.~\ref{fig:simulation}(a) over the volume, we obtain the total
transferred mass of $\delta M=7.79\times 10^{-24}$ kg. Dividing this by
the photon number of the light pulse, we then obtain the value of
$2.17$ eV/$c^2$ for the transferred mass per photon.
Within the relative error of $9\times10^{-4}$, this equals the MP
quasiparticle value, given by $(n_\mathrm{p}^\mathrm{(L)}n_\mathrm{g}^\mathrm{(L)}-1)\hbar\omega_0^\mathrm{(L)}/c^2$.
For a more detailed discussion of the correspondence between
the MP quasiparticle model and the OCD simulations, see Ref.~\cite{Partanen2017e}.

In illustrating the excess mass density $\rho_\mathrm{MDW}^\mathrm{(L)}=\rho_\mathrm{a}^\mathrm{(L)}-\rho_\mathrm{a0}^\mathrm{(L)}$ in Ref.~\cite{Partanen2017e}, we used an approximative formula $\rho_\mathrm{MDW}^\mathrm{(L)}\approx\rho_\mathrm{a0}^\mathrm{(L)}v_\mathrm{a}^\mathrm{(L)}/v_\mathrm{g}^\mathrm{(L)}$, which is not a fair \emph{local} approximation of $\rho_\mathrm{MDW}^\mathrm{(L)}$ for \emph{dispersive} media, even if the volume integrals of densities are closely equal. For a \emph{nondispersive} medium, the approximation above is extremely accurate also locally when the kinetic energy of the atoms in the L frame is negligible \cite{Partanen2017c}. 

Figure \ref{fig:simulation}(b) depicts the atomic velocity as a function of the position, corresponding to the MDW in Fig.~\ref{fig:simulation}(a). The atomic velocity follows the Gaussian form of the pulse as the mass density of the atomic MDW in Fig.~\ref{fig:simulation}(a). However, in contrast to the total mass density of the MDW, the atomic velocity becomes zero at the nodes of the electromagnetic field, in which the optical force is also zero.
The momentum of atoms in the MDW is obtained in the L frame by volume integrating the classical momentum density component of $\mathbf{T}_\mathrm{mat}^\mathrm{(L)}$ in Eq.~\eqref{eq:MATSEML}
at an arbitrary time.

Figure \ref{fig:simulation}(c) shows the atomic displacements corresponding to the MDW in Fig.~\ref{fig:simulation}(a), again,
as a function of the position.
On the left of the light pulse,
the atomic displacement has saturated to a constant value of
$r_\mathrm{a,max}=4.0\times10^{-19}$ m.
Within the relative error of $10^{-4}$, we obtain
$r_\mathrm{a,max}=\delta M/(\rho_\mathrm{a0}^\mathrm{(L)}A)$. The leading edge of the
optical pulse is propagating to the right approximately at the position $z=2.5$ $\mu$m. Therefore, to the right of $z=2.5$ $\mu$m, the atomic displacement is zero. The optoelastically driven MDW is manifested by the fact that the atoms are more densely spaced at the position of the light pulse as the atoms on the left of the pulse have been displaced forward and the atoms
on the right of the pulse are still at their equilibrium positions.

\begin{figure}
\centering
\includegraphics[width=0.95\columnwidth]{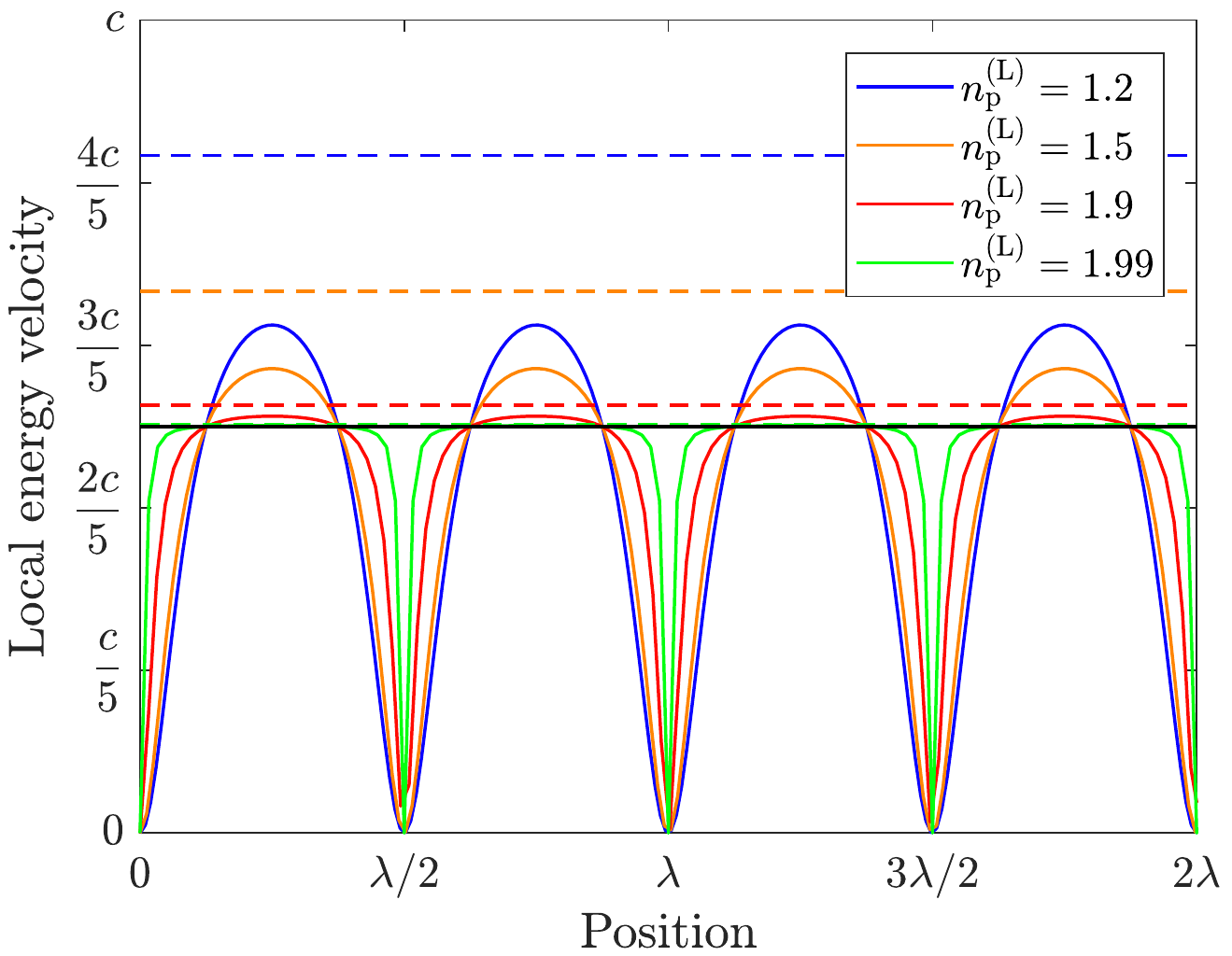}
\vspace{-0.2cm}
\caption{\label{fig:velocity}
Local energy velocity of light in the L frame as a function of the position in units of the wavelength of light in the medium for phase refractive indices $n_\mathrm{p}^\mathrm{(L)}=1.2$, $1.5$, $1.9$, $1.99$, when the group refractive index is fixed to $n_\mathrm{g}^\mathrm{(L)}=2$. The solid curves depict the local velocities for each refractive index and the dashed lines with the same colors are the corresponding phase velocities. The horizontal black solid line presents the group velocity. Note that the horizontal scaling of each graph depends on the refractive index of the medium through the central wavelength of light in the medium, given by $\lambda=\lambda^\mathrm{(L)}=\lambda_0^\mathrm{(L)}/n_\mathrm{p}^\mathrm{(L)}$, where $\lambda_0^\mathrm{(L)}$ is the central wavelength in vacuum.}
\vspace{-0.3cm}
\end{figure}

\begin{figure*}
\centering
\includegraphics[width=0.95\textwidth]{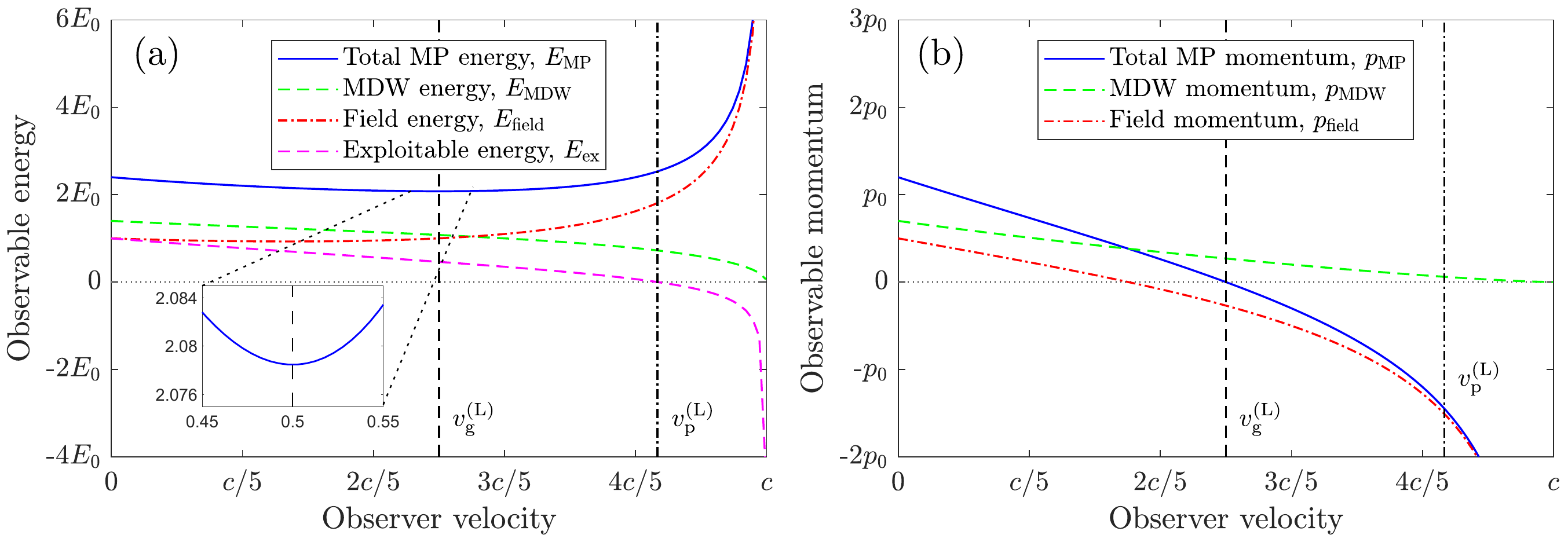}
\vspace{-0.2cm}
\caption{\label{fig:gframe}
(a) The MDW energy, the field energy, the total MP energy, and the total exploitable energy and (b) the field momentum, the MDW momentum, and the total MP momentum as a function of the velocity of an inertial observer moving in the L frame parallel to the propagation velocity of light. The energies and momenta in this figure are volume integrals of the corresponding densities. The energy is given in units of $E_0=E_\mathrm{ex}^\mathrm{(L)}$, which is the exploitable energy of light in the L frame corresponding to observer velocity $v=0$. The momentum is given in units of $p_0=E_0/c$. The phase refractive index is $n_\mathrm{p}^\mathrm{(L)}=1.2$ and the group refractive index is $n_\mathrm{g}^\mathrm{(L)}=2$ in the L frame, for which $v=0$. The vertical dashed line corresponds to the group velocity and the vertical dash-dotted line corresponds to the phase velocity. From this figure, we conclude that the total MP energy obtains its minimum value and the MP momentum becomes zero at the group velocity. Thus, the inertial frame propagating with the group velocity of light is the rest frame of light in accordance with the definition of the rest frame in the special theory of relativity.}
\vspace{-0.2cm}
\end{figure*}

\subsection{Local energy velocity of light}

The energy density and momentum density components of the SEM tensor of the field in Eq.~\eqref{eq:EMSEML} have different position and time dependencies in a dispersive medium. Therefore, it is interesting to study their ratio, which is the local propagation velocity of the energy density of the field $\mathbf{v}_\mathrm{field}^\mathrm{(L)}=c^2\mathbf{G}_\mathrm{field}^\mathrm{(L)}/W_\mathrm{field}^\mathrm{(L)}$. It can be shown that this velocity is in the L frame very accurately equal to the corresponding quantity $\mathbf{v}_\mathrm{MDW}^\mathrm{(L)}$ of the MDW SEM tensor in Eq.~\eqref{eq:MDWSEML}. The different time and position dependencies of the local energy and momentum densities are a direct consequence of the interference of different frequency components. The energies of forward and backward propagating frequency components are added up to obtain the total energy, while the magnitudes of the corresponding momenta must be subtracted to obtain the total momentum. The unequal propagation velocities of different frequency components in a dispersive medium then lead to different time- and position dependencies of the energy and momentum densities in the real space.

Figure \ref{fig:velocity} illustrates the local energy velocity of light as a function of the position in units of the wavelength of light in the medium for phase refractive indices $n_\mathrm{p}^\mathrm{(L)}=1.2$, $1.5$, $1.9$, $1.99$, when the group refractive index is fixed to $n_\mathrm{g}^\mathrm{(L)}=2$. The illustration is made for the total distance of two wavelengths. It is seen that the local energy velocity becomes zero at the field nodes and obtains its maximum values in the middle between the nodes. In the limit of a nondispersive medium, where the phase refractive index becomes equal to the group refractive index, the local energy velocity becomes equal to the group velocity apart from the single points of the field nodes. Therefore, the local energy velocity in Fig.~\ref{fig:velocity} is an obvious generalization of the local propagation velocity of light, which is constant in a nondispersive medium.

\section{\label{sec:observer}Energy and momentum of light measured by a general inertial observer}

In this section, we study the energy and momentum of light measured by a general inertial observer moving in the L frame. The energy and momentum of light and their field and MDW parts are obtained as volume integrals of the corresponding densities for a Gaussian light pulse whose electric field is given in Eq.~\eqref{eq:electricfield}. The resulting analytic formulas are given in Appendix \ref{apx:observer}. In the illustrations of this section, the phase and group refractive indices of the medium in the L frame are assumed to be $n_\mathrm{p}^\mathrm{(L)}=1.2$ and $n_\mathrm{g}^\mathrm{(L)}=2$.

Figure \ref{fig:gframe}(a) presents the energies of the electromagnetic field, the MDW, and the coupled MP state of light as a function of the velocity of an inertial observer moving in the L frame parallel to the light. It also shows the total exploitable energy of light, calculated using Eq.~\eqref{eq:Wex}. It is seen that the total energy of the coupled MP state of light obtains its minimum value for an observer moving at the group velocity, i.e., in the rest frame of light. The energy of the field is a monotonically increasing function and its energy obtains the minimum value in the L frame, corresponding to the observer velocity equal to zero. The energy of the MDW is a monotonically decreasing function and it asymptotically becomes zero when the observer velocity approaches the vacuum velocity of light. The total exploitable energy of light is a monotonically decreasing function and it becomes zero at the phase velocity in agreement with the well-known Doppler shift. That the total exploitable energy of light becomes negative above the phase velocity is expected to be related to the fact that, for particles propagating faster than the phase velocity of light in the medium, it becomes energetically favorable to emit Cherenkov or Askaryan radiation.

Figure \ref{fig:gframe}(b) illustrates the momenta of the field, the MDW, and the coupled MP state of light as a function of the velocity of the observer in the L frame. All momenta decrease monotonically; their values become smaller or more negative when the observer is moving faster. The momentum of the field changes its sign well before the group velocity. The momentum of the MDW is always positive and it asymptotically approaches zero when the observer velocity approaches the vacuum velocity of light. This result for the momentum of the MDW is intuitive since it must necessarily stay positive as the atomic velocities resulting from the optical force are larger in the forward direction compared to the atomic velocities in the equilibrium state of the medium. The momentum of the coupled MP state is seen to become zero at the group velocity. The zero total momentum of light at the group velocity agrees with the results of the MP quasiparticle model presented in Ref.~\cite{Partanen2017e}.

In summary, from Figs.~\ref{fig:gframe}(a) and \ref{fig:gframe}(b), we conclude that the total MP energy obtains its minimum value and the MP momentum becomes zero at the group velocity. Thus, the inertial frame propagating with the group velocity of light is the \emph{rest frame of light} in accordance with the definition of the rest frame in the special theory of relativity.

One can also conclude, on the basis of Fig.~\ref{fig:gframe}, that the total energy, momentum, and the total transferred mass $\delta M^\mathrm{(L)}=\int\rho_\mathrm{MDW}^\mathrm{(L)}d^3r$ fulfill the single MP quasiparticle theory as defined in Ref.~\cite{Partanen2017e}. Transforming the total energy of the coupled MP state from the L frame, $E_\mathrm{MP}^\mathrm{(L)}=E_\mathrm{ex}^\mathrm{(L)}+\delta M^\mathrm{(L)}c^2$, to the frame moving with the phase velocity of light (F frame) using Eq.~\eqref{eq:intEmp}, we obtain $E_\mathrm{MP}^\mathrm{(F)}=\gamma_\mathrm{p}^\mathrm{(L)}(E_\mathrm{ex}^\mathrm{(L)}-v_\mathrm{p}^\mathrm{(L)}p_\mathrm{MP}^\mathrm{(L)})+\gamma_\mathrm{p}^\mathrm{(L)}\delta M^\mathrm{(L)}c^2$. Comparing this with Fig.~\ref{fig:gframe}(a) we conclude that, in the inertial frame moving with $v_\mathrm{p}^\mathrm{(L)}$, the first term on the right-hand side representing the total exploitable energy in the F frame becomes zero, thus giving $p_\mathrm{MP}^\mathrm{(L)}=n_\mathrm{p}^\mathrm{(L)}E_\mathrm{ex}^\mathrm{(L)}/c$. Second, from Fig.~\ref{fig:gframe}, we conclude that the total momentum of the MP state of light is zero in the frame moving with the group velocity (R frame). From Eq.~\eqref{eq:intPmp}, we obtain $p_\mathrm{MP}^\mathrm{(R)}=\gamma_\mathrm{\mathbf{v}_\mathrm{g}}^\mathrm{(L)}(p_\mathrm{MP}^\mathrm{(L)}-v_\mathrm{g}^\mathrm{(L)}E_\mathrm{ex}^\mathrm{(L)}/c^2-v_\mathrm{g}^\mathrm{(L)}\delta M^\mathrm{(L)})$. Setting $p_\mathrm{MP}^\mathrm{(R)}$ to zero and using the value of $p_\mathrm{MP}^\mathrm{(L)}$ obtained above, gives $\delta M^\mathrm{(L)}=(n_\mathrm{p}^\mathrm{(L)}n_\mathrm{g}^\mathrm{(L)}-1)E_\mathrm{ex}^\mathrm{(L)}/c^2$, again in agreement with the MP quasiparticle model of Ref.~\cite{Partanen2017e}.

\section{\label{sec:discussion}Discussion}

\subsection{SEM tensors for nondispersive media}

The SEM tensors for a nondispersive medium are obtained from the SEM tensors of the present work by setting the phase and group refractive indices equal. In distinction to previous work in Ref.~\cite{Partanen2019a}, we have accounted for the kinetic energy of the medium through the power-conversion density in the dynamical equations of the field and the medium. Therefore, there is a difference between the definitions of the SEM tensors in the present work and in Ref.~\cite{Partanen2019a}. This difference in the definitions, which is small in the L frame and essential in the G frame, is discussed in more detail in Appendix \ref{apx:comparison}.

\subsection{Experimental evidence}

A previous experiment that is accurate enough for observing a difference between the phase and group refractive indices in the measurement of optical forces was reported by Jones \emph{et al.} \cite{Jones1978}. In this experiment, the radiation force of  light reflected from a mirror immersed in various liquids was found to be $n_\mathrm{p}^\mathrm{(L)}$ times the force of the same light on a mirror in vacuum. Therefore, there is full agreement between the force measured in the experiment by Jones \emph{et al.} \cite{Jones1978} and the change of the total momentum of the MP state per unit time when the light is reflected from a mirror in a liquid. Another experiment, which is capable of studying the momentum of light in dispersive media is that by Campbell \emph{et al.}~\cite{Campbell2005}. This experiment studied the recoil of atoms in a dilute Bose-Einstein condensate gas and showed the same $n_\mathrm{p}^\mathrm{(L)}$ proportionality of the momentum of light as the experiment by Jones \emph{et al.} \cite{Jones1978}. So far, there are no reported experiments on the transferred mass of the MDW. However, possible experimental setups for the measurement of the displacement of atoms by a light pulse have been suggested \cite{Partanen2017e,Partanen2018a,Partanen2019d,Partanen2019f}.

\subsection{Energy density of light in dispersive media}

There are very few theoretical studies of the energy density of the field in a dispersive medium describing \emph{both} the time- and position-dependence of the energy density \cite{Philbin2011,Vorobyev2012,Vorobyev2013,Ruppin2002,Yaghjian2007}. The theoretical challenge is partly related to the description of losses, but the enigma is present also for lossless media.
Typically, textbooks of electromagnetism \cite{Jackson1999,Landau1984} present the energy density of a dispersive medium in a form \emph{averaged} over the harmonic cycle, which corresponds to Eq.~\eqref{eq:energyfieldaverageL} of the present work, where the averaged energy exchange density $\big\langle W_\phi^\mathrm{(L)}\big\rangle$ is typically neglected. This time-averaged form cannot be used in the \emph{local} conservation laws in Eqs.~\eqref{eq:continuityfieldG} and \eqref{eq:momentumconservationfieldG}, which are crucial for the covariant description of the field-medium interaction. From this perspective, the chosen approach of defining the local energy density of the field using the conservation law of energy with the well-known Poynting vector is the most natural, and it was previously used, e.g., by Philbin \cite{Philbin2011}. This approach leads to additional energy density terms related to the interference of partial waves, as expected. If Philbin had used the generalized optical force density and the associated power-conversion density, he would have ended up to the present theory.

\subsection{Momentum of light in previous theories}

In the literature, the two most widely used momentum densities of light are the Minkowski and Abraham momentum densities. The Minkowski momentum density is commonly defined as $\mathbf{G}_\mathrm{M}^\mathrm{(L)}=\mathbf{D}^\mathrm{(L)}\times\mathbf{B}^\mathrm{(L)}$ and the Abraham momentum density as $\mathbf{G}_\mathrm{A}^\mathrm{(L)}=\mathbf{E}^\mathrm{(L)}\times\mathbf{H}^\mathrm{(L)}/c^2$ \cite{Barnett2010a}. For \emph{nondispersive} media with $n_\mathrm{p}^\mathrm{(L)}=n_\mathrm{g}^\mathrm{(L)}=n^\mathrm{(L)}$, these momentum densities correspond to single-photon expectation values in the L frame, given by $p_\mathrm{M}^\mathrm{(L)}=n^\mathrm{(L)}\hbar\omega_0^\mathrm{(L)}/c$ and $p_\mathrm{A}^\mathrm{(L)}=\hbar\omega_0^\mathrm{(L)}/(n^\mathrm{(L)}c)$.
For \emph{dispersive} media, Garrison \emph{et al.}~\cite{Garrison2004}
have calculated the single-photon expectation
value of the volume integral of $\mathbf{G}_\mathrm{M}^\mathrm{(L)}$, and obtained $p_\mathrm{M}^\mathrm{(L)}=(n_\mathrm{p}^\mathrm{(L)})^2\hbar\omega_0^\mathrm{(L)}/(n_\mathrm{g}^\mathrm{(L)}c)$. Correspondingly, the Abraham momentum density $\mathbf{G}_\mathrm{A}^\mathrm{(L)}$ gives for dispersive media $p_\mathrm{A}^\mathrm{(L)}=\hbar\omega_0^\mathrm{(L)}/(n_\mathrm{g}^\mathrm{(L)}c)$. Neither of these values is equal to the single-photon momentum value $p^\mathrm{(L)}=n_\mathrm{p}^\mathrm{(L)}\hbar\omega_0^\mathrm{(L)}/c$ following from the de Broglie hypothesis, from the measurements of Jones \emph{et al.}~\cite{Jones1978} and Campbell \emph{et al.}~\cite{Campbell2005}, or from the present theory. In the present theory, the MP momentum $p_\mathrm{MP}^\mathrm{(L)}=n_\mathrm{p}^\mathrm{(L)}\hbar\omega_0^\mathrm{(L)}/c$ is obtained for the single-photon expectation value calculated using the total MP momentum density $\mathbf{G}_\mathrm{MP}^\mathrm{(L)}=\rho_\mathrm{a}^\mathrm{(L)}\mathbf{v}_\mathrm{a}^\mathrm{(L)}+\mathbf{E}^\mathrm{(L)}\times\mathbf{H}^\mathrm{(L)}/c^2$. A more detailed comparison with previous theories is left to another context.

\section{\label{sec:conclusions}Conclusions}

Starting from the generalized optical force density and the momentum density of the field, we have formulated the covariant theory of light in a dispersive medium using the conservation laws and the dynamical equations of the field and the medium. The present work extends the previously developed stress-energy-momentum (SEM) tensor formalism of the mass-polariton (MP) theory of light \cite{Partanen2019a,Partanen2019b} to include the effects of dispersion. The MP theory is also made fundamentally more complete by including the energy exchange between the field and the medium through the power-conversion density. The integrals of the energy and momentum densities of the present work reproduce the results of the previously introduced MP quasiparticle model of a dispersive medium \cite{Partanen2017e}. Since the kinetic energy of the medium is in the present work included in the energy density of the medium, there is also a related correction in the energy density of the field. Therefore, the field and medium SEM tensors of the present work are not identical to the corresponding SEM tensors of previous works on the MP theory in the G frame, where the kinetic energy density of the medium is not small.

Since both of the energy and momentum conversion source terms are accounted for in the energy and momentum conservation laws, all SEM tensors of the present work are symmetric and correspond to the most fundamental definition of the SEM tensor. Consequently, even for the field and medium subsystems, we have found no need to introduce asymmetric SEM tensors, such as the Minkowski SEM tensor, which have been studied in previous literature \cite{Penfield1967,Kemp2017,Pfeifer2007}. Using asymmetric SEM tensors in previous works can be seen to be partly related to the negligence of the optical-force-driven atomic rest energy transfer associated with the atomic MDW.

In addition to being symmetric, all SEM tensors of the present work also satisfy the Lorentz transformation of second-rank tensors. Since the four-divergence of the SEM tensor of the coupled MP state of light is zero, the volume integrals of the energy and momentum density components of the MP SEM tensor form a four-vector, which is the four-momentum of light in agreement with the special theory of relativity. The four-divergences of the SEM tensors of the field and medium subsystems are within the sign difference equal to the generalized optical four-force density introduced in the present work.

The final outcome of the theoretical derivation is more sophisticated than in the MP theory for nondispersive media, as anticipated from the feature-rich interference patterns of partial waves propagating at different phase velocities. One of the most prominent features of this interference in dispersive media is the appearance of fluctuational terms in the key physical quantities. In addition to the constant phase and group velocities, which do not depend on the position and time, the theory also includes local velocities of the field and the MDW. The energy-density-weighted averages of these position- and time-dependent velocities are equal to the group velocity of light. Our theoretical work has also produced a number of other side results, from which we want to mention the interpretation of the integral of the MP SEM tensor contraction normalized by the photon number as the Lagrangian of the MP quasiparticle.

For negative-index metamaterials, the present theory has been applied in a preprint \cite{Partanen2021c}. The derivation of the MP theory in a dispersive medium from the Lagrangian densities of the field and the medium is left as a topic of a future work. This would be a generalization of the corresponding derivation for a nondispersive medium presented in an earlier work \cite{Partanen2019a}. The detailed description of the MP theory of light in the presence of free electric charges, free electric currents, and losses is also left as a topic of a future work. The present theory enables detailed study of the optical effects at material interfaces by including to it the well-known term originating from the gradient of the refractive index. However, this term must be generalized for a dispersive medium first. The present classical theory must also be extended to the quantum domain to treat the quantum mechanical spin of light.


\begin{acknowledgments}
This work has been funded by the Academy of Finland under Contract No.~318197 
and European Union's Horizon 2020 Marie Sk\l{}odowska-Curie Actions (MSCA) 
individual fellowship DynaLight under Contract No.~846218. Mathematica
has been extensively used to verify the equations of the present work.
\end{acknowledgments}

\appendix

\section{\label{apx:quasiparticlemodel}Summary of the MP quasiparticle model for dispersive media}

The present SEM tensor analysis of light in dispersive media is based on the key properties of the MP quasiparticle analysis and OCD simulations of light in dispersive media presented in Ref.~\cite{Partanen2017e}. The MP quasiparticle model is a kinematic relativistic model, which one can use to study the Lorentz transformations of the energy and momentum of the field and the medium between different inertial frames. This model has two key properties: \emph{First}, the total exploitable energy of light vanishes in the F frame propagating at the phase velocity. This is in accordance with the experimentally accurately verified Doppler shift of light in dispersive media \cite{ChenJi2011}. \emph{Second}, the total momentum of the coupled MP state of light becomes zero in the R frame propagating at the group velocity. This inertial frame is then the rest frame of light as defined in the special theory of relativity. The results of the MP quasiparticle model for dispersive media are given for the L frame in the last column of Table \ref{tbl:quasiparticle}. The relation of the MP model to the generalized optical force density in Eq.~\eqref{eq:AbrahamforceL} is briefly discussed below.

\begin{table*}[ht]
 \setlength{\tabcolsep}{4.5pt}
 \renewcommand{\arraystretch}{1.4}
 \caption{\label{tbl:quasiparticle}
 The relations of the energies $E_\mathrm{MP}^\mathrm{(L)}$, $E_\mathrm{field}^\mathrm{(L)}$, and $E_\mathrm{MDW}^\mathrm{(L)}$ and the momenta $\mathbf{p}_\mathrm{MP}^\mathrm{(L)}$, $\mathbf{p}_\mathrm{field}^\mathrm{(L)}$, and $\mathbf{p}_\mathrm{MDW}^\mathrm{(L)}$ of the MP state, the field, and the MDW in the L frame calculated using the OCD simulations and the MP quasiparticle model of Ref.~\cite{Partanen2017e}. In contrast to the present work, the extremely small energy exchange density $W_\phi^\mathrm{(L)}$ and energy $E_\phi^\mathrm{(L)}$ in the L frame have been set to zero in Ref.~\cite{Partanen2017e}. The integrands of $E_\mathrm{MP}^\mathrm{(L)}$ and $E_\mathrm{field}^\mathrm{(L)}$ in this table are not the actual time- and position-dependent energy densities, but, to simplify comparison with Ref.~\cite{Partanen2017e}, they are written in terms of the Brillouin form of Eq.~\eqref{eq:energyfieldaverageL} with time average over the harmonic cycle removed as in Ref.~\cite{Partanen2017e}. This gives the same value for $E_\mathrm{field}^\mathrm{(L)}$ as the \emph{integral} of the actual energy density of the field described in the present work. With defining the photon number $N_0=E_\mathrm{ex}^\mathrm{(L)}/\hbar\omega_0^\mathrm{(L)}$, the values in the OCD and MP columns are pairwise equal when in the \emph{L frame} very small energy exchange between the field and the kinetic energy of the medium is set to zero.}
\begin{tabular}{ccc}
   \hline\hline\\[-11pt]
   Quantity & OCD & MP \\[4pt]
   \hline\\[-12pt]
   $E_\mathrm{MP}^\mathrm{(L)}=E_\mathrm{field}^\mathrm{(L)}+E_\mathrm{MDW}^\mathrm{(L)}$ & $\displaystyle\int\Big\{\frac{1}{2}\Big[\frac{d(\omega_0^\mathrm{(L)}\epsilon^\mathrm{(L)})}{d\omega_0^\mathrm{(L)}}|\mathbf{E}^\mathrm{(L)}|^2+\frac{d(\omega_0^\mathrm{(L)}\mu^\mathrm{(L)})}{d\omega_0^\mathrm{(L)}}|\mathbf{H}^\mathrm{(L)}|^2\Big]-W_\phi^\mathrm{(L)}+\rho_\mathrm{MDW}^\mathrm{(L)}c^2\Big\}d^3r$ & $\displaystyle n_\mathrm{p}^\mathrm{(L)}n_\mathrm{g}^\mathrm{(L)}N_0\hbar\omega_0^\mathrm{(L)}$\\[4pt]
   $E_\mathrm{field}^\mathrm{(L)}$ & $\displaystyle\int\Big\{\frac{1}{2}\Big[\frac{d(\omega_0^\mathrm{(L)}\epsilon^\mathrm{(L)})}{d\omega_0^\mathrm{(L)}}|\mathbf{E}^\mathrm{(L)}|^2+\frac{d(\omega_0^\mathrm{(L)}\mu^\mathrm{(L)})}{d\omega_0^\mathrm{(L)}}|\mathbf{H}^\mathrm{(L)}|^2\Big]-W_\phi^\mathrm{(L)}\Big\}d^3r$ & $N_0\hbar\omega_0^\mathrm{(L)}$\\
   $E_\mathrm{MDW}^\mathrm{(L)}$ & $\displaystyle\int\rho_\mathrm{MDW}^\mathrm{(L)}c^2d^3r$ & $\displaystyle(n_\mathrm{p}^\mathrm{(L)}n_\mathrm{g}^\mathrm{(L)}-1)N_0\hbar\omega_0^\mathrm{(L)}$\\[4pt]
   $\mathbf{p}_\mathrm{MP}^\mathrm{(L)}$       & $\displaystyle\int\Big(\frac{\mathbf{E}^\mathrm{(L)}\times\mathbf{H}^\mathrm{(L)}}{c^2}+\rho_\mathrm{a}^\mathrm{(L)}\mathbf{v}_\mathrm{a}^\mathrm{(L)}\Big)d^3r$ & $\displaystyle\frac{n_\mathrm{p}^\mathrm{(L)}N_0\hbar\omega_0^\mathrm{(L)}}{c}\hat{\mathbf{v}}_\mathrm{p}$ \\[4pt]
   $\mathbf{p}_\mathrm{field}^\mathrm{(L)}$      & $\displaystyle\int\frac{\mathbf{E}^\mathrm{(L)}\times\mathbf{H}^\mathrm{(L)}}{c^2}d^3r$ & $\displaystyle\frac{N_0\hbar\omega_0^\mathrm{(L)}}{n_\mathrm{g}^\mathrm{(L)}c}\hat{\mathbf{v}}_\mathrm{p}$ \\[4pt]
   $\mathbf{p}_\mathrm{MDW}^\mathrm{(L)}$             & $\displaystyle\int\rho_\mathrm{a}^\mathrm{(L)}\mathbf{v}_\mathrm{a}^\mathrm{(L)}d^3r$ & $\displaystyle\Big(n_\mathrm{p}^\mathrm{(L)}-\frac{1}{n_\mathrm{g}^\mathrm{(L)}}\Big)\frac{N_0\hbar\omega_0^\mathrm{(L)}}{c}\hat{\mathbf{v}}_\mathrm{p}$ \\[8pt]
   $\delta M^\mathrm{(L)}=E_\mathrm{MDW,rest}^\mathrm{(L)}/c^2$ & $\displaystyle\int\rho_\mathrm{MDW,rest}^\mathrm{(L)}d^3r$ & $\displaystyle(n_\mathrm{p}^\mathrm{(L)}n_\mathrm{g}^\mathrm{(L)}-1)\frac{N_0\hbar\omega_0^\mathrm{(L)}}{c^2}$ \\[4pt]
   \hline\hline
 \end{tabular}
\end{table*}

In Ref.~\cite{Partanen2017e}, we furthermore compared the MP quasiparticle model results with the OCD simulations of the momentum density and the nonequilibrium mass density of the medium, moving under the influence of the optical force. These numerical simulations were carried out by solving Newton's equation of the medium for a given generalized expression of the optical force \cite{Partanen2017e}. The results of the OCD simulations are summarized in the middle column of Table \ref{tbl:quasiparticle}. For \emph{nondispersive} media, all integrands in Table \ref{tbl:quasiparticle} present as such the densities of the integrated physical quantities. For \emph{dispersive} media, discussed in the present work, the densities are generally more complicated as one should expect from the anticipated complicated interference of partial waves propagating at different phase velocities.

The MP quasiparticle model of Ref.~\cite{Partanen2017e} gives in Eq.~(27) of Ref.~\cite{Partanen2017e} the field part of the MP quasiparticle momentum $p_\mathrm{field}^\mathrm{(L)}=\hbar\omega_0^\mathrm{(L)}/(n_\mathrm{g}^\mathrm{(L)}c)$ and the MDW part
$p_\mathrm{MDW}^\mathrm{(L)}=(n_\mathrm{p}^\mathrm{(L)}-1/n_\mathrm{g}^\mathrm{(L)})\hbar\omega_0^\mathrm{(L)}/c$. 
This gives the quasiparticle momentum ratio $p_\mathrm{MDW}^\mathrm{(L)}/p_\mathrm{field}^\mathrm{(L)}=n_\mathrm{p}^\mathrm{(L)}n_\mathrm{g}^\mathrm{(L)}-1$. If this same momentum ratio is assumed to apply for the momentum densities, i.e., $G_\mathrm{MDW}^\mathrm{(L)}/G_\mathrm{field}^\mathrm{(L)}=n_\mathrm{p}^\mathrm{(L)}n_\mathrm{g}^\mathrm{(L)}-1$, we can write the momentum density of the MDW in the L frame as
\begin{equation}
 \mathbf{G}_\mathrm{MDW}^\mathrm{(L)}=(n_\mathrm{p}^\mathrm{(L)}n_\mathrm{g}^\mathrm{(L)}-1)\mathbf{G}_\mathrm{field}^\mathrm{(L)}=\frac{n_\mathrm{p}^\mathrm{(L)}n_\mathrm{g}^\mathrm{(L)}-1}{c^2}\mathbf{E}^\mathrm{(L)}\times\mathbf{H}^\mathrm{(L)}.
 \label{eq:GMDW}
\end{equation}
In the second equality, we have used Eq.~\eqref{eq:fieldmomentumdensityL}.
Newton's equation of motion relates the force density experienced by the nonrelativistic medium atoms in Eq.~\eqref{eq:GMDW} to the time derivative of the momentum density of atoms. From this condition, we then obtain the expression of the generalized optical force density given in Eq.~\eqref{eq:AbrahamforceL}. Note that, in the case of nondispersive media with $n_\mathrm{p}^\mathrm{(L)}=n_\mathrm{g}^\mathrm{(L)}$, the optical force density corresponding to Eq.~\eqref{eq:AbrahamforceL} has also been derived directly from the principle of least action as detailed in Ref.~\cite{Partanen2019b}. Such a derivation is also expected to be possible in the present case of dispersive media. Together with the derivation of the expression of the generalized optical force in Eq.~\eqref{eq:AbrahamforceL} from the known interactions between the optical field and the induced dipoles, accounting also for the dipole-dipole interactions, this is an interesting topic for a future work.

\section{\label{apx:Newtonsolution}Solution of Newton's equation of motion in the L frame}

In this appendix, we present an approximate solution to Newton's equation of motion of the medium under the influence of the generalized optical force. In the L frame, the magnitude of the left-hand side of Newton's equation in Eq.~\eqref{eq:NewtonG} can be written as
\begin{align}
 &n_\mathrm{a}^\mathrm{(L)}\frac{d(\gamma_{\mathbf{v}_\mathrm{a}}^\mathrm{(L)}m_0|\mathbf{v}_\mathrm{a}^\mathrm{(L)}|)}{dt}\nonumber\\
 &=n_\mathrm{a}^\mathrm{(L)}\Big(\frac{\partial(\gamma_{\mathbf{v}_\mathrm{a}}^\mathrm{(L)}m_0|\mathbf{v}_\mathrm{a}^\mathrm{(L)}|)}{\partial t}+|\mathbf{v}_\mathrm{a}^\mathrm{(L)}|\frac{\partial(\gamma_{\mathbf{v}_\mathrm{a}}^\mathrm{(L)}m_0|\mathbf{v}_\mathrm{a}^\mathrm{(L)}|)}{\partial s}\Big)\nonumber\\
 &=n_\mathrm{a,min}^\mathrm{(L)}\frac{\partial(\gamma_{\mathbf{v}_\mathrm{a}}^\mathrm{(L)}m_0|\mathbf{v}_\mathrm{a}^\mathrm{(L)}|)}{\partial t}.
 \label{eq:materialderivative}
\end{align}
The first equality in Eq.~\eqref{eq:materialderivative} expresses the total time derivative in terms of partial derivatives with respect to time and position, i.e., it is known as the material derivative. The second equality is specific to the MP theory of light relating the local minimum number density $n_\mathrm{a,min}^\mathrm{(L)}$ to the disturbed number density $n_\mathrm{a}^\mathrm{(L)}$ \cite{Partanen2019b}. The local minimum number density $n_\mathrm{a,min}^\mathrm{(L)}$ is defined as an envelope function formed from the minimum values of $n_\mathrm{a}^\mathrm{(L)}$ in a harmonic cycle, i.e., $n_\mathrm{a}^\mathrm{(L)}$ at points where the field vectors and the resulting optical force are zero. In a nondispersive medium, $n_\mathrm{a,min}^\mathrm{(L)}$ is equal to the equilibrium number density $n_\mathrm{a0}^\mathrm{(L)}$, but in a dispersive medium it deviates from $n_\mathrm{a0}^\mathrm{(L)}$ as described below. This deviation unambiguously follows from the optical force, and thus, it also directly emerges from the OCD simulations of Sec.~\ref{sec:simulations}. Using the last equality of Eq.~\eqref{eq:materialderivative}, we can express the disturbed number density of the medium as
\begin{align}
 n_\mathrm{a}^\mathrm{(L)}
 &=\dfrac{n_\mathrm{a,min}^\mathrm{(L)}\dfrac{\partial(\gamma_{\mathbf{v}_\mathrm{a}}^\mathrm{(L)} m_0|\mathbf{v}_\mathrm{a}^\mathrm{(L)}|)}{\partial t}}{\dfrac{\partial(\gamma_{\mathbf{v}_\mathrm{a}}^\mathrm{(L)}m_0|\mathbf{v}_\mathrm{a}^\mathrm{(L)}|)}{\partial t}+|\mathbf{v}_\mathrm{a}^\mathrm{(L)}|\dfrac{\partial(\gamma_{\mathbf{v}_\mathrm{a}}^\mathrm{(L)}m_0|\mathbf{v}_\mathrm{a}^\mathrm{(L)}|)}{\partial s}}\nonumber\\
 &=\dfrac{n_\mathrm{a,min}^\mathrm{(L)}}{1-\dfrac{n_\mathrm{p}^\mathrm{(L)}|\mathbf{v}_\mathrm{a}^\mathrm{(L)}|}{c}}.
 \label{eq:rhoa0}
\end{align}
In the second equality of Eq.~\eqref{eq:rhoa0}, we have used $\frac{\partial}{\partial s}(\gamma_{\mathbf{v}_\mathrm{a}}^\mathrm{(L)}m_0|\mathbf{v}_\mathrm{a}^\mathrm{(L)}|)/\frac{\partial}{\partial t}(\gamma_{\mathbf{v}_\mathrm{a}}^\mathrm{(L)}m_0|\mathbf{v}_\mathrm{a}^\mathrm{(L)}|)=\frac{\partial}{\partial s}|\mathbf{v}_\mathrm{a}^\mathrm{(L)}|/\frac{\partial}{\partial t}|\mathbf{v}_\mathrm{a}^\mathrm{(L)}|=-n_\mathrm{p}^\mathrm{(L)}/c$, which applies in the \emph{assumed monochromatic field limit}. The right-hand side of Eq.~\eqref{eq:rhoa0} is equal to the right-hand side of Eq.~\eqref{eq:rhoasol}.

In the L frame, the atoms at equilibrium are not moving, and thus, all atomic velocities are entirely determined by the generalized optical force density in Eq.~\eqref{eq:AbrahamforceL}. Using this expression for the force density, writing the left-hand side of Newton's equation in Eq.~\eqref{eq:NewtonG} using the last form of Eq.~\eqref{eq:materialderivative}, integrating both sides of Newton's equation over time, and setting the integration constant to zero, it follows that
\begin{align}
 \gamma_{\mathbf{v}_\mathrm{a}}^\mathrm{(L)}\mathbf{v}_\mathrm{a}^\mathrm{(L)}
 &=\frac{n_\mathrm{p}^\mathrm{(L)}n_\mathrm{g}^\mathrm{(L)}-1}{n_\mathrm{a,min}^\mathrm{(L)}m_0c^2}\mathbf{E}^\mathrm{(L)}\times\mathbf{H}^\mathrm{(L)}\nonumber\\
 &\hspace{0.4cm}-\frac{1}{n_\mathrm{a,min}^\mathrm{(L)}}\int_{-\infty}^t\gamma_{\mathbf{v}_\mathrm{a}}^\mathrm{(L)}|\mathbf{v}_\mathrm{a}^\mathrm{(L)}|\frac{\partial n_\mathrm{a,min}^\mathrm{(L)}}{\partial t'}dt'.
 \label{eq:atomicvelocity}
\end{align}
Equation \eqref{eq:atomicvelocity} can be solved for $\mathbf{v}_\mathrm{a}^\mathrm{(L)}$ by setting the second term on the right hand side of this equation to zero. This approximation can be later justified using the solutions of the atomic number densities and velocities to compare the mutual magnitudes of the two terms of Eq.~\eqref{eq:atomicvelocity}. The condition at which the relative error of our approximations becomes zero is describe in more detail below. Then, using Eq.~\eqref{eq:phasevelocityL}, we obtain $\mathbf{v}_\mathrm{a}^\mathrm{(L)}$, corresponding to Eq.~\eqref{eq:atomicvelocitysol} as
\begin{equation}
 \mathbf{v}_\mathrm{a}^\mathrm{(L)}
 =\frac{\dfrac{(n_\mathrm{p}^\mathrm{(L)}n_\mathrm{g}^\mathrm{(L)}-1)W_\mathrm{ex,nd}^\mathrm{(L)}}{n_\mathrm{a,min}^\mathrm{(L)}m_0c^2}\mathbf{v}_\mathrm{p}^\mathrm{(L)}}{\sqrt{1+\Bigg[\dfrac{(n_\mathrm{p}^\mathrm{(L)}n_\mathrm{g}^\mathrm{(L)}-1)W_\mathrm{ex,nd}^\mathrm{(L)}}{n_\mathrm{p}^\mathrm{(L)}n_\mathrm{a,min}^\mathrm{(L)}m_0c^2}\Bigg]^2}}.
 \label{eq:atomicvelocitysolA}
\end{equation}

The expression of $n_\mathrm{a,min}^\mathrm{(L)}$ appearing in the expression of $n_\mathrm{a}^\mathrm{(L)}$ in Eq.~\eqref{eq:rhoa0} and $\mathbf{v}_\mathrm{a}^\mathrm{(L)}$ in Eq.~\eqref{eq:atomicvelocitysolA} is yet to be determined. Setting the last form of Eq.~\eqref{eq:rhoa0} equal to Eq.~\eqref{eq:rhoa} provides a partial differential equation from which it is complicated to find a solution for $n_\mathrm{a,min}^\mathrm{(L)}$. However, we can obtain a solution for $n_\mathrm{a,min}^\mathrm{(L)}$ using the last form of Eq.~\eqref{eq:rhoa0} and approximating Eq.~\eqref{eq:rhoa} by writing $\nabla\cdot\mathbf{r}_\mathrm{a}^\mathrm{(L)}\approx\int_{-\infty}^t\nabla\cdot\mathbf{v}_\mathrm{a}^\mathrm{(L)}dt'\approx\int_{-\infty}^t\nabla\cdot(\gamma_{\mathbf{v}_\mathrm{a}}^\mathrm{(L)}\mathbf{v}_\mathrm{a}^\mathrm{(L)})dt'\approx\frac{n_\mathrm{p}^\mathrm{(L)}n_\mathrm{g}^\mathrm{(L)}-1}{n_\mathrm{a,min}^\mathrm{(L)}m_0c^2}\int_{-\infty}^t\nabla\cdot(\mathbf{E}^\mathrm{(L)}\times\mathbf{H}^\mathrm{(L)})dt'=-\frac{n_\mathrm{p}^\mathrm{(L)}n_\mathrm{g}^\mathrm{(L)}-1}{n_\mathrm{a,min}^\mathrm{(L)}m_0c^2}\big[W_\mathrm{ex,nd}^\mathrm{(L)}+(n_\mathrm{g}^\mathrm{(L)}/n_\mathrm{p}^\mathrm{(L)}-1)\big\langle W_\mathrm{ex,nd}^\mathrm{(L)}\big\rangle\big]$. Here we have used $\mathbf{v}_\mathrm{a}^\mathrm{(L)}\approx\gamma_{\mathbf{v}_\mathrm{a}}^\mathrm{(L)}\mathbf{v}_\mathrm{a}^\mathrm{(L)}$, for which we then apply Eq.~\eqref{eq:atomicvelocity} and take $n_\mathrm{a,min}^\mathrm{(L)}$ outside the integration and differentiation. For the resulting integral, we have used Eq.~\eqref{eq:derivation} below. Substituting the final result of this calculation to the right hand side of Eq.~\eqref{eq:rhoa}, which is set equal to the right hand side of Eq.~\eqref{eq:rhoa0} at a position where the fields and consequently $W_\mathrm{ex,nd}^\mathrm{(L)}$ and $\mathbf{v}_\mathrm{a}^\mathrm{(L)}$ are zero at a given time, and solving for $n_\mathrm{a,min}^\mathrm{(L)}$ gives
\begin{equation}
 n_\mathrm{a,min}^\mathrm{(L)}=n_\mathrm{a0}^\mathrm{(L)}+(n_\mathrm{p}^\mathrm{(L)}n_\mathrm{g}^\mathrm{(L)}-1)\Big(\frac{n_\mathrm{g}^\mathrm{(L)}}{n_\mathrm{p}^\mathrm{(L)}}-1\Big)\frac{\big\langle W_\mathrm{ex,nd}^\mathrm{(L)}\big\rangle}{m_0c^2}.
 \label{eq:rhoamin}
\end{equation}
This corresponds to Eq.~\eqref{eq:namin}. Thus, we have now derived the approximative solution of Newton's equation of motion of the medium in the L frame, presented in Sec.~\ref{sec:Newtonsolution}. The accuracy of the approximations is discussed below.

When $n_\mathrm{a}^\mathrm{(L)}$ and $\mathbf{v}_\mathrm{a}^\mathrm{(L)}$ from Eqs.~\eqref{eq:rhoasol} and \eqref{eq:atomicvelocitysol} are substituted into Newton's equation in Eq.~\eqref{eq:NewtonG}, in the conservation law of the number of atoms in Eq.~\eqref{eq:numberconservationG}, or in the conservation laws of energy and momentum in Eqs.~\eqref{eq:masscontinuityG} and \eqref{eq:momentumcontinuityG}, it is numerically found that these equations are satisfied exactly at all positions and times in the monochromatic field limit independently of the used field strength and the equilibrium number density of the medium. However, even if the approximations are locally very accurate, their global justification deserves a second thought. In the derivations above, we have set the second term of Eq.~\eqref{eq:atomicvelocity} to zero. When $n_\mathrm{a,min}^\mathrm{(L)}$ and $\mathbf{v}_\mathrm{a}$ are used inside this integral, the relative error of this approximation is found to become negligible when the following condition is satisfied:
\begin{equation}
 \frac{|n_\mathrm{a,min}^\mathrm{(L)}-n_\mathrm{a0}^\mathrm{(L)}|}{n_\mathrm{a0}^\mathrm{(L)}}
 \!=\!\Big|(n_\mathrm{p}^\mathrm{(L)}n_\mathrm{g}^\mathrm{(L)}-1)\Big(\frac{n_\mathrm{g}^\mathrm{(L)}}{n_\mathrm{p}^\mathrm{(L)}}-1\Big)\frac{\big\langle W_\mathrm{ex,nd}^\mathrm{(L)}\big\rangle}{n_\mathrm{a0}^\mathrm{(L)}m_0c^2}\Big|\!\ll 1.
 \label{eq:accuracycondition}
\end{equation}
We see that the left-hand side of Eq.~\eqref{eq:accuracycondition} is zero for nondispersive media, and accordingly, Eq.~\eqref{eq:accuracycondition}  is necessarily satisfied. For dispersive media, the rest energy density of the medium, $\rho_\mathrm{a0}^\mathrm{(L)}c^2=n_\mathrm{a0}^\mathrm{(L)}m_0c^2$, is several orders of magnitude larger than $\big\langle W_\mathrm{ex,nd}^\mathrm{(L)}\big\rangle$ for any realistic light field propagating in any realistic condensed material. Thus, the left-hand side of Eq.~\eqref{eq:accuracycondition} is extremely small and  the approximate solution of Newton's equation derived above is highly accurate also for light propagating in dispersive condensed media. In the case of low-density media, the phase and group refractive indices are close to unity, which makes the left-hand side of Eq.~\eqref{eq:accuracycondition} still small. Therefore, we can conclude that the present theory is applicable to a broad range of solids, liquids, and gases.

\section{\label{apx:powerconversion}Calculation of the energy exchange density in the L frame}

The energy exchange density in the L frame, $W_\phi^\mathrm{(L)}$, is defined in Eq.~\eqref{eq:WphiL} as the time integral of the power conversion density $\phi_\mathrm{opt}^\mathrm{(L)}$. This time integral can be calculated as
\begin{align}
 W_\phi^\mathrm{(L)}
 &=\int_{-\infty}^t\mathbf{f}_\mathrm{opt}^\mathrm{(L)}\cdot\mathbf{v}_\mathrm{a}^\mathrm{(L)}dt'\nonumber\\
 &=\int_{-\infty}^tn_\mathrm{a,min}^\mathrm{(L)}\frac{\partial(\gamma_{\mathbf{v}_\mathrm{a}}^\mathrm{(L)}m_0|\mathbf{v}_\mathrm{a}^\mathrm{(L)}|)}{\partial t}|\mathbf{v}_\mathrm{a}^\mathrm{(L)}|dt'\nonumber\\
 &=\int_{-\infty}^t\frac{\partial[(\gamma_{\mathbf{v}_\mathrm{a}}^\mathrm{(L)}-1)m_0n_\mathrm{a,min}^\mathrm{(L)}c^2]}{\partial t}dt'\nonumber\\
 &\hspace{0.4cm}-\int_{-\infty}^t(\gamma_{\mathbf{v}_\mathrm{a}}^\mathrm{(L)}-1)m_0c^2\frac{\partial n_\mathrm{a,min}^\mathrm{(L)}}{\partial t}dt'\nonumber\\
 &=(\gamma_{\mathbf{v}_\mathrm{a}}^\mathrm{(L)}-1)\rho_\mathrm{a,min}^\mathrm{(L)}c^2
 -\int_{-\infty}^t\!\!(\gamma_{\mathbf{v}_\mathrm{a}}^\mathrm{(L)}\!-\!1)m_0c^2\frac{\partial n_\mathrm{a,min}^\mathrm{(L)}}{\partial t}dt'\nonumber\\
 &\approx(\gamma_{\mathbf{v}_\mathrm{a}}^\mathrm{(L)}-1)\rho_\mathrm{a,min}^\mathrm{(L)}c^2.
 \label{eq:derivation3}
\end{align}
In the first equality, we have used Eq.~\eqref{eq:powerconversion}, in the second equality, we have used Eq.~\eqref{eq:materialderivative}, and in the third equality we have used the chain rule of differentiaion. In the fourth equality, in the calculation of the first term, we have used the boundary condition that $\mathbf{v}_\mathrm{a}^\mathrm{(L)}$ becomes zero at the negative infinity. In the last step, we have set the second term on the fourth line of Eq.~\eqref{eq:derivation3} to zero. For the condition at which the relative error of this approximation becomes zero, see Appendix \ref{apx:Newtonsolution}.

\section{\label{apx:sem}Definitions of the SEM tensor in previous literature}

The SEM tensor of the present work in Eq.~\eqref{eq:emt} is based on the classical definition of the energy and momentum densities and stresses \cite{Landau1989}. Here we briefly review selected other SEM tensor definitions existing in previous literature with the goal to put the used definition into a wider perspective.

\subsection{Hilbert SEM tensor}
The Hilbert SEM tensor appears naturally in the general theory of relativity. It is obtained from the action principle by varying the action $S=\int\mathcal{L}\sqrt{-g}d^4x$ with respect to the components of the metric tensor $g^{\mu\nu}$. Here $\mathcal{L}$ is the Lagrangian density of the field and the medium and $g=\mathrm{Det}(g_{\mu\nu})$ is the determinant of the matrix representation of the metric tensor. The resulting expression for the Hilbert SEM tensor is given by \cite{Misner1973,Landau1989}
\begin{equation}
 (T_\mathrm{H})_{\mu\nu}=\frac{2}{\sqrt{-g}}\frac{\delta(\sqrt{-g}\mathcal{L})}{\delta g^{\mu\nu}}=2\frac{\delta\mathcal{L}}{\delta g^{\mu\nu}}-g_{\mu\nu}\mathcal{L}.
 \label{eq:HSEM}
\end{equation}
The Hilbert SEM tensor in Eq.~\eqref{eq:HSEM} is both symmetric and gauge-invariant. In the general theory of relativity, the symmetry of the Hilbert SEM tensor follows naturally from the unavoidable symmetry of the metric tensor. At least for the total isolated system of the field and the medium, it can be expected that the Hilbert SEM tensor is equivalent to the classical definition of the SEM tensor when \emph{all forms} of energy and momentum densities and stresses are included.

\subsection{Canonical SEM tensor}
The SEM tensor reviewed in this section has been given the name canonical SEM tensor because it is obtained from an appropriate Lagrangian density via Noether's theorem. In the pertinent Lagrangian density, the electromagnetic field is presented by using the components of the electromagnetic four-potential $A_\alpha$ as generalized coordinates. Then, the variation of the action with respect to the four-potential components results in the expression of the canonical SEM tensor, given by \cite{Landau1989}
\begin{align}
 (T_\mathrm{C})^{\mu\nu} =\frac{\partial\mathcal{L}}{\partial(\partial_\mu A_\alpha)}\partial^\nu A_\alpha-g^{\mu\nu}\mathcal{L}.
 \label{eq:CSEM}
\end{align}
This canonical SEM tensor is generally neither symmetric nor gauge independent. The gauge dependence emerges from the well-known gauge-dependence of the electromagnetic four-potential. The derivation of the canonical SEM tensor ensures that this tensor is divergenceless in the absence of free charges and currents as $\partial_\nu(T_\mathrm{C})^{\mu\nu}=0$, but the asymmetry of $(T_\mathrm{C})^{\mu\nu}$ leads to the known violation of the classical conservation law of angular momentum in the form discussed in Sec.~\ref{sec:conservation} with the definition of the angular momentum tensor in Sec.~\ref{sec:amt}. In previous literature \cite{Mita2000,Bliokh2013b}, it has been considered that the spin of light is related to the asymmetry of the canonical SEM tensor, so that when the spin-related terms are added to it, one ends up to the symmetric Belinfante-Rosenfeld SEM tensor described below.

\subsection{Belinfante-Rosenfeld SEM tensor}

The Belinfante-Rosenfeld symmetrization \cite{Belinfante1940,Rosenfeld1940,Ramos2015,Schroder1968} aims to generate a symmetric SEM tensor from the generally asymmetric canonical SEM tensor in Eq.~\eqref{eq:CSEM}. This procedure is based on the observation that adding to any tensor a tensor of the form $\partial_\lambda\Psi^{\mu\nu\lambda}$, where $\Psi^{\mu\nu\lambda}$ is antisymmetric with respect to its last two indices, does not change the value of the four-divergence of the tensor since it satisfies an identity $\partial_\nu\partial_\lambda\Psi^{\mu\nu\lambda}=0$ \cite{Landau1989}. Accordingly, the Belinfante-Rosenfeld SEM tensor is given by
\begin{equation}
 (T_\mathrm{BR})^{\mu\nu}=(T_\mathrm{C})^{\mu\nu}+\partial_\lambda\Psi^{\mu\nu\lambda},\hspace{0.5cm}\Psi^{\mu\nu\lambda}=-\Psi^{\mu\lambda\nu}.
 \label{eq:BRSEM0}
\end{equation}
To find the tensor $\Psi^{\mu\nu\lambda}$ to be added to $(T_\mathrm{C})^{\mu\nu}$, in the Belinfante-Rosenfeld symmetrization procedure, one first defines the spin-current tensor $S^{\lambda\nu\mu}$ and then uses it to write $\Psi^{\mu\nu\lambda}$ as follows \cite{Belinfante1940,Bliokh2013b}
\begin{align}
 \partial_\lambda S^{\mu\nu\lambda} &=(T_\mathrm{C})^{\mu\nu}-(T_\mathrm{C})^{\nu\mu},\nonumber\\
 \Psi^{\mu\nu\lambda} &=\frac{1}{2}(S^{\nu\lambda\nu}+S^{\mu\lambda\nu}-S^{\mu\nu\lambda}).
\end{align}
At first sight, this symmetrization procedure is far from being obvious. However, Belinfante and Rosenfeld showed that the resulting SEM tensor $(T_\mathrm{BR})^{\mu\nu}$ is indeed equivalent to the Hilbert SEM tensor in Eq.~\eqref{eq:HSEM}. From this perspective, the Belinfante-Rosenfeld SEM tensor is not a separate SEM tensor alternative, but the two commonly used SEM tensors are the Hilbert SEM tensor and the canonical SEM tensor above, and the Belinfante-Rosenfeld symmetrization procedure presents a connection between these two tensors.

For light propagating inside a medium, the optical-force-mediated coupling between the field and the medium complicates the situation in such a way that the Lagrangian density of the field becomes generally dependent on the four-velocity of the medium \cite{Partanen2019b}. Thus, the variation of the action with respect to the four-potential and keeping the four-velocity of the medium constant does not, without further considerations, produce such a canonical SEM tensor for the field whose four-divergence would give the true optical force density experienced by the medium. Thus, the Belinfante-Rosenfeld type symmetrization may require reconsideration when it is used to describe SEM tensors of general interacting subsystems. This is an interesting topic for a separate work.

\section{\label{apx:energydensity}Calculation of the energy density of the field in the L frame}

Here we calculate the energy density of the field in the L frame in Eq.~\eqref{eq:Wfieldintegral}. The second term of Eq.~\eqref{eq:Wfieldintegral} is the opposite of the energy exchange density calculated in Eq.~\eqref{eq:derivation3}.
The first term of Eq.~\eqref{eq:Wfieldintegral} can be calculated using the Gaussian light pulse described in Sec.~\ref{sec:simulations}, for which the electric field is given in Eqs.~\eqref{eq:electricfieldgeneral} and \eqref{eq:electricfield}. For this term, using the electric field in Eq.~\eqref{eq:electricfield} and the monochromatic field limit approximation of the magnetic field described below Eq.~\eqref{eq:electricfield}, we obtain by analytic integration
\begin{widetext}
\begin{align}
 &-\int_{-\infty}^t\nabla\cdot(\mathbf{E}^\mathrm{(L)}\times\mathbf{H}^\mathrm{(L)})dt'\nonumber\\
 &=W_\mathrm{ex,nd}^\mathrm{(L)}\frac{n_\mathrm{g}^\mathrm{(L)}}{n_\mathrm{p}^\mathrm{(L)}}
 +\frac{\sqrt{\pi}(\omega_0^\mathrm{(L)})^3n_\mathrm{p}^\mathrm{(L)}(n_\mathrm{g}^\mathrm{(L)}\!-\!n_\mathrm{p}^\mathrm{(L)})\varepsilon_0\mathcal{E}^2}{4c\Delta k_0^\mathrm{(L)}}e^{-(k_0^\mathrm{(L)}/\Delta k_0^\mathrm{(L)})^2}
 \Big\{e^{2ik_0^\mathrm{(L)}(n_\mathrm{g}^\mathrm{(L)}-n_\mathrm{p}^\mathrm{(L)})z}\mathrm{erfi}\Big[\frac{k_0^\mathrm{(L)}}{\Delta k_0^\mathrm{(L)}}-in_\mathrm{g}^\mathrm{(L)}\Delta k_0^\mathrm{(L)}\Big(z-\frac{ct}{n_\mathrm{g}^\mathrm{(L)}}\Big)\Big]
 \nonumber\\
 &\hspace{0.4cm}+e^{-2ik_0^\mathrm{(L)}(n_\mathrm{g}^\mathrm{(L)}-n_\mathrm{p}^\mathrm{(L)})z}\mathrm{erfi}\Big[\frac{k_0^\mathrm{(L)}}{\Delta k_0^\mathrm{(L)}}+in_\mathrm{g}^\mathrm{(L)}\Delta k_0^\mathrm{(L)}\Big(z-\frac{ct}{n_\mathrm{g}^\mathrm{(L)}}\Big)\Big]\Big\}
 \xrightarrow{\Delta k_0^\mathrm{(L)}\rightarrow0}
 W_\mathrm{ex,nd}^\mathrm{(L)}\frac{n_\mathrm{g}^\mathrm{(L)}}{n_\mathrm{p}^\mathrm{(L)}}
 +\big(\big\langle W_\mathrm{ex,nd}^\mathrm{(L)}\big\rangle\!-\!W_\mathrm{ex,nd}^\mathrm{(L)}\big)\Big(\frac{n_\mathrm{g}^\mathrm{(L)}}{n_\mathrm{p}^\mathrm{(L)}}-1\Big).
 \label{eq:derivation}
\end{align}
Here $\mathrm{erfi}(x)$ is the imaginary error function. The approximation used for the latter terms of Eq.~\eqref{eq:derivation} is obtained in the monochromatic field limit, $\Delta k_0^\mathrm{(L)}\rightarrow0$.

\begin{table*}[ht]
 \setlength{\tabcolsep}{4.5pt}
 \renewcommand{\arraystretch}{2.4}
 \caption{\label{tbl:transformations}
 Summary of the Lorentz transformations of the scalar and three-dimensional vector quantities of the present work from the G frame to the G$'$ frame. The G$'$ frame is moving in the G frame at a constant relative velocity $\mathbf{v}$. The number current $\boldsymbol{\Gamma}$ corresponding to the number density $n$ moving with velocity $\mathbf{u}$ is defined as $\boldsymbol{\Gamma}=n\mathbf{u}$. Thus, the Lorentz transformation of the velocity $\mathbf{u}$ can be obtained by dividing the Lorentz transformation of $\boldsymbol{\Gamma}$ side by side by the Lorentz transformation of $n$, resulting in $\mathbf{u}'=-(\mathbf{v}\ominus\mathbf{u})$, where $\ominus$ denotes the relativistic velocity subtraction, given in the caption of Table \ref{tbl:parameters}. The velocities $\mathbf{v}_\mathrm{a}$, $\mathbf{v}_\mathrm{a0}$, $\mathbf{v}_\mathrm{p}$, and $\mathbf{v}_\mathrm{g}$ satisfy this transformation. The quantities $(\phi/c,\mathbf{f})$, $(ct,\mathbf{r})$, $(\omega/c,\mathbf{k})$, $(nc,\boldsymbol{\Gamma})$, and $(E/c,\mathbf{p})$ are four-vectors. The number currents corresponding to the number densities $n_\mathrm{a}$, $n_\mathrm{a0}$, and $n_\mathrm{a,min}$ of the present work are $\boldsymbol{\Gamma}_\mathrm{a}=n_\mathrm{a}\mathbf{v}_\mathrm{a}$, $\boldsymbol{\Gamma}_\mathrm{a0}=n_\mathrm{a0}\mathbf{v}_\mathrm{a0}$, and $\boldsymbol{\Gamma}_\mathrm{a,min}=n_\mathrm{a,min}\mathbf{v}_\mathrm{a0}$. The transformations of the mass densities $\rho_\mathrm{a}=n_\mathrm{a}E_\mathrm{a}/c^2$, $\rho_\mathrm{a0}=n_\mathrm{a0}E_\mathrm{a0}/c^2$, and $\rho_\mathrm{a,min}=n_\mathrm{a,min}E_\mathrm{a0}/c^2$ follow from the transformations of the number density and single-atom energy \cite{Penfield1967}, using the nonequilibrium and equilibrium single-atom energies $E_\mathrm{a}=\gamma_{\mathbf{v}_\mathrm{a}}m_0c^2$ and $E_\mathrm{a0}=\gamma_{\mathbf{v}_\mathrm{a0}}m_0c^2$ and momenta $\mathbf{p}_\mathrm{a}=\gamma_{\mathbf{v}_\mathrm{a}}m_0\mathbf{v}_\mathrm{a}$ and $\mathbf{p}_\mathrm{a0}=\gamma_{\mathbf{v}_\mathrm{a0}}m_0\mathbf{v}_\mathrm{a0}$. The subscripts $\parallel$ and $\perp$ in this table denote parallel and perpendicular components to the velocity $\mathbf{v}$, defined for the generic vector $\mathbf{F}$ by $\mathbf{F}_\parallel=(\mathbf{F}\cdot\hat{\mathbf{v}})\hat{\mathbf{v}}$ and $\mathbf{F}_\perp=\mathbf{F}-(\mathbf{F}\cdot\hat{\mathbf{v}})\hat{\mathbf{v}}$, where $\hat{\mathbf{v}}$ is the unit vector parallel to $\mathbf{v}$.}
\begin{tabular}{|c|}
   \hline
   Electric field and magnetic flux density \\[4pt]
   \hline
 $\mathbf{E}'=\mathbf{E}_\parallel+\gamma_\mathbf{v}(\mathbf{E}_\perp+\mathbf{v}\times\mathbf{B})$\\
 $\mathbf{B}'=\mathbf{B}_\parallel+\gamma_\mathbf{v}\Big(\mathbf{B}_\perp-\dfrac{\mathbf{v}\times\mathbf{E}}{c^2}\Big)$\\[4pt]
   \hline
 \end{tabular}
 \hspace{0.2cm}
 \begin{tabular}{|c|}
   \hline
   Electric flux density and magnetic field \\[4pt]
   \hline
 $\mathbf{D}'=\mathbf{D}_\parallel+\gamma_\mathbf{v}\Big(\mathbf{D}_\perp+\dfrac{\mathbf{v}\times\mathbf{H}}{c^2}\Big)$\\
 $\mathbf{H}'=\mathbf{H}_\parallel+\gamma_\mathbf{v}(\mathbf{H}_\perp-\mathbf{v}\times\mathbf{D})$\\[4pt]
   \hline
 \end{tabular}
\hspace{-1.5cm}
 \newline
 \vspace*{0.2cm}
 \newline 
 \begin{tabular}{|c|}
   \hline
   Power-conversion and force densities \\[4pt]
   \hline
 $\phi'=\gamma_\mathbf{v}(\phi-\mathbf{v}\cdot\mathbf{f})$\\
 $\mathbf{f}'=\mathbf{f}_\perp+\gamma_\mathbf{v}\Big(\mathbf{f}_\parallel-\dfrac{\mathbf{v}\phi}{c^2}\Big)$\\[4pt]
   \hline
 \end{tabular}
 \hspace{0.2cm}
 \begin{tabular}{|c|}
   \hline
   Time and position \\[4pt]
   \hline
 $t'=\gamma_\mathbf{v}\Big(t-\dfrac{\mathbf{v}\cdot\mathbf{r}}{c^2}\Big)$\\
 $\mathbf{r}'=\mathbf{r}_\perp+\gamma_\mathbf{v}(\mathbf{r}_\parallel-\mathbf{v}t)$\\[4pt]
   \hline
 \end{tabular}
 \hspace{0.2cm}
 \begin{tabular}{|c|}
   \hline
   Angular frequency and wave vector \\[4pt]
   \hline
 $\omega'=\gamma_\mathbf{v}(\omega-\mathbf{v}\cdot\mathbf{k})$\\
 $\mathbf{k}'=\mathbf{k}_\perp+\gamma_\mathbf{v}\Big(\mathbf{k}_\parallel-\dfrac{\mathbf{v}\omega}{c^2}\Big)$\\[4pt]
   \hline
 \end{tabular}
 \newline
 \vspace*{0.2cm}
 \newline 
 \hspace*{-1cm}
 \begin{tabular}{|c|}
   \hline
   Number density and number current \\[4pt]
   \hline
 $n'=\gamma_\mathbf{v}\Big(n-\dfrac{\mathbf{v}\cdot\boldsymbol{\Gamma}}{c^2}\Big)$\\
 $\boldsymbol{\Gamma}'=\boldsymbol{\Gamma}_\perp+\gamma_\mathbf{v}(\boldsymbol{\Gamma}_\parallel-\mathbf{v}n)$\\[4pt]
   \hline
 \end{tabular}
 \hspace{0.2cm}
 \begin{tabular}{|c|}
   \hline
   Energy and momentum \\[4pt]
   \hline
 $E'=\gamma_\mathbf{v}(E-\mathbf{v}\cdot\mathbf{p})$\\
 $\mathbf{p}'=\mathbf{p}_\perp+\gamma_\mathbf{v}\Big(\mathbf{p}_\parallel-\dfrac{\mathbf{v}E}{c^2}\Big)$\\[4pt]
   \hline
 \end{tabular}
 \hspace{0.2cm}
 \begin{tabular}{|c|}
   \hline
   Angular and boost momenta \\[4pt]
   \hline
 $\mathbf{J}'=\mathbf{J}_\parallel+\gamma_\mathbf{v}(\mathbf{J}_\perp+\mathbf{v}\times\mathbf{N})$\\
 $\mathbf{N}'=\mathbf{N}_\parallel+\gamma_\mathbf{v}\Big(\mathbf{N}_\perp-\dfrac{\mathbf{v}\times\mathbf{J}}{c^2}\Big)$\\[4pt]
   \hline
 \end{tabular}
 \vspace{-0.1cm}
\end{table*}

\section{\label{apx:stresstensor}Calculation of the stress tensor of the field in the L frame}

Here we calculate the stress tensor of the field in the L frame in Eq.~\eqref{eq:stressfieldintegral}. By using the Gaussian light pulse described in Sec.~\ref{sec:simulations}, the integral in Eq.~\eqref{eq:stressfieldintegral} can be calculated analytically with the same approximations as in Appendix \ref{apx:energydensity}, as
\begin{align}
 \boldsymbol{\mathcal{T}}_\mathrm{field}^\mathrm{(L)}
 &=-\int_{-\infty}^z\Big(\mathbf{f}_\mathrm{opt}^\mathrm{(L)}+\frac{\partial\mathbf{G}_\mathrm{field}^\mathrm{(L)}}{\partial t}\Big)dz'\otimes\hat{\mathbf{v}}_\mathrm{p}^\mathrm{(L)}
 =-\frac{n_\mathrm{p}^\mathrm{(L)}n_\mathrm{g}^\mathrm{(L)}}{c^2}\int_{-\infty}^z\frac{\partial}{\partial t}(\mathbf{E}^\mathrm{(L)}\times\mathbf{H}^\mathrm{(L)})dz'\otimes\hat{\mathbf{v}}_\mathrm{p}^\mathrm{(L)}\nonumber\\
 &=\frac{n_\mathrm{p}^\mathrm{(L)}}{c}(\mathbf{E}^\mathrm{(L)}\times\mathbf{H}^\mathrm{(L)})\otimes\hat{\mathbf{v}}_\mathrm{p}^\mathrm{(L)}
 -\frac{\sqrt{\pi}(\omega_0^\mathrm{(L)})^3(n_\mathrm{p}^\mathrm{(L)})^2(n_\mathrm{g}^\mathrm{(L)}-n_\mathrm{p}^\mathrm{(L)})\varepsilon_0\mathcal{E}^2}{4c\,n_\mathrm{g}^\mathrm{(L)}\Delta k_0^\mathrm{(L)}}e^{-[n_\mathrm{p}^\mathrm{(L)}k_0^\mathrm{(L)}/(n_\mathrm{g}^\mathrm{(L)}\Delta k_0^\mathrm{(L)})]^2}\nonumber\\
 &\hspace{0.4cm}\times 
 \Big\{e^{2i\omega_0^\mathrm{(L)}(1-n_\mathrm{p}^\mathrm{(L)}/n_\mathrm{g}^\mathrm{(L)})t}\mathrm{erfi}\Big[\frac{n_\mathrm{p}^\mathrm{(L)}k_0^\mathrm{(L)}}{n_\mathrm{g}^\mathrm{(L)}\Delta k_0^\mathrm{(L)}}-in_\mathrm{g}^\mathrm{(L)}\Delta k_0^\mathrm{(L)}\Big(z-\frac{ct}{n_\mathrm{g}^\mathrm{(L)}}\Big)\Big]
 \nonumber\\
 &\hspace{0.4cm}+e^{-2i\omega_0^\mathrm{(L)}(1-n_\mathrm{p}^\mathrm{(L)}/n_\mathrm{g}^\mathrm{(L)})t}\mathrm{erfi}\Big[\frac{n_\mathrm{p}^\mathrm{(L)}k_0^\mathrm{(L)}}{n_\mathrm{g}^\mathrm{(L)}\Delta k_0^\mathrm{(L)}}+in_\mathrm{g}^\mathrm{(L)}\Delta k_0^\mathrm{(L)}\Big(z-\frac{ct}{n_\mathrm{g}^\mathrm{(L)}}\Big)\Big]\Big\}\otimes\hat{\mathbf{v}}_\mathrm{p}^\mathrm{(L)}\nonumber\\
 &\xrightarrow{\Delta k_0^\mathrm{(L)}\rightarrow0}
 W_\mathrm{ex,nd}^\mathrm{(L)}\hat{\mathbf{v}}_\mathrm{p}^\mathrm{(L)}\otimes\hat{\mathbf{v}}_\mathrm{p}^\mathrm{(L)}
 -\big(\big\langle W_\mathrm{ex,nd}^\mathrm{(L)}\big\rangle-W_\mathrm{ex,nd}^\mathrm{(L)}\big)\Big(\frac{n_\mathrm{g}^\mathrm{(L)}}{n_\mathrm{p}^\mathrm{(L)}}-1\Big)\hat{\mathbf{v}}_\mathrm{p}^\mathrm{(L)}\otimes\hat{\mathbf{v}}_\mathrm{p}^\mathrm{(L)}
 \label{eq:derivation2}
\end{align}
\end{widetext}
The approximation used for the latter terms of Eq.~\eqref{eq:derivation2} is obtained in the monochromatic field limit, $\Delta k_0^\mathrm{(L)}\rightarrow0$. For the term $(\mathbf{E}^\mathrm{(L)}\times\mathbf{H}^\mathrm{(L)})\otimes\hat{\mathbf{v}}_\mathrm{p}^\mathrm{(L)}$, we have used the relation $\mathbf{E}^\mathrm{(L)}\times\mathbf{H}^\mathrm{(L)}=W_\mathrm{ex,nd}^\mathrm{(L)}\mathbf{v}_\mathrm{p}^\mathrm{(L)}$ following from Eq.~\eqref{eq:phasevelocityL} and $\mathbf{v}_\mathrm{p}^\mathrm{(L)}=\hat{\mathbf{v}}_\mathrm{p}^\mathrm{(L)}c/n_\mathrm{p}^\mathrm{(L)}$.

\section{\label{apx:transformations}Lorentz transformations of the quantities in the SEM tensor components}

The Lorentz transformations of the quantities used in the present work from the G frame to the G$'$ frame are summarized in Table \ref{tbl:transformations}. The Lorentz transformations of the electric and magnetic fields and flux densities are also directly related to the Lorentz transformations of the second-rank electromagnetic field tensor in Eq.~\eqref{eq:Ftensor} and the electromagnetic displacement tensor in Eq.~\eqref{eq:Dtensor} \cite{Landau1989,Jackson1999}.

\section{\label{apx:observer}Energy and momentum of light measured by a general inertial observer}

The energies and momenta of the field, MDW, and the coupled MP state of light as a function of the velocity of an inertial observer moving in the L frame parallel to the propagation velocity of light are obtained as volume integrals of the corresponding densities. In the limit described in Eq.~\eqref{eq:accuracycondition}, we then obtain
\begin{align}
 E_\mathrm{MP} &=\gamma_v\Big(1-\dfrac{v}{n_\mathrm{g}^\mathrm{(L)}c}\Big)n_\mathrm{p}^\mathrm{(L)}n_\mathrm{g}^\mathrm{(L)}E_0,\nonumber\\
 p_\mathrm{MP} &=\gamma_v\Big(1-\dfrac{n_\mathrm{g}^\mathrm{(L)}v}{c}\Big)n_\mathrm{p}^\mathrm{(L)}p_0,
\end{align}
\begin{align}
 E_\mathrm{field} &=\gamma_v\dfrac{1-\dfrac{2v}{n_\mathrm{g}^\mathrm{(L)}c}+\dfrac{n_\mathrm{p}^\mathrm{(L)}v^2}{n_\mathrm{g}^\mathrm{(L)}c^2}}{1-\dfrac{v}{n_\mathrm{g}^\mathrm{(L)}c}}E_0,\nonumber\\
 p_\mathrm{field} &=\gamma_v\dfrac{\dfrac{1}{n_\mathrm{g}^\mathrm{(L)}}-\dfrac{(n_\mathrm{g}^\mathrm{(L)}+n_\mathrm{p}^\mathrm{(L)})v}{n_\mathrm{g}^\mathrm{(L)}c}+\dfrac{v^2}{n_\mathrm{g}^\mathrm{(L)}c^2}}{1-\dfrac{v}{n_\mathrm{g}^\mathrm{(L)}c}}p_0,
\end{align}
\begin{align}
 E_\mathrm{MDW} &=\gamma_v\dfrac{1-\dfrac{2v}{n_\mathrm{g}^\mathrm{(L)}c}}{1-\dfrac{v}{n_\mathrm{g}^\mathrm{(L)}c}}(n_\mathrm{p}^\mathrm{(L)}n_\mathrm{g}^\mathrm{(L)}-1)E_0,\nonumber\\
 p_\mathrm{MDW} &=\gamma_v\dfrac{\dfrac{1}{n_\mathrm{g}^\mathrm{(L)}}-\dfrac{v}{c}+\dfrac{v^2}{n_\mathrm{g}^\mathrm{(L)}c^2}}{1-\dfrac{v}{n_\mathrm{g}^\mathrm{(L)}c}}(n_\mathrm{p}^\mathrm{(L)}n_\mathrm{g}^\mathrm{(L)}-1)p_0.
\end{align}
Here $E_0=E_\mathrm{ex}^\mathrm{(L)}$ is the exploitable energy of light in the L frame corresponding to observer velocity $v=0$ and $p_0=E_0/c$. The exploitable energy of light is given by
\begin{equation}
 E_\mathrm{ex}=\gamma_v\Big(1-\dfrac{n_\mathrm{p}^\mathrm{(L)}v}{c}\Big)E_0.
\end{equation}

\section{\label{apx:comparison}Comparison to SEM tensors of previous works on the MP theory of light}

To enable direct comparison of the SEM tensors between the present work and the previous work in Ref.~\cite{Partanen2019a}, it is convenient to present the SEM tensor of the field $\mathbf{T}_\mathrm{field,nd}$ of the present work for nondispersive media by writing it in a form
\begin{align}
 &\mathbf{T}_\mathrm{field,nd}\nonumber\\
 &=\bigg[\begin{array}{cc}
  \frac{1}{2}(\mathbf{E}\!\cdot\!\mathbf{D}\!+\!\mathbf{H}\!\cdot\!\mathbf{B}) & \frac{1}{c}(\mathbf{E}\!\times\!\mathbf{H})^T\\
  \frac{1}{c}\mathbf{E}\!\times\!\mathbf{H} & \frac{1}{2}(\mathbf{E}\!\cdot\!\mathbf{D}\!+\!\mathbf{H}\!\cdot\!\mathbf{B})\mathbf{I}-\mathbf{E}\!\otimes\!\mathbf{D}-\mathbf{H}\!\otimes\!\mathbf{B}
 \end{array}\bigg]\nonumber\\
 &\hspace{0.5cm}-\Delta\mathbf{T}.
 \label{eq:EMSEMGnd}
\end{align}
Here the first term on the right is the SEM tensor of the field in Eq.~(6) of Ref.~\cite{Partanen2019a} and the power-conversion-related difference term $\Delta\mathbf{T}$ is given by
\begin{equation}
 \Delta\mathbf{T} =\dfrac{W_\phi}{c^2}\bigg[\begin{array}{cc}
  c^2 & c\mathbf{v}_\mathrm{a0}^T\\
  c\mathbf{v}_\mathrm{a0} & \dfrac{\gamma_{\mathbf{v}_\mathrm{a}}\mathbf{v}_\mathrm{a}-\gamma_{\mathbf{v}_\mathrm{a0}}\mathbf{v}_\mathrm{a0}}{\gamma_{\mathbf{v}_\mathrm{a}}-\gamma_{\mathbf{v}_\mathrm{a0}}}\otimes\mathbf{v}_\mathrm{a0}
 \end{array}\bigg].
\end{equation}
In the L frame, where $\mathbf{v}_\mathrm{a0}^\mathrm{(L)}=\mathbf{0}$, $\Delta\mathbf{T}$ is equal to $\mathbf{T}_\mathrm{int}$ in Eq.~\eqref{eq:TintL} when we set $n_\mathrm{p}^\mathrm{(L)}=n_\mathrm{g}^\mathrm{(L)}$.

Correspondingly, the SEM tensor of the MDW of the present work is obtained for nondispersive media from the SEM tensor of the MDW in Eq.~(7) of Ref.~\cite{Partanen2019a} by writing
\begin{align}
 \mathbf{T}_\mathrm{MDW,nd}&=
 \bigg[\begin{array}{cc}
  \rho_\mathrm{MDW,l}c^2 & \rho_\mathrm{MDW,l}c\mathbf{v}_\mathrm{l}^T\\
  \rho_\mathrm{MDW,l}c\mathbf{v}_\mathrm{l} & \rho_\mathrm{MDW,l}\mathbf{v}_\mathrm{a}\otimes\mathbf{v}_\mathrm{l}
 \end{array}\bigg]
 +\Delta\mathbf{T}.
 \label{eq:MDWSEMGnd}
\end{align}
The first term on the right in Eq.~\eqref{eq:MDWSEMGnd} corresponds to the SEM tensor of the MDW in the last form of Eq.~(7) of Ref.~\cite{Partanen2019a}. In  Eq.~\eqref{eq:MDWSEMGnd}, we denote the MDW mass density used in Ref.~\cite{Partanen2019a} by $\rho_\mathrm{MDW,l}$. It is related to the MDW mass density of the present work as
\begin{equation}
 \rho_\mathrm{MDW,l}c^2=\rho_\mathrm{MDW}c^2-W_\phi=\gamma_{\mathbf{v}_\mathrm{a}}m_0c^2(n_\mathrm{a}-n_\mathrm{a0}).
 \label{eq:rhomdwl}
\end{equation}
This relation of $\rho_\mathrm{MDW,l}$ to the atomic number densities was not explicitly given in Ref.~\cite{Partanen2019a}. To enable detailed comparison of Eq.~\eqref{eq:MDWSEMGnd} with the MDW SEM tensor in Eq.~\eqref{eq:MDWSEMG}, one must express $\mathbf{v}_\mathrm{l}$ in Eq.~\eqref{eq:MDWSEMGnd} in terms of the atomic velocities and number densities as
\begin{equation}
 \mathbf{v}_\mathrm{l}=\frac{n_\mathrm{a}\mathbf{v}_\mathrm{a}-n_\mathrm{a0}\mathbf{v}_\mathrm{a0}}{n_\mathrm{a}-n_\mathrm{a0}}.
\end{equation}
One can show using Eqs.~\eqref{eq:rhoasol}, \eqref{eq:atomicvelocitysol}, and \eqref{eq:namin} that $\mathbf{v}_\mathrm{l}$ is equal to the phase and group velocities of light in a nondispersive medium.

Since the term $\Delta\mathbf{T}$ appears with different sign in $\mathbf{T}_\mathrm{field,nd}$ and $\mathbf{T}_\mathrm{MDW,nd}$ in Eqs.~\eqref{eq:EMSEMGnd} and \eqref{eq:MDWSEMGnd}, it cancels from the SEM tensor of the coupled MP state of light, which is the sum $\mathbf{T}_\mathrm{MP,nd}=\mathbf{T}_\mathrm{field,nd}+\mathbf{T}_\mathrm{MDW,nd}$. Thus, $\mathbf{T}_\mathrm{MP,nd}$ is equal to the MP SEM tensor in Eq.~(14) of Ref.~\cite{Partanen2019a}. For facilitating the comparison of the results of Ref.~\cite{Partanen2019a} and the present work, we also point out that the Lorentz transformation of the modified MDW mass density in Eq.~\eqref{eq:rhomdwl} above, used in Ref.~\cite{Partanen2019a}, is a special case of the Lorentz transformations of the number densities and the atomic energy presented in Table \ref{tbl:transformations} leading to Eq.~(36) of Ref.~\cite{Partanen2019a}.

The MP SEM tensor satisfies the Lorentz transformation of second-rank tensors in Eq.~\eqref{eq:LorentzT}. The same applies to the SEM tensors of the field and the MDW in the present work. In contrast, due to the difference $\pm\Delta\mathbf{T}$ in the SEM tensor definitions of the field and the MDW in the present work and in Ref.~\cite{Partanen2019a}, the SEM tensors of the field and the MDW given in Ref.~\cite{Partanen2019a} did not separately satisfy the Lorentz transformation of second-rank tensors in Eq.~\eqref{eq:LorentzT}. The difference $\pm\Delta\mathbf{T}$ also explains why the SEM tensors of the field and the MDW in Ref.~\cite{Partanen2019a} were asymmetric in the G frame, while the corresponding SEM tensors of the present work are symmetric in all inertial frames.

In summary, the comparison of the present results with Ref.~\cite{Partanen2019a} emphasizes the need to retain the strict definition of the SEM tensor in Eq.~\eqref{eq:emt} not only for the coupled state of light but also separately for its field and medium parts. Thereby, the \emph{symmetry of the SEM tensors} of the total system and subsystems can be preserved in all inertial frames.

\end{document}